\documentclass[12pt]{article}
\usepackage{amssymb,amsmath,epsfig}

\renewcommand{\theequation}{\arabic{section}.\arabic{equation}}

\begin{document}

\title{\bf Wave Properties of Isothermal \\Magneto-Rotational Fluids}

\author{\small{M. Sharif \thanks{msharif@math.pu.edu.pk} and Umber Sheikh}\\
\small{Department of Mathematics, University of the Punjab,}\\
\small{Quaid-e-Azam Campus Lahore-54590, Pakistan.}}
\date{}
\maketitle

\begin{abstract}
In this paper, the isothermal plasma wave properties in the
neighborhood of the pair production region for the Kerr black hole
magnetosphere are discussed. We have considered the Fourier
analyzed form of the perturbed general relativistic
magnetohydrodynamical equations whose determinant leads to a
dispersion relation. For the special scenario, the $x$-component
of the complex wave vectors are numerically calculated. Respective
components of the propagation vector, attenuation vector, phase
and group velocities are shown in graphs. We have particularly
investigated the existence of a Veselago medium and wave behavior
(modes of waves dispersion).
\end{abstract}

\section{Introduction}

The sky is full of fascinating cosmic objects including planets,
stars and compact objects like black holes. A black hole is a region
of space where the pull of gravity is so strong the even light
cannot escape. Black holes are described in the framework of
relativistic astrophysics. General Relativity leads to insight and
understanding of black hole physics.

Rotating black holes exist in the nuclei of galaxies. The
gravitational pull of the black hole drags the material (highly
magnetized plasma) from nearby stars and forms an accretion disk as
it spirals inwards and finally disappears into the black hole.
Highly magnetized plasma surrounding the black hole horizon is
considered as a magneto-rotational fluid governed by the general
relativistic magnetohydrodynamical (GRMHD) equations due to its
motion under immense gravity. The magnetohydrodynamical waves
produced in these magneto-rotational fluids not only help to confirm
the existence of a black hole but also transmit the information
inside the magnetosphere. In addition, these show the response of
black holes to the external perturbations. The 3+1 ADM formalism
\cite{ADM} is frequently used to discuss the space plasmas into the
arena of black holes. Thorne and Macdonald \cite{TM}-\cite{TPM}
dealt the electromagnetic fields of the black hole theory with 3+1
formalism. Later, this formalism was used by Holcomb and Tajima
\cite{HT}, Holcomb \cite{Ho} and Dettmann et al. \cite{De} to study
wave propagation in the early universe. Buzzi et al.
\cite{BHT1,BHT2} used the 3+1 GRMHD equations to study one
dimensional transverse and longitudinal waves in two-component
plasma in the vicinity of the Schwarzschild horizon.

There exists large body of literature \cite{Pu}-\cite{Ko} which
indicates keen interest in plasmas, plasma processes and plasma
waves for the magnetospheres of compact objects. Punsley \cite{Pu}
discussed the electromagnetism and plasma waves created in the
ingoing and outgoing perfect MHD ergospheric winds. Beskin et al.
\cite{Bg} gave the comprehensive analysis of the pulsar
magnetosphere and discussed the pair creation region around a black
hole. Park and Vishniac \cite{PaV} included the effects of plasma
accretion flows on the mass and angular momentum-loss rates in
addition to the pure electromagnetic extraction of energy from the
rotating black hole. Takahashi et al. \cite{T1} focused on the
Alfven surface. By considering the cold limit, trans-Alfvenic
solutions are characterized and MHD inflow is discussed in relation
to the extraction of energy from the rotating black hole and the
accretion of winds or jets.

Ruffini and Wilson \cite{RW} proposed the idea of energy extraction
from the Kerr black hole. Blandford and Znajek \cite{BZ} were the
first to construct a global model for the magnetosphere of a Kerr
black hole in the force-free limit. They demonstrated that energy
could be extracted under certain conditions in the form of a
Poynting flux. They applied their model to radio loud AGNs, and
proposed that extragalactic jets are the consequence of this energy
extraction mechanism \cite{Beg}. Several authors \cite{several}
discussed different aspects of Blandford - Znajek process.

Energy extraction is discussed by different authors \cite{Li1}
assuming the magnetic field lines connected with the transition
region instead of connecting with remote loads inside the accretion
disk. Zhang \cite{Z1,Z2} formulated the 3+1 black hole theory for
stationary symmetric GRMHD and applied it to cold plasma filled Kerr
magnetosphere. Using the interface conditions defined at the pair
production region, he obtained the numerical solutions which show
that outflow of energy flux is possible for the specific angular
frequency and specific $x$-component of the wave vector.

Lee et al. \cite{Lee1}-\cite{Lee3} considered the Blandford-Znajek
process as one of the viable models of powering the gamma-ray bursts
effectively. The gamma ray bursts is a consequence of interaction of
magnetic field and plasma. The observations of a broad Fe K$\alpha$
line in the bright Seyfert 1 galaxy MCG-6-30-15 and M87
\cite{Wilm}-\cite{Ju} suggest evidences of the extraction of
rotational energy from a Kerr black hole by a magnetic field. Thus,
the popularity of the GRMHD behavior of plasma has increased and the
study of GRMHD waves become more important. Koide et al. \cite{Ko}
discussed plasma flowing into a rapidly rotating black hole by using
numerical simulations via GRMHD.

Sharif et al. \cite{S15a} considered the cold and isothermal plasmas
in the vicinity of a Schwarzschild event horizon. Plasma wave
properties are investigated using 3+1 formalism. The results are
generalized by considering the Schwarzschild geometry as a
restricted Kerr model. We \cite{Ut}-\cite{S7} have investigated cold
plasma wave properties (real and complex wave numbers) and
isothermal plasma wave properties (only real wave numbers) \cite{S8}
for Kerr planar analogue.

This paper is devoted to discuss the isothermal plasma wave
properties by using complex wave numbers. We consider an isothermal
plasma in the vicinity of the Kerr magnetosphere's pair production
region. The corresponding dispersion relations are obtained by using
Fourier analyzed perturbed GRMHD equations. We then evaluate wave
numbers which are used to find $x$-components of propagation and
attenuation vectors, phase and group velocities. These quantities
help to investigate the properties of plasma.

The organization of the paper is given as follows: Next section
provides the background metric with flow assumptions and respective
physical interpretation. Section \textbf{3} contains the Fourier
analyzed GRMHD equations for isothermal plasma in Kerr planar
analogue. Section \textbf{4} is devoted to the numerical solutions
of the dispersion relation including Figures and related discussion.
We shall summarize this discussion in Section \textbf{5}.

\section{3+1 Background and Relative GRMHD Equations}

The general line element in ADM 3+1 formalism can be written as
\cite{Z2}
\begin{eqnarray}{\setcounter{equation}{1}}\label{1.10.1}
ds^2&=&-\alpha^2dt^2+\eta_{ij}(dx^i+\beta^idt)(dx^j+\beta^jdt),
\end{eqnarray}
where $\alpha$ is the time lapse and $\bar{\beta}$ is the
three-dimensional shift vector and $\eta_{ij}$ are the components of
three-dimensional hypersurface metric. The quantities
$\alpha,~\beta_i,~\eta_{ij},~(i,j=1,2,3)$ are dependent on
coordinates $(t,~x_i)$.

We consider the planar analogue of the Kerr metric (Eq.(2.2) of
\cite{S8})
\begin{equation}\label{st2}
ds^2=-dt^2+(dx+\beta(z)dt)^2+dy^2+dz^2.
\end{equation}
Here the lapse function is set to unity and the shift vector $\beta$
is along $x$-direction only. The details of the analogue metric are
mentioned in Section \textbf{2} of \cite{S8}. In this planar
analogue (Eq.(\ref{st2})), we consider a rotating isothermal plasma
admitting the following equation of state
\begin{equation}
\label{ip} \mu=\frac{\rho+p}{\rho_0}=constant,
\end{equation}
where $\mu,~\rho_0,~\rho$ and $p$ (all are taken to be constants)
are the specific enthalpy, rest and moving mass densities and
pressure of the fluid respectively. This highly magnetized plasma
includes the effects of gravity (due to nearest correspondence to
the black hole). Thus perfect GRMHD equations express it completely.

For the line element given by Eq.(\ref{1.10.1}), the perfect GRMHD
equations (Maxwell's equations of magnetic field evolution, local
conservation laws of mass, momentum and energy in force-free
magnetosphere respectively) are \cite{Ut}
\begin{eqnarray}
\label{pm2}
&&\frac{d\textbf{B}}{d\tau}+\frac{1}{\alpha}\textbf{B}.\nabla
\bar{\beta}+\theta
\textbf{B}=\frac{1}{\alpha}\nabla\times(\alpha\textbf{V}\times\textbf{B}),
\end{eqnarray}
\begin{eqnarray}
&&\frac{D\rho_0}{D\tau}+\rho_0\gamma^2\textbf{V}.\frac{D\textbf{V}}{D\tau}+
\frac{\rho_0}{\alpha}\left\{\frac{g_{,t}}{2g}
+\nabla.(\alpha\textbf{V}-\bar{\beta})\right\}=0,\\
&&\left\{\left(\rho_0\gamma^2\mu+\frac{\textbf{B}^2}{4\pi}\right)\eta_{ij}
+\rho_0\gamma^4\mu
V_iV_j-\frac{1}{4\pi}B_iB_j\right\}\frac{DV^j}{D\tau}
+\rho_0\gamma^2V_i\frac{D\mu}{D\tau}\nonumber\\
&&-\left(\frac{\textbf{B}^2}{4\pi}\eta_{ij}
-\frac{1}{4\pi}B_iB_j\right){V^j}_{|k}V^k =-\rho_0\gamma^2\mu
\left\{a_i-\frac{1}{\alpha}\bar{\beta}_{j|i}V^j
-(\pounds_t\eta_{ij})V^j\right\}\nonumber\\
&&-p_{|i}+\frac{1}{4\pi}(\textbf{V}\times \textbf{B})_i
\nabla.(\textbf{V}\times \textbf{B}) -\frac{1}{8\pi\alpha^2}(\alpha
\textbf{B})^2_{|i}+\frac{1}{4\pi\alpha}(\alpha
B_i)_{|j}B^j\nonumber\\
&&-\frac{1}{4\pi\alpha}(\textbf{B}\times\{\textbf{V}\times [\nabla
\times(\alpha \textbf{V}\times \textbf{B})-(\textbf{B}.\nabla )
\bar{\beta}]\nonumber\\
&&+(\textbf{V}\times\textbf{B}).\nabla \bar{\beta}
\})_i,\\
\label{pm3}
&&\frac{d}{d\tau}(\rho_0\mu\gamma^2)-\frac{d
p}{d\tau}
+\theta\left((\rho_0\mu\gamma^2-p)+\frac{1}{8\pi}((\textbf{V}\times\textbf{B})^2
+\textbf{B}^2))\right)\nonumber\\
&&+\frac{1}{2\alpha}\{\rho_0\mu\gamma^2V^iV^j+p\eta^{ij}
-\frac{1}{4\pi}((\textbf{V}\times\textbf{B})^i
(\textbf{V}\times\textbf{B})^j+B^iB^j)\nonumber\\
&&+\frac{1}{8\pi}((\textbf{V}\times\textbf{B})^2
+\textbf{B}^2)\eta^{ij}\}\pounds_t\eta_{ij}
+\rho_0\mu\gamma^2\textbf{V}.\textbf{a}+\rho_0\mu\gamma^4\textbf{V}.
(\textbf{V}.\nabla)\textbf{V}\nonumber\\
&&-\frac{1}{2\pi}\textbf{a}.((\textbf{V}\times\textbf{B})\times\textbf{B})
-\frac{1}{4\pi}(\nabla\times(\textbf{V}\times\textbf{B})).\textbf{B}
-\frac{1}{\alpha}\{\rho_0\mu\gamma^2\textbf{V}.(\textbf{V}.\nabla)\bar{\beta}
\nonumber\\
&&+p(\nabla.\bar{\beta})-\frac{1}{4\pi}(\textbf{V}\times\textbf{B}).
((\textbf{V}\times\textbf{B}).\nabla)\bar{\beta}
-\frac{1}{4\pi}\textbf{B}.(\textbf{B}.\nabla)\bar{\beta}\nonumber\\
&&+\frac{1}{8\pi}((\textbf{V}\times\textbf{B})^2
+\textbf{B}^2))(\nabla.\bar{\beta})\}+\frac{1}{4\pi}
\left(\frac{1}{\alpha}(\textbf{V}\times\textbf{B}).\{(\textbf{V}\times\textbf{B})
.\nabla\}\bar{\beta}
\right.\nonumber\\
&&\left.+(\textbf{V}\times\textbf{B}).(\textbf{a}\times\textbf{B})
+2(\textbf{V}\times\textbf{B}).\frac{d}{d\tau}(\textbf{V}\times\textbf{B})
\bar{\beta}\right.\nonumber\\
&&\left.+\theta(\textbf{V}\times\textbf{B}).(\textbf{V}\times\textbf{B})
+\textbf{B}.\frac{d\textbf{B}}{d\tau}\right)=0,
\end{eqnarray}
where $\textbf{B},~\textbf{V},~\textbf{a}$ respectively denote the
three dimensional magnetic field, velocity and acceleration of the
fluid, $\theta$ gives the expansion of the fluid and
$\frac{d}{d\tau}$ is time derivative measured by the fiducial
observer, $\frac{D}{D\tau}$ is the time derivative with respect to
motion of the fluid, $\pounds_t$ denotes the Lie derivative with
respect to time parameter $t$. The quantity $g$ is the determinant
of the three dimensional hypersurface metric. The respective form of
Eqs.(\ref{pm2})-(\ref{pm3}) for the Kerr geometry is given in
\textbf{Appendix A}.

\section{Plasma Discussion and Physical Interpretation}

We are going to perform a local analysis to the plasma waves. For
this purpose, we have assumed that the plasma living in the
spacetime (\ref{st2}) admits a free fall along $z$-direction due to
the black hole's gravity and along $x$-direction due to rotation of
the black hole (in direction of shift vector of our analogue
spacetime). The fluid's velocity ($\textbf{V}$) and magnetic field
($\textbf{B}$) measured by our fiducial observer (a natural observer
associated with the spacetime and locally measures the quantities)
can be described as follows:
\begin{eqnarray}{\setcounter{equation}{1}}
\textbf{V}=V(z)\textbf{e}_\textbf{x}+u(z)\textbf{e}_\textbf{z},\quad\
\textbf{B}=B\{\lambda(z)\textbf{e}_\textbf{x}
+\textbf{e}_\textbf{z}\},
\end{eqnarray}
where $B$ is a constant.

In our stationary symmetric background, the $x$-component of the
velocity vector can be written in the form \cite{Z2} $V=C+\lambda
u,$ where $C\equiv\beta+V_F$ with $V_F$ as an integration
constant. We assume the value of the shift function \cite{Z2}
$\beta=\tanh(z)-1$ and $V_F=1$. Further, $B^2=8\pi$ and
$\lambda=1$ are taken so that the magnetic field becomes constant.
Thus the $x$-component takes the form $V=1+\beta+u=\tanh(z)+u.$
Substituting the value of $V$ in the conservation law of rest-mass
for three-dimensional hypersurface, i.e.,
\begin{equation}\label{clm}
\rho_0 \gamma u=\mu(\rho+p)\gamma u=A~(constant)
\end{equation}
with the assumption that rest-mass density is constant and
$A/\rho_0=1$, we get an equation of the form
$$3u^2+2u\tanh(z)+\tanh^2(z)-1=0$$
quadratic in $u$. The solutions of this equation ($u_1$ and $u_2$)
and the corresponding values of $V$ ($V_1$ and $V_2$) are given as
follows (Eqs.(5.2)-(5.3) of \cite{S8}):
\begin{eqnarray}\label{u1}
\left.
\begin{array}{ccc}
u_1 & = & -\frac{1}{3}\tanh(z)-\frac{1}{3}\sqrt{3-2\tanh^2(z)}, \\
V_1 & = & \frac{2}{3}\tanh(z)-\frac{1}{3}\sqrt{3-2\tanh^2(z)},
\end{array}\right\}\\
\label{u2} \left.
\begin{array}{ccc}
u_2 & = & -\frac{1}{3}\tanh(z)+\frac{1}{3}\sqrt{3-2\tanh^2(z)}, \\
V_2 & = & \frac{2}{3}\tanh(z)+\frac{1}{3}\sqrt{3-2\tanh^2(z)}.
\end{array}\right\}
\end{eqnarray}
The corresponding Poynting vector ($\textbf{S}$) is
\begin{equation}\label{P}
\textbf{S}=\frac{1}{4\pi}\textbf{E}\times\textbf{B}
=2\tanh(z)(\textbf{e}_\textbf{x}-\textbf{e}_\textbf{z}),
\end{equation}
where $\textbf{E}$ is the electric field as measured by the fiducial
observer. This electric field can be determined by the perfect flow
assumption, i.e.,
\begin{equation}
\textbf{E}=-\textbf{V}\times\textbf{B}.
\end{equation}

We have assumed that plasma is produced in the pair production
region lying at $z=0$. A perturbation method is used in which we
perturb the equilibrium by a very small amount, for example, the
background plasma mass density is not the same as $\rho_0$
everywhere but changes slightly with $\delta\rho$. The perturbed
quantities are assumed to be small and we have considered these as
first order quantities. Perturbation assumptions and equations are
given in \textbf{Appendix A}.

The perturbations are produced due to plasma production in the pair
production region. In this region, electron-positron pairs are
produced and taken apart by strong forces of gravity. The electrons
move towards the outer end of the magnetosphere while the positron
move towards the event horizon. Thus the displaced electron results
a positive charge density which is produced at the position from
where electrons are displaced. These positive charges then attract
the electrons. Electrons move back but they overshoot it, and come
back but overshoot again and thus oscillation is produced. Same is
the case of positrons which produce oscillations in the region
towards the event horizon of the black hole.

Here we check the properties of the plasma as a consequence of the
plasma perturbations. For this purpose, we have specifically tried
to over check the regions in the neighborhood of the pair production
region, i.e. $z=0.$ It is important to note that the production of
plasma does not make any change in the background geometry. It only
disturbs the fluid dynamics. Thus, all geometric terms are
considered to 0th order.

\section{Fourier Analyzed Perturbed GRMHD Equations}

We have assumed that the perturbations produced are simple harmonic
waves. We shall discuss the wave properties with the normal mode
analysis (Fourier decomposition of waves). The following assumptions
will be used for the sake of Fourier analysis (Eq.(3.3) of
\cite{S8}).
\begin{eqnarray}{\setcounter{equation}{1}}\label{ps}
\tilde{\rho}(t,x,z)=c_1e^{-i(\omega
t-k_xx-k_zz)},&~&\tilde{p}(t,x,z)=c_6e^{-i(\omega
t-k_xx-k_zz)},\nonumber\\
v_x(t,x,z)=c_2e^{-i(\omega t-k_xx-k_zz)},&~&
v_z(t,x,z)=c_3e^{-i(\omega t-k_xx-k_zz)},\nonumber\\
b_x(t,x,z)=c_4e^{-i(\omega t-k_xx-k_zz)},&~&
b_z(t,x,z)=c_5e^{-i(\omega t-k_xx-k_zz)}.
\end{eqnarray}
Here $c_i,~(i=1,2,...,6)$ are constants, $\textbf{k}=(k_x,0,k_z)$
is the wave vector and $\omega$ is the angular frequency of the
waves.

The wave vector leads to the following quantities:
\begin{itemize}
  \item $Re(\textbf{k})$ gives a vector, called propagation vector,
  that shows three dimensional propagation of waves.
  \item $Im(\textbf{k})$ gives a vector, called attenuation vector,
  that shows damping and growth of waves in three dimensions.
  \item The phase velocity vector gives the speed of phase in
  three dimensions and can be calculated by the formula
  $\frac{\omega}{Re(\textbf{k})}.$
  \item The group velocity vector gives the speed of wave packet
  in three dimensions and can be calculated by the formula
  $\frac{d\omega}{dRe(\textbf{k})}.$
\end{itemize}

We consider the perturbed perfect GRMHD equations written in 3+1
formalism for the Kerr planar analogue with isothermal state of
plasma (Eq.(\ref{ip})) given by Eqs.(\ref{a3})-(\ref{b3}). The
Fourier analyzed perturbed form of these equations is given as
follows (Eqs.(4.6)-(4.12) of \cite{S8}):
\begin{eqnarray}\label{a4}
&&\iota k_z c_2-(\iota k_z\lambda+\lambda')c_3-c_4(\iota k_z
u-\iota \omega+u')+c_5(\vartheta\iota k_z+\vartheta')=0,\\
\label{b4} &&\iota k_x c_2-\iota k_x\lambda c_3+\iota
c_5(\vartheta k_x+k_zu-\omega)=0,\\
\label{c4} &&k_xc_4=-k_zc_5,\\\label{d4}
&&c_1[\iota\rho(-\omega+\vartheta
k_x+uk_z)-(p'u+pu'+pu\gamma^2\varphi)]+c_2(\rho+p)\nonumber\\
&&\times\left[-\eta\gamma^2V+\iota k_z\gamma^2uV+\iota
k_x\pi+\gamma^2u\{(\pi+\gamma^2V^2)V'+2\gamma^2uVu'\}\right]\nonumber\\
&&+c_3(\rho+p)\left[-\eta\gamma^2u+\iota k_x\gamma^2uV+(\iota
k_z-(1-2\gamma^2u^2)\frac{u'}{u})\varrho
+2\gamma^4u^2VV'\right]\nonumber\\
&&+c_6[\iota p(-\omega+\vartheta k_x+uk_z)+(p'u+pu'+pu\gamma^2\varphi)]=0,\\
\label{e4} &&c_1\rho\gamma^2u\{\pi V'+\gamma^2uVu'\}
+c_6p[\gamma^2u\{\pi V'+\gamma^2uVu'\}+\iota
k_x]\nonumber\\
&&+c_2\left[-\eta\left\{(\rho+p)\gamma^2\pi
+\psi\right\}+\zeta\left\{(\rho+p)\gamma^2(1
+\gamma^2V^2)-\psi\right\}\right.\nonumber\end{eqnarray}
\begin{eqnarray}
&&\left.+(\rho+p)\gamma^4u\{(\pi+3\gamma^2V^2)uu'+4\pi VV'\}\right]
+c_3\left[-\eta\left\{(\rho+p)\gamma^4uV
\right.\right.\nonumber\\
&&\left.\left.-\psi\lambda \right\}+\zeta\left\{(\rho+p)\gamma^4uV
+\psi\lambda\right\}+(\rho+p)\gamma^2[2\gamma^2(\varrho
+\gamma^2u^2)uVu'\right.\nonumber\\
&&\left.+\{(\varrho+\gamma^2u^2)(\pi+\gamma^2V^2)
-\gamma^2V^2\}V']+\psi u\lambda'\right]+\psi c_4\{-\iota
k_z(1-u^2)\nonumber\\
&&+uu'\}+\psi c_5\left\{-\lambda'-u\vartheta'+\iota
k_x(1-V^2)-2\iota uVk_z\right\}=0,\\
\label{f4} &&c_1\gamma^2\rho[u\{\varrho
u'+\gamma^2VuV'\}-V\beta']+c_6[\gamma^2p\{\varrho uu'+
\gamma^2Vu^2V'-V\beta'\}\nonumber\\
&&+p'+p\iota
k_z]+c_2\left[-\eta\left\{(\rho+p)\gamma^4uV-\psi\lambda\right\}+\zeta
\left\{(\rho+p)\gamma^4uV\right.\right.\nonumber\\
&&\left.\left.+\psi\lambda
\right\}+(\rho+p)\gamma^2\{\gamma^2u^2V'(\pi+3\gamma^2V^2)
-\beta'(\pi+\gamma^2V^2)+2V\gamma^2uu'\right.\nonumber\\
&&\left.\times(\varrho+\gamma^2u^2)\}\right]
+c_3\left[-\eta\left\{(\rho+p)\gamma^2\varrho+\psi\lambda^2\right\}
+\zeta\left\{(\rho+p)\gamma^2\varrho
-\psi\lambda^2\right\}\right.\nonumber\\
&&\left.+(\rho+p)\gamma^2[u'(1+\gamma^2 u^2)(1+4\gamma^2
u^2)+2u\gamma^2\{(\varrho+\gamma^2u^2)VV'-V\beta'\}]\right.\nonumber\\
&&\left.-\psi\lambda\lambda'u\right]+\psi c_4\left\{\iota
k_z\lambda (1-u^2)+\lambda'-\lambda uu'\right\}+\psi c_5\{2\lambda
uV\iota k_z+\lambda u\vartheta'\nonumber\\
&&-\lambda \iota k_x(1-V^2)]=0,\\
\label{g4} &&c_1\gamma^2[\rho'u-\rho(\iota\omega+u'+2u
\gamma^2\varphi-\gamma^2uV\beta'+\iota
k_x\vartheta+\iota k_zu)]\nonumber\\
&&+c_6[-\iota\omega
p(\gamma^2-1)+\gamma^2\{p'u+p(u'+2u\gamma^2\varphi-\gamma^2uV\beta'+\iota
k_x\vartheta+\iota
k_zu)\}]\nonumber\\
&&+c_2[-\iota\omega\{2(\rho+p)\gamma^4V-\psi\chi\}+\iota
k_x[(\rho+p)\gamma^2\{1+2\gamma^2V\vartheta\}
-\psi\vartheta\chi]\nonumber\\
&&+\iota k_zu\{2(\rho+p)\gamma^4V+\psi\chi\}
+(\rho+p)\gamma^2u\{2\gamma^2V'+6\gamma^4V\varphi-\beta'(\pi\nonumber\\
&&+\gamma^2V^2)\}-\psi\lambda']+c_3[-\iota\omega\{2(\rho+p)\gamma^4u+\psi\lambda
\chi\}+\iota
k_x\vartheta\{2(\rho+p)\gamma^4u\nonumber\\
&&+\psi\lambda\chi\}+\iota
k_z\{(\rho+p)\gamma^2(\varrho+\gamma^2u^2)-\psi\lambda u\chi\}
+(\rho+p)\gamma^2\{-\frac{u'}{u}+2\gamma^2uu'\nonumber\\
&&+6\gamma^4u^2\varphi+\gamma^2\varphi
-V\beta'\varrho\}+\psi\lambda'\{\lambda-u\chi\}]
+c_4\psi[u\{\lambda'-\chi u'\}+\iota k_z\chi\nonumber\\
&&\times(1-u^2)]+c_5\psi \{-\lambda'V+u\chi\vartheta'-\iota
k_x\chi(1-V^2)+2\iota k_z uV\}=0.
\end{eqnarray}
where
$\vartheta=V-\beta,~\varphi=VV'+uu',\chi=u\lambda-V,~\psi=\frac{B^2}{4\pi},
~\varrho=1+\gamma^2u^2,\pi=1+\gamma^2V^2,~\zeta=\iota(k_xV+k_zu)~
\textrm{and}~\eta=\iota(\omega+\beta k_x)$. Here prime denotes the
derivative of the quantity with respect to the variable $z$.

\section{Numerical Solutions}

Equation (\ref{c4}) gives $k_z=-\frac{c_4}{c_5}k_x$ and we will
assume $k_z=-k_x$ for the sake of simplicity. Using this relation
and the assumptions given in paragraphs 1 and 2 of \textbf{Section
3}, we simplify Eqs.(\ref{a4}), (\ref{b4}), (\ref{d4})-(\ref{g4}).
To obtain a dispersion relation, we use the method given in
\cite{Das}. We equate the determinant of the coefficients of
constants ($c_1,~c_2,~c_3,~c_4,~c_5,~c_6$) for Eqs.(\ref{a4}),
(\ref{b4}), (\ref{d4})-(\ref{g4}) to zero. It gives an equation of
the type

\begin{eqnarray}\label{relation}{\setcounter{equation}{1}}
&&A_1(\omega,z){k_x}^6+A_2(\omega,z){k_x}^5+A_3(\omega,z){k_x}^4+
A_4(\omega,z){k_x}^3\nonumber\\
&&+A_5(\omega,z){k_x}^2+A_6(\omega,z)k_x+A_7(\omega,z)=0,
\end{eqnarray}
sextic in $k_x$ which cannot be solved exactly to obtain the values
of $k_x$. Thus we use the software \emph{Mathematica} to solve it by
using the fluid flow velocities (\ref{u1} and \ref{u2}). Related
Mathematica codes are given in \textbf{Appendix B}.

We assume that the pair production region is exactly at $z=0.$ Our
region of consideration is $-5\leq z\leq5$ for wave analysis. This
region is taken only to discuss the waves near the pair production
region as much as possible. Since the flow variables have large
variations in the region $-1<z<1$, we avoid this region and solve
the dispersion relation for $-5\leq z\leq-1$ and $1\leq z\leq5$
and thus we have two meshes (just like the near and far zones in
em-radiation). The mesh $-5\leq z\leq-1$ indicates the
neighborhood of the pair production region towards the event
horizon and the mesh $1\leq z\leq5$ represents the neighborhood of
the pair production region towards the outer end of the
magnetosphere. We deal with each mesh separately. For a numerical
solution, we take the step-length $0.2$ for $z$ and $\omega$ and
find the value of $k_x$ at each point
$(z_i,\omega_j),~(i=1,...,21,~j=1,...51)$ of the mesh.

Our dispersion relation (Eq.(\ref{relation})) leads to six complex
values of $k_x$ at each point. Thus $k_x=k_{Rx}+\iota k_{Ix}$. The
real part $k_{Rx}$ represents the $x$-component of the propagation
vector which leads to the $x$-components of the phase and group
velocities. The imaginary part $k_{Ix}$ represents the $x$-component
of the attenuation vector. The $x$-component of the propagation
vector gives the $x$-components of the phase velocity
($v_{px}=\frac{\omega}{k_{Rx}}$) and group velocity
($v_{gx}=\frac{d\omega}{dk_{Rx}}$) of the waves. Each root of the
dispersion relation is separated and numerical interpolation is used
to estimate the surface. The $x$-components of the propagation
vector, attenuation vector, phase and group velocities can be
evaluated from these interpolation functions.

It is mentioned here that we investigate the $x$-component of the
above quantities. The equation $k_x=-k_z$ indicates that the
behavior of the $z$-component of the wave vector is opposite to that
of the $x$-component. If the $x$-component of the wave is normally
dispersed in one region, the $z$-component is anomalously dispersed
and vice versa. When the $x$-components of the propagation vector,
attenuation vector, phase and group velocity vectors increase, their
corresponding $z$-components show a decrease and vice versa. The
components of the propagation and attenuation vectors, the phase and
group velocities lead us to the wave behavior of the Kerr black hole
magnetosphere and the properties of a Veselago medium.

Here, we give a criteria which is used to discuss the two properties
of the medium.
\begin{itemize}
\item A medium is a Veselago if the propagation vector is in opposite
direction to the Poynting vector \cite{Veselago}. In the usual
medium, both these vectors admit same direction.
\item If the phase velocity is greater than the group
velocity, dispersion of waves is normal, anomalous otherwise
\cite{Ac}.
\end{itemize}
We shall use these facts to infer our results.

We have considered two velocities (given by Eqs.(\ref{u1}) and
(\ref{u2})) for the rotating plasma. For each velocity, we obtain
six complex values of the $x$-component of the wave vector. For
fluid flow with velocity components given by (\ref{u1}), Figures
\textbf{1-6} show the dispersion relations in the neighborhood of
the pair production region towards the horizon whereas Figures
\textbf{7-12} indicate these in the neighborhood of the pair
production region towards the outer end of the magnetosphere. For
the velocity components given by Eq.(\ref{u2}), Figures
\textbf{13-18} and \textbf{19-24}, respectively show the dispersion
relations in the neighborhood of the pair production region towards
the event horizon and towards the outer end of the magnetosphere.
All the Figures have graphs \textbf{A}, \textbf{B}, \textbf{C} and
\textbf{D} representing the $x$-components of the propagation and
attenuation vectors, phase and group velocities respectively.

In the following, we shall discuss some of the figures which lead to
important results. A summary of the discussion of results obtained
from the remaining figures is given in the last section.

In Figure \textbf{1A}, the $x$-component of the propagation vector
takes negative values in the regions $-5\leq z\leq-4.598$ and
$-4.4<z\leq-1$ while it has positive values for $-4.598<z\leq-4.4$.
For the region towards the event horizon, the $x$-component of the
Poynting vector, given by Eq.(\ref{P}), admits negative values. When
we compare these two, the $x$-components of the propagation and
Poynting vectors turn out to be in the opposite direction for the
region $-4.598<z\leq-4.4$ and thus the medium is a Veselago. Figure
\textbf{1B} indicates that the $x$-component of the attenuation
vector increases with the increase in angular frequency and
decreases in $z$ in the region $-2.0\leq z\leq
-1.0,~0.175\leq\omega\leq10$. This shows damping of waves with
increase in angular frequency and wave growth with increase in $z$.
The remaining region shows random damping and growth of waves.

Similarly, in Figure \textbf{7A}, the $x$-component of the
propagation vector is negative throughout the region. The
$x$-component of the Poynting vector, given by Eq.(\ref{P}), admits
positive values for the region away from the event horizon. The
opposite direction of the $x$-components of the propagation and the
Poynting vectors indicates the presence of a Veselago medium.

Table 1 gives the figures along with the region where we have found
a Veselago medium, i.e., the the propagation vector is in opposite
direction to the attenuation vector.

The negative phase velocity propagation is a property of the usual
medium as well as Veselago medium. In this medium, the right hand
rule of electromagnetism changes into the left hand rule. This
medium has been demonstrated experimentally \cite{SSS} (the material
formed is called left-handed material or meta-material) and is
characterized by simultaneous negative magnetic permeability and
electric permittivity. It has Poynting vector of a time-harmonic
plane wave directed opposite with respect to the wave vector. This
exhibits unusual properties like support of backward waves and
anomalous negative refraction of plane monochromatic electromagnetic
waves. Thus, backward waves are supported by the plasma in these
regions.

Figures \textbf{1C} and \textbf{1D} indicate that the $x$-component
of the group velocity is greater than the phase velocity in the
region $-1.75\leq z\leq-1.0,~0.5\leq \omega\leq10$ which shows
anomalous dispersion of waves. Rest of the region admits random
points of normal and anomalous dispersion of waves. Table 2 contains
figures with regions of the normal dispersion of waves. The regions
of normal dispersion indicates that the waves can pass through them
without any restriction. Thus the energy flux can pass through these
regions (via waves).

\begin{table}
\begin{tabular}{|c|c|}
  \hline
  \textbf{Figure} & \textbf{Regions of Veselago Medium} \\
  \hline
  \textbf{2A} & $-1.8\leq z\leq-1$ \\
  \hline
  \textbf{3A} & $-2.05\leq z\leq -1.0$ \\
  \hline
  \textbf{4A} & $-2.1\leq z\leq -1.0$ \\
  \hline
  \textbf{5A} & $-5\leq z\leq-1$ \\
  \hline
  \textbf{6A} & $-5\leq z\leq-4.6$ and $-4.375\leq z\leq -1$ \\
  \hline
  \textbf{8A} & $1\leq z\leq5$ \\
  \hline
  \textbf{14A}, \textbf{15A}, \textbf{16A}, \textbf{17A} and \textbf{18A} & $-5\leq z\leq-1$ \\
  \hline
  \textbf{19A} & $0<z\leq4.425$ and $4.585<z\leq5$ \\
  \hline
  \textbf{24A} & $4.42<z\leq4.595$ \\
  \hline
\end{tabular}
\caption{Regions admitting a Veselago medium in respective figures}
\end{table}
\begin{table}
\begin{tabular}{|c|c|}
  \hline
  \textbf{Figure} & \textbf{Regions of Normal Dispersion} \\
  \hline
  \textbf{2C} and \textbf{2D} & $-1.8\leq z\leq -1.0$  \\
  \hline
  \textbf{7C} and \textbf{7D} & $1\leq z\leq5$ \\
  \hline
  \textbf{8C} and \textbf{8D} & $0.5<\omega\leq5$ \\
  \hline
  \textbf{10C} and \textbf{10D} & $0.15\leq\omega\leq10$ \\
  \hline
  \textbf{12C} and \textbf{12D} & $1\leq z\leq 5,~0.3\leq\omega\leq0.401$ \\
  \hline
  \textbf{13C} and \textbf{13D} & $0.1\leq\omega\leq 0.5$ \\
  \hline
  \textbf{14C} and \textbf{14D} & $0\leq\omega<0.5$ \\
  \hline
  \textbf{15C} and \textbf{15D} & \rm{Except~for~negligible~angular~frequency} \\
  \hline
  \textbf{16C} and \textbf{16D} & $-5.0\leq z\leq-1.025,~0.4\leq\omega\leq10$ \\
  \hline
  \textbf{18C} and \textbf{18D} & $1.01\leq\omega\leq2$ \\
  \hline
  \textbf{19C} and \textbf{19D} & $1\leq z\leq2,~0<\omega\leq10$ and $2\leq z\leq2.8,~0.75\leq\omega\leq10$ \\
  \hline
  \textbf{20C} and \textbf{20D} & $1\leq z\leq1.8,~0.5\leq\omega\leq10$ \\
  \hline
\end{tabular}
\caption{Regions of normal dispersion of waves in respective
figures}
\end{table}

\section{Summary}

This paper is devoted to discuss the properties of isothermal plasma
waves by using the perturbed GRMHD equations in 3+1 formalism. These
non-linear ordinary differential equations are simplified by using
the Fourier decomposition of waves (Fourier analysis of waves). We
have computed the dispersion relation for the sinusoidal waves and
solved it to check the properties of the medium. This has been done
in the neighborhood of the pair production region of the Kerr
magnetosphere.

A dispersion relation is found by using \emph{Mathematica} and wave
numbers are deduced in terms of angular frequency. The real part of
the $x$-component of the wave vector gives the $x$-component of the
phase and group velocities which lead to the properties of plasma.
The imaginary part of the $x$-component of the wave vector indicates
the $x$-component of the attenuation vector. All these quantities
are shown in the graphs.

The following two objectives are investigated in this work:
\begin{enumerate}
\item The behavior of GRMHD waves under the influence of
gravity and magnetospheric wind is analyzed (especially dispersion
of waves is considered). This helps us to detect whether the
extraction of energy is possible from the black hole.
\item The existence of a Veselago medium in the black hole regime
is checked.
\end{enumerate}

To obtain these objectives, we considered two values of the flow
velocity. The dispersion relations for each velocity are calculated
and complex $x$-component of the wave vectors are evaluated by using
the relation $k_x=-k_z$. This relation gives us the $z$-component of
the wave vector dependent upon the $x$-component. The direction of
the Poynting and the wave vectors in the region gives an estimate of
the refractive index (positive if both have the same direction and
negative if both have the opposite direction). The effect of
rotation of the black hole yields $k_x$ and the gravitational effect
gives $k_z$. The results can be summarized as follows:

In Figure 1, the medium is usual near the pair production region and
shows anomalous dispersion for most of the part. Figure 2 indicates
that normal dispersion of waves occurs in a Veselago medium near the
pair production region. Figure 3 shows random points of normal and
anomalous dispersion in a Veselago medium living near the pair
production region. Figure 4 shows that there exists a Veselago
medium with random normal and anomalous points of dispersion. In
Figure 5, a Veselago medium is found with random points of normal
and anomalous dispersion in the neighborhood of the pair production
region. Figure 6 shows that there exists a Veselago medium near the
pair production region. Random points of normal and anomalous
dispersion are found away from the pair production region.

Figure 7 indicates the presence of a Veselago medium along with the
normal dispersion of waves. Figure 8 represents most of the region
admitting normal dispersion for the waves with high frequencies in a
Veselago medium. In Figure 9, anomalous dispersion of waves in a
usual medium is found throughout the region whereas Figure 10 shows
that the waves with higher angular frequency are normally dispersed
in a usual medium. Figure 11 represents a usual medium admitting
anomalous dispersion of waves in most of the region. In Figure 12,
small regions admitting normal dispersion of waves are found in
usual medium.

In Figure 13, a very small region shows normal dispersion of waves
in usual medium. In Figure 14, presence of a Veselago medium is
confirmed. Normal dispersion occurs for the waves with low angular
frequency, anomalous otherwise. A Veselago medium admitting normal
dispersion, except for low frequency waves, is found in Figure 15.
In Figure 16, a Veselago medium exists throughout the region.
Moreover, a small region near the pair production region admits
anomalous dispersion. Most of the region shows normal dispersion.
Figure 17 indicates anomalous dispersion of waves in a Veselago
medium for most of the region. In Figure 18 random points of
normal and anomalous dispersion are found in a Veselago medium.

In Figure 19, a large and a small region exhibit the properties of a
Veselago medium. Moreover, near the pair production region normal
dispersion of waves exists. Figure 20 shows that most of the region
near the pair production region admits normal dispersion of waves in
a usual medium. In Figure 21, random points of normal and anomalous
dispersion are found. Further, small regions of a Veselago medium
are present. In Figure 22, we obtain the usual medium near the pair
production region. Random points of normal and anomalous dispersion
are found. Figure 23 indicates that the medium is usual admitting
anomalous dispersion of waves near the pair production region.
Figure 24 shows anomalous dispersion of waves in the usual medium
near the pair production region.

From the Figure analysis, we arrive at the following conclusions.
\begin{enumerate}
\item We can conclude that the magnetosphere allows the waves to
disperse normally towards the event horizon as well as the outer
end of the magnetosphere. Thus negative energy flux can fall
inside the black hole and rotational energy can be extracted in
response.
\item It does not matter whether the region lies towards the event
horizon or towards the outer end of the magnetosphere, the
presence of a Veselago medium only depends on the rotation of the
background.
\end{enumerate}

When we compare the results with the previous literature, we obtain
the following:
\begin{itemize}
\item On comparing our results with \cite{S7}, only Figures 7 and 14
show that energy can be extracted by inflow of negative energy
influx. In this work, we have obtained Figures 15, 7, 10 and 20
which show normal dispersion of waves. This indicates that there are
more chances of waves to move towards the outer end of the
magnetosphere (Figures 7, 10 and 20). Consequently there are more
chances of the extraction of energy.
\item Here we obtain clear cut conditions to show that
extraction of energy is possible. In \cite{S8}, only inflow of the
waves towards the event horizon is possible and their outflow
towards the outer end of the magnetosphere is not possible. Thus the
exchange of energy from the event horizon with the accreting
material was expected to be allowed. However, here we have obtained
clear evidences that the energy can be extracted from the event
horizon, passes through the accretion disk and can move towards the
outer end of the magnetosphere as given by observational
identifications in \cite{Wilm}-\cite{Ju}. This does not negate the
work given in \cite{S8}, but generalizes that idea.
\end{itemize}

%\vspace{0.5cm}
\newpage
{\bf Acknowledgment}

\vspace{0.5cm}

We appreciate the Higher Education Commission Islamabad, Pakistan,
for its financial support during this work through the {\it
Indigenous PhD 5000 Fellowship Program Batch-II}. Fruitful
discussion with Prof. Asghar Qadir is also appreciated.

\renewcommand{\theequation}{A\arabic{equation}}
\setcounter{equation}{0}
\section*{Appendix A}
This Appendix includes the GRMHD equations and their perturbed
form by using some assumptions. The component form of these
equations is also given.

The set of perfect GRMHD equations for isothermal plasma living in
Kerr planar analogue (with rotation assumed to be along
x-direction) can be written as (Eqs.(4.1)-(4.5) of \cite{S8})
\begin{eqnarray}{\setcounter{equation}{1}}
\label{a1} &&\frac{d\textbf{B}}{d\tau}+(\textbf{B}.\nabla)\beta
=\nabla\times(\textbf{V}\times\textbf{B}),\\
\label{b1}
&&\nabla.\textbf{B}=0,\\
\label{c1}
&&\frac{D(\rho+p)}{D\tau}+(\rho+p)\gamma^2\textbf{V}.\frac{D\textbf{V}}{D\tau}+
(\rho+p)\nabla.\textbf{V}=0,\\
\label{d1} &&\left[\left\{(\rho+p)\gamma^2
+\frac{\textbf{B}^2}{4\pi}\right\}\delta_{ij}+(\rho+p)\gamma^4
V_iV_j-\frac{1}{4\pi}B_iB_j\right]\frac{dV^j}{d\tau}\nonumber\\
&&+(\rho+p)\gamma^2V_{i,k}V^k+(\rho+p)\gamma^4V_iV_j{V^j}_{,k}V^k
=(\rho+p)\gamma^2\beta_{j,i}V^j\nonumber\\
&&-p_{,i}+\frac{1}{4\pi}(B_{i,j}-B_{j,i})B^j
-\frac{1}{4\pi}\left\{\textbf{B}\times\left(\textbf{V}\times
\frac{d\textbf{B}}{d\tau}\right)\right\}_i,\\
\label{e1} &&\gamma^2\textbf{V}.\frac{D}{D\tau}(\rho+p)
+2(\rho+p)\gamma^4\textbf{V}.\frac{D\textbf{V}}{D\tau}
-\frac{dp}{d\tau}+(\rho+p)\gamma^2(\nabla.\textbf{V})\nonumber\\
&&-(\rho+p)\gamma^2\textbf{V}.(\textbf{V}.\nabla)\beta
+\frac{1}{4\pi}\left[(\textbf{V}\times\textbf{B}).(\nabla
\times\textbf{B})\right.\nonumber\\
&&\left.+(\textbf{V}\times\textbf{B}).\frac{d}{d\tau}(\textbf{V}\times\textbf{B})
+(\textbf{V}\times\textbf{B}).\{(\textbf{V}\times\textbf{B}).\nabla\}\beta\right]=0,
\end{eqnarray}
with $\textbf{V}$, $\textbf{B}$ and $\textbf{E}$ are fiducial
observer (FIDO) measured fluid velocity, magnetic and electric
fields respectively, $\frac{d}{d \tau}\equiv
\frac{\partial}{\partial t}-\beta.\nabla$ is the FIDO measured
rate of change of any three-dimensional vector in absolute space,
$\gamma$ is the Lorentz factor and
$\frac{D}{D\tau}\equiv\frac{d}{d\tau}+\textbf{V}.\nabla
=\frac{\partial}{\partial t}+(\textbf{V}-\beta).\nabla$ is the
time derivative moving along the fluid.

The perturbed variables take the following form
\begin{eqnarray}
\label{pv}
\rho=\rho^0+\rho\tilde{\rho},&~&p=p^0+p\tilde{p},\nonumber\\
\textbf{V}=\textbf{V}^0+\textbf{v},&~&
\textbf{B}=\textbf{B}^0+B\textbf{b},
\end{eqnarray}
where the dimensionless perturbed quantities are
\begin{eqnarray}
\label{p} \tilde{\rho}&\equiv&\frac{\delta
\rho}{\rho}=\tilde{\rho}(t,x,z),\quad
\tilde{p}\equiv\frac{\delta p}{p}=\tilde{p}(t,x,z),\nonumber\\
\textbf{v}&\equiv& \delta
\textbf{V}=v_x(t,x,z)\textbf{e}_\textbf{x}+v
_z(t,x,z)\textbf{e}_\textbf{z},\nonumber\\
\textbf{b}&\equiv& \frac{ \delta
\textbf{B}}{B}=b_x(t,x,z)\textbf{e}_\textbf{x}
+b_z(t,x,z)\textbf{e}_\textbf{z}.
\end{eqnarray}

When we introduce the perturbations from Eq.(\ref{pv}), the
linearized GRMHD Eqs.(\ref{a1})-(\ref{e1}) become
\begin{eqnarray}\label{a2}
&&\left\{(\frac{\partial}{\partial t}-\beta.\nabla)(\delta
\textbf{B})\right\}=\nabla\times(\textbf{v}\times\textbf{B})
+\nabla\times(\textbf{V}\times\delta\textbf{B})-(\delta\textbf{B}.\nabla)\beta,\\
\label{b2}
&&\nabla.(\delta \textbf{B})=0,\\
\label{c2} &&\left\{\frac{\partial}{\partial
t}+(\textbf{V}-\beta).\nabla\right\}(\delta\rho+\delta
p)+(\rho+p)\gamma^2\textbf{V}.\left\{\frac{\partial}{\partial
t}+(\textbf{V}-\beta).\nabla\right\}\textbf{v}\nonumber\\
&&+(\rho+p)(\nabla.\textbf{v})+(\delta\rho+\delta
p)(\nabla.\textbf{V})+(\delta\rho+\delta
p)\gamma^2\textbf{V}.(\textbf{V}.\nabla)\textbf{V}
\nonumber\\
&&=-2(\rho+p)\gamma^2(\textbf{V}.\textbf{v})(\textbf{V}.\nabla)\ln
\gamma-(\rho+p)\gamma^2(\textbf{V}.\nabla\textbf{V}).\textbf{v}\nonumber\\
&&+(\rho+p)\textbf{v}.\nabla \ln u,\\
\label{d2}
&&\left[\left\{(\rho+p)\gamma^2+\frac{\textbf{B}^2}{4\pi}\right\}\delta_{ij}
+(\rho+p)\gamma^4 V_iV_j-\frac{1}{4\pi}B_i
B_j\right]\left(\frac{\partial}{\partial t}-\beta.\nabla
\right)v^j\nonumber\\
&&+(\rho+p)\gamma^2v_{i,j}V^j+(\rho+p)\gamma^4V_iv_{j,k}V^jV^k+\frac{1}{4\pi}
\left[\textbf{B}\times\left\{\textbf{V}\times\frac{d(\delta\textbf{B)}}
{d\tau}\right\}\right]_i\nonumber\\
&&-\frac{1}{4\pi}\left\{(\delta B_i)_{,j}-(\delta
B_j)_{,i}\right\}B^j=-(\delta p)_{,i}+\gamma^2[(\delta\rho+\delta
p)V^j\nonumber\\
&&+2(\rho+p)\gamma^2(\textbf{V}.\textbf{v})V^j+(\rho+p)v^j]\beta_{j,i}
+\frac{1}{4\pi}(B_{i,j}-B_{j,i})\delta B^j\nonumber\\
&&-(\rho+p)\gamma^4(v_iV^j+v_jV^i)V_{k,j}V^k-\gamma^2\{(\delta
\rho+\delta
p)V^j\nonumber\\
&&+2(\rho+p)\gamma^2(\textbf{V}.\textbf{v})V^j+(\rho+p)
v^j\}V_{i,j}-\gamma^4V_i\{(\delta\rho+\delta
p)V^j\nonumber\\
&&+4(\rho+p)\gamma^2(\textbf{V}.\textbf{v})V^j
+(\rho+p)v^j\}V_{j,k}V^k,
\end{eqnarray}
\begin{eqnarray}
\label{e2} &&\gamma^2\left\{\frac{\partial}{\partial
t}+(\textbf{V}-\beta).\nabla\right\}(\delta\rho+\delta
p)+2(\rho+p)\gamma^4\textbf{V}.\left\{\frac{\partial}{\partial
t}\right.\nonumber\\
&&\left.+(\textbf{V}-\beta).\nabla\right\}\textbf{v}-(\rho+p)\gamma^2\textbf{v}.\nabla
\ln u+(\rho+p)\gamma^4\textbf{V}.(\textbf{v}.\nabla)\textbf{V}\nonumber\\
&&+6(\rho+p)\gamma^6(\textbf{V}.\textbf{v})\textbf{V}.(\textbf{V}.\nabla)\textbf{V}
+2(\delta\rho+\delta
p)\gamma^4\textbf{V}.(\textbf{V}.\nabla)\textbf{V}\nonumber\\
&& +2(\rho+p)\gamma^4\textbf{v}.(\textbf{V}.\nabla)\textbf{V}
-2(\rho+p)\gamma^4(\textbf{V}.\textbf{v})\textbf{V}.\nabla \ln
u+2(\rho+p)\gamma^4\nonumber\\
&&\times(\textbf{V}.\textbf{v})
(\nabla.\textbf{V})-\frac{\partial}{\partial
t}(\delta{p})+(\delta\rho+\delta
p)\gamma^2(\nabla.\textbf{V})+(\rho+p)\gamma^2(\nabla.\textbf{v})
\nonumber\\
&&-\gamma^2(\beta.\nabla)(\delta\rho+\delta
p)+2(\rho+p)\gamma^4(\textbf{V}.\textbf{v})(\beta.\nabla \ln
u)-6(\rho+p)\nonumber\\
&&\times\gamma^6(\textbf{V}.\textbf{v})\textbf{V}
.(\beta.\nabla)\textbf{V}
-2(\rho+p)\gamma^4\textbf{v}.(\beta.\nabla)\textbf{V}-2(\delta\rho+\delta
p)\gamma^4\textbf{V}.(\beta.\nabla)\textbf{V}
\nonumber\\
&&-(\delta\rho+\delta
p)\gamma^2\textbf{V}.(\textbf{V}.\nabla)\beta-(\rho+p)\gamma^2\textbf{V}.
(\textbf{v}.\nabla)\beta\nonumber\\
&&-2(\rho+p)\gamma^4(\textbf{V}.\textbf{v})
\textbf{V}.(\textbf{V}.\nabla)\beta+\frac{1}{4\pi}\left[(\textbf{v}\times\textbf{B}).(\nabla\times
\textbf{B})\right.\nonumber\\
&&\left.+(\textbf{V}\times\delta\textbf{B}).(\nabla
\times\textbf{B})+(\textbf{V}\times\textbf{B})
.(\nabla\times\delta\textbf{B})\right.\nonumber\\
&&\left.+(\textbf{V}
\times\textbf{B}).\left\{\frac{d\textbf{v}}{d\tau}\times\textbf{B}+\textbf{V}
\times \frac{d \delta \textbf{B}}{d\tau}\right\}\right]=0.
\end{eqnarray}
These equations show the change in the behavior of moving mass,
velocity and magnetic field of the fluid when the perturbations
are involved. The component form of Eqs.(\ref{a2})-(\ref{b2}) are
given as follows.
\begin{eqnarray}\label{a3}
&&\frac{db_x}{d\tau}+Vb_{x,x}+ub_{x,z}=-u'b_x
+\vartheta'b_z+v_{x,z}-\lambda v_{z,z}-\lambda'v_z,\\
\label{b3} &&\frac{db_z}{d\tau}+Vb_{z,x}+ub_{z,z}=\lambda
v_{z,x}-v_{x,x},\\\label{c3} &&b_{x,x}+b_{z,z}=0,\\
\label{d3}
&&\rho\frac{d\tilde{\rho}}{d\tau}+p\frac{d\tilde{p}}{d\tau}+\rho
V\tilde{\rho}_{,x}+pV\tilde{p}_{,x}+\rho
u\tilde{\rho}_{,z}+pu\tilde{p}_{,z}
-(\tilde{\rho}-\tilde{p})\{p'u+pu'\nonumber\\
&&+pu\gamma^2(\varphi)\}+(\rho+p)\gamma^2\left(V\frac{dv_x}{d\tau}
+u\frac{dv_z}{d\tau}\right)+(\rho+p)\{\pi v_{x,x}+\varrho v_{z,z}\nonumber\\
&&+uV\gamma^2(v_{x,z}+v_{z,x})\}=-(\rho+p)\gamma^2u[(\pi+\gamma^2V^2)V'+2\gamma^2uVu']v_x\nonumber\\
&&+(\rho+p)[(1-2\gamma^2u^2)\varrho\frac{u'}{u}-2\gamma^4u^2VV']v_z,\\
\label{e3}
&&\left\{(\rho+p)\gamma^2\pi+\psi\right\}\frac{dv_x}{d\tau}
+\left\{(\rho+p)\gamma^4uV-\psi\lambda
\right\}\frac{dv_z}{d\tau}-2\psi uVb_{z,z}\nonumber\\
&&+\left\{(\rho+p)\gamma^2\pi-\psi\right\}(Vv_{x,x}+uv_{x,z})
+\left\{(\rho+p)\gamma^4uV+\psi\lambda \right\}\nonumber\\
&&\times(Vv_{z,x}+uv_{z,z})+\psi\{(1-V^2)b_{z,x}-(1-u^2)b_{x,z}\}=-\psi uu'b_x\nonumber\\
&&+\psi\{\lambda'+u\vartheta'\}b_z
-p\tilde{p}_{,x}-(\rho\tilde{\rho}+p\tilde{p})\gamma^2
u\{(1+\gamma^2 V^2)V'+\gamma^2uVu'\}\nonumber
\end{eqnarray}
\begin{eqnarray}
&&-(\rho+p)\gamma^4 u\{(1+4\gamma^2V^2)uu'+4\pi
VV'\}v_x-[(\rho+p)\gamma^2\nonumber\\
&&\times[\{(1+2\gamma^2 u^2)(1+2\gamma^2 V^2)-\gamma^2
V^2\}V'\nonumber\\
&&+2\gamma^2(1+2\gamma^2 u^2)uVu']+\psi u\lambda']v_z,\\\label{f3}
&&\left\{(\rho+p)\gamma^2\varrho
+\psi\lambda^2\right\}\frac{dv_z}{d\tau}
+\left\{(\rho+p)\gamma^4uV
-\psi\lambda \right\}\frac{dv_x}{d\tau}\nonumber\\
&&+\left\{(\rho+p)\gamma^2\varrho -\psi\lambda^2\right\}(Vv_{z,x}
+uv_{z,z})+2\psi\lambda uVb_{z,z}\nonumber\\
&&+\left\{(\rho+p)\gamma^4uV +\psi\lambda
\right\}(Vv_{x,x}+uv_{x,z})-\psi\lambda \{(1-V^2)b_{z,x}\nonumber\\
&&-(1-u^2)b_{x,z}\}=-\psi(\lambda'-\lambda uu')b_x
-\psi\lambda u\vartheta'b_z-\rho\tilde{\rho}\gamma^2\{uu'\varrho\nonumber\\
&&+\gamma^2u^2VV'-V\beta'\}
-[p\tilde{p}_{,z}+p'\tilde{p}+p\tilde{p}\gamma^2\{uu'\varrho+\gamma^2u^2VV'-V\beta'\}]\nonumber\\
&&-(\rho+p)\gamma^2[\gamma^2u^2(1+4\gamma^2V^2)V'-(1+2\gamma^2
V^2)\beta'\nonumber\\
&&+2\gamma^2uV(1+2\gamma^2
u^2)u']v_x-[(\rho+p)\gamma^2\{-2\gamma^2uV\beta'+\varrho(1+4\gamma^2u^2)u'\nonumber\\
&&+2\gamma^2(\varrho+\gamma^2u^2)uVV'
\}-\psi\lambda u\lambda']v_z,\\
\label{g3} &&\gamma^2\rho\frac{\partial \tilde{\rho}}{\partial
t}+p(\gamma^2-1)\frac{\partial \tilde{p}}{\partial
t}+\tilde{\rho}\gamma^2\{\rho'u+\rho u'+2\rho u
\gamma^2\varphi-\rho
uV\beta'\}+\gamma^2\rho\tilde{\rho}_{,x}\vartheta\nonumber\\
&&
+\gamma^2u\rho\tilde{\rho}_{,z}+\tilde{p}\gamma^2\{up'+u'p+2pu\gamma^2\varphi
-p\gamma^2uV\beta'\}+\gamma^2p\tilde{p}_{,x}\vartheta
+\gamma^2up\tilde{p}_{,z}\nonumber\\
&&+\frac{\partial v_x}{\partial
t}\{2(\rho+p)\gamma^4V-\psi\chi\}+\frac{\partial v_z}{\partial
t}\{2(\rho+p)\gamma^4u+\psi\lambda \chi\}\nonumber\\
&&+v_{x,x}[(\rho+p)\gamma^2\{1+2\gamma^2V\vartheta\}
-\psi\vartheta\chi]+v_{x,z}u[2(\rho+p)\gamma^4V
+\psi\chi]\nonumber\\
&&+v_{z,x}\vartheta[2(\rho+p)\gamma^4u
+\psi\lambda\chi]+v_{z,z}\{(\rho+p)(\varrho+\gamma^2u^2)
-\psi\lambda u\chi\}\nonumber\\
&&+\psi\chi \{(1-u^2)b_{x,z}-(1-V^2)b_{z,x}
+2uVb_{z,z})\}\nonumber\\
&&+\psi ub_x(\lambda'-\chi u') +\psi
b_z(-\lambda'V+u\chi\vartheta')+v_x[(\rho+p)\gamma^2u\{2\gamma^2V'\nonumber\\
&&+6\gamma^4V\varphi-\beta'(\pi+\gamma^2V^2)\}
-\psi\lambda']+v_z[\psi\lambda'(\lambda-u\chi)
\nonumber\\
&&+(\rho+p)\gamma^2\{-\frac{u'}{u}+2\gamma^2uu'+6\gamma^4u^2\varphi+\gamma^2\varphi-V\beta'\varrho\}]=0.
\end{eqnarray}
We have used the conservation law of rest-mass for three-dimensional
hypersurface given by Eq.(\ref{clm}) to simplify Eq.(\ref{d3}).

\renewcommand{\theequation}{B\arabic{equation}}
\setcounter{equation}{0}
\section*{Appendix B}

Here we include the \emph{Mathematica} program used to calculate a
dispersion relation. This dispersion relation is obtained from the
real part of the determinant.

In the start, we must specify relative assumptions given in Section
\textbf{3} to simplify the dispersion relations. In the following,
we only consider the velocity components given by Eq.(\ref{u1}). We
shall change the values of $u,~V,~u',~V'$ to obtain the dispersion
relations for the velocity components given by Eq.(\ref{u2}).

We have used the notations $T,~S,~k,~l$ for $\tanh(z),~\sec
h(z),~k_x,~k_z$ respectively. The components $g_{ij}$ are the
coefficients of $c_j$'s in respective equations Eqs.(\ref{a4}),
(\ref{c4}), (\ref{d4})-(\ref{g4}).\\
$\\
\rho = 1/2 \\
\rho' = D[\rho,z]\\
  p = 1/2\\
  p' = D[p, z]\\
  l = -k\\
  T = m\\
  S = Sqrt[1-m^2]\\
  \beta =(T-1)\\
  \beta'=S^2\\
  u = 1/3~(-T-(3~S^2+T^2))\\
  u' = 1/3~(-S^2+(2~S^2~T)/(3~S^2+T^2))\\
  V = 1/3~(2~T-(3~S^2+T^2))\\
  V' = 1/3~(2~S^2+(2~S^2~T)/(3~S^2+T^2))\\
 \lambda = 1\\
  \lambda' = D[\lambda, z]\\
  \gamma = 1/(1 - u^2 - V^2)\\
  \rm{Simplify}[\gamma]\\
  \\
b=Det[\left(
  \begin{array}{cccccc}
    g_{11} & g_{12} & g_{13} & g_{14} & g_{15} & g_{16} \\
    g_{21} & g_{22} & g_{23} & g_{24} & g_{25} & g_{26} \\
    g_{31} & g_{32} & g_{33} & g_{34} & g_{35} & g_{36} \\
    g_{41} & g_{42} & g_{43} & g_{44} & g_{45} & g_{46} \\
    g_{51} & g_{52} & g_{53} & g_{54} & g_{55} & g_{56} \\
    g_{61} & g_{62} & g_{63} & g_{64} & g_{65} & g_{66} \\
  \end{array}
\right)]\\
\\
\rm{Expand}[\%]\\
$\\
The determinant of the coefficients of the constants
$c_1,~c_2,~c_3,~c_4,~c_5$ and $c_6$ from Eqs.(\ref{a4}), (\ref{b4}),
(\ref{d4})-(\ref{g4}) gives a dispersion relation sextic in $k_x$.
This equation cannot be solved analytically for the exact solutions
and is solved numerically. Here we have given commands used to solve
the dispersion relation only for the region towards the event
horizon. Similar commands with the replacement of $\{z,-5,-1,0.2\}$
by $\{z,1,5,0.2\}$ can be used to calculate the roots for the region
away from the event horizon. The commands are given as
follows:\\
$\\
\rm{R}=\rm{NSolve}[b==0,k];\\
\\
\rm{Table}[\{z,\omega,Part[R,1]\},\{z,-5,-1,0.2\},\{\omega,0,10,0.2\}]\\
\rm{Table}[\{z, \omega,Part[R,2]\},\{z,-5,-1,0.2\},\{\omega,0,10,0.2\}]\\
\rm{Table}[\{z,\omega,Part[R,3]\},\{z,-5,-1,0.2\},\{\omega,0,10,0.2\}]\\
\rm{Table}[\{z,\omega,Part[R,4]\},\{z,-5,-1,0.2\},\{\omega,0,10,0.2\}]\\
\rm{Table}[\{z,\omega,Part[R,5]\},\{z,-5,-1,0.2\},\{\omega,0,10,0.2\}]\\
$\\
Afterwards, each root is separated in the form of arrays for further
manipulation. Each root is approximated for each point of the
two-dimensional meshes $-5\leq z\leq-1,~0\leq\omega\leq 10$ and
$1\leq z\leq5,~0\leq\omega\leq 10$, with equal step lengths $0.2$
for $z$ and $\omega$. The numerical values in each region are used
to approximate interpolation functions.

The data for each root is converted and separately dealt in
Mathematica by the following commands. The graph of the root
admitting real values at each point of the mesh can be generated by
the following procedure to obtain Figure 1 (e.g.).\\
$\\\rm{datapts}=\{\{a_{11},a_{12},...a_{1~51}\},\{a_{21},a_{22},...a_{2~51}\}
,...\{a_{26~1},a_{26~2},...a_{26~51}\}\};\\
\\
\rm{r}=\rm{ListInterpolation}[\rm{datapts},\{\{-5,-1\},\{0,10\}\}]\\
\\
\rm{p}=\rm{Re}[\rm{r}[\rm{x,y}]];\\
\\
\rm{q}=\rm{Im}[\rm{r}[\rm{x,y}]];\\
\\
\rm{p1}=\rm{Plot3D}[\rm{p},\{z,-5,-1\},\{\omega,0,10\},
\rm{AxesLabel}\rightarrow\{"z","\omega", "~"\}]\\
\\
\rm{p2}=\rm{Plot3D}[\rm{q},\{z,-5,-1\},\{\omega,0,10\},
\rm{AxesLabel}\rightarrow\{"z", "\omega", "~"\}]\\
\\
\rm{p3}=\rm{Plot3D}[\frac{\textrm{y}}{\textrm{p}},\{z,-5,-1\},\{\omega,0,10\},
\rm{AxesLabel}\rightarrow\{"z","\omega","~"\}]\\
\\
\rm{l}=\rm{D}[\rm{r}[\rm{x,y}],\rm{y}];\\
\\
\rm{p4}=\rm{Plot3D}[\frac{1}{\rm{Re[l]}},\{z,-5,-1\},\{\omega,0,10\},
\rm{AxesLabel}\rightarrow\{"z","\omega","~"\}]\\
$\\

These commands generate $x$-components of the propagation and
attenuation vectors, phase and group velocities for the specific set
of data points in the form of graphs. These graphs are used to
investigate the properties of the medium.

\newpage

$\begin{array}{ccccc}
&A&B&C&D\\
1&\epsfig{file=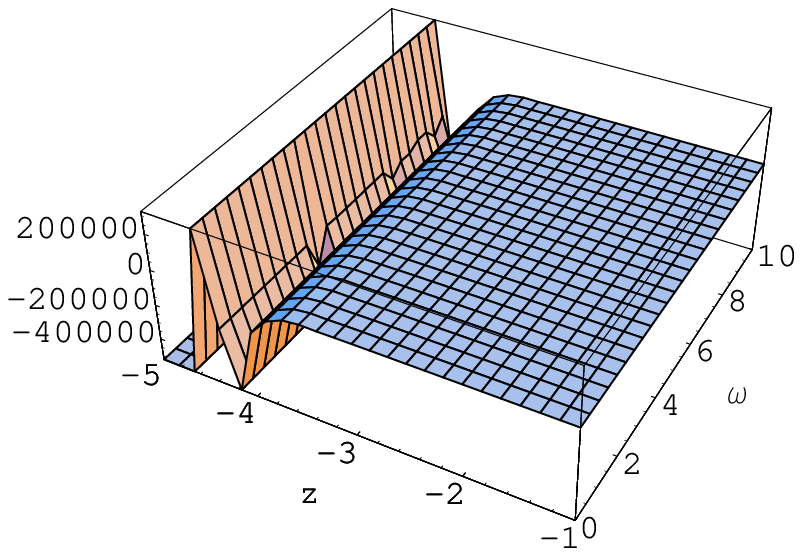,width=0.215\linewidth}
&\epsfig{file=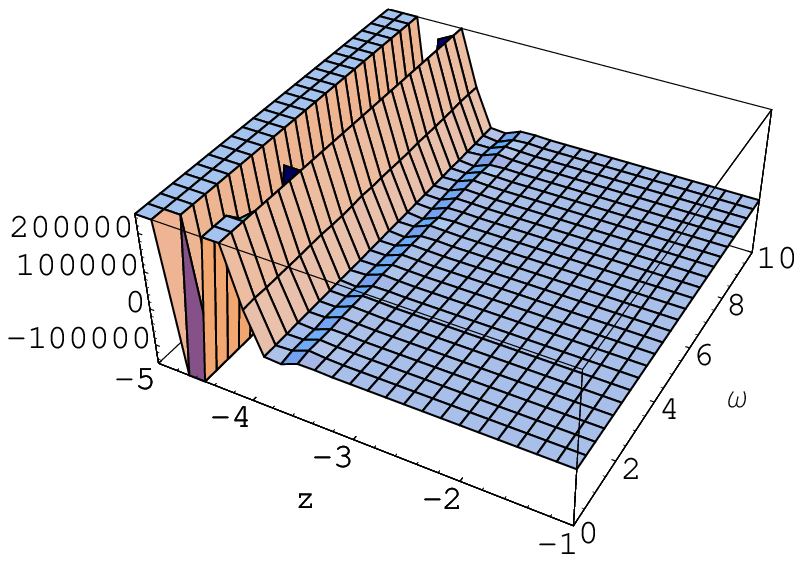,width=0.215\linewidth}
&\epsfig{file=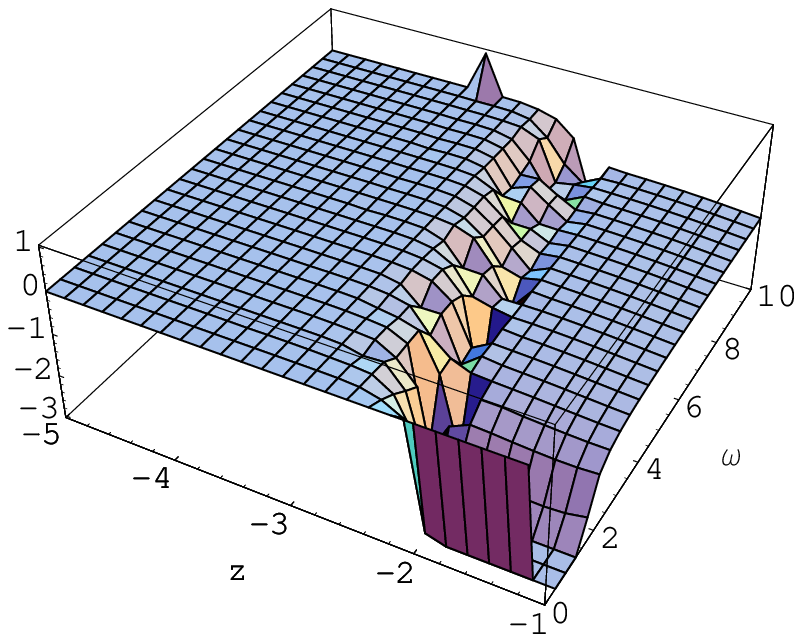,width=0.215\linewidth}
&\epsfig{file=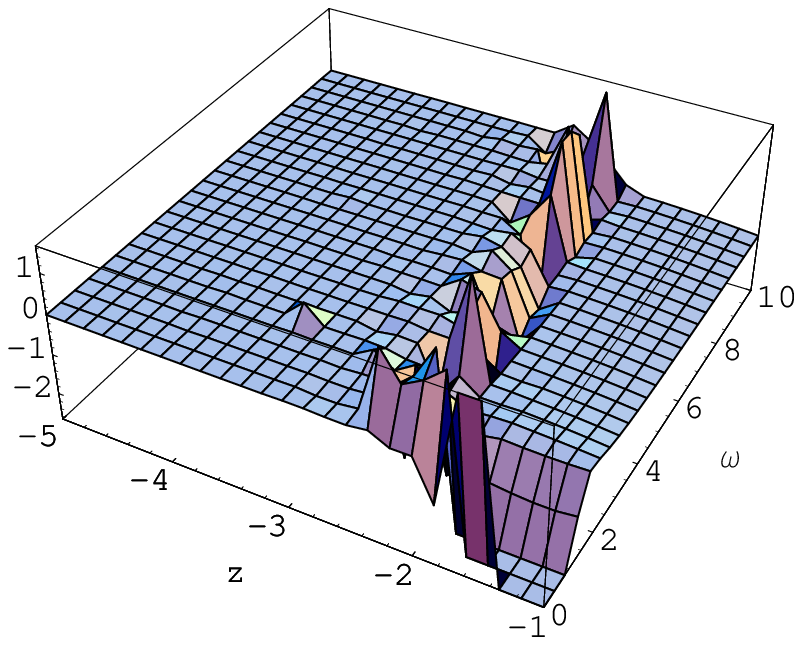,width=0.215\linewidth}\\
2&\epsfig{file=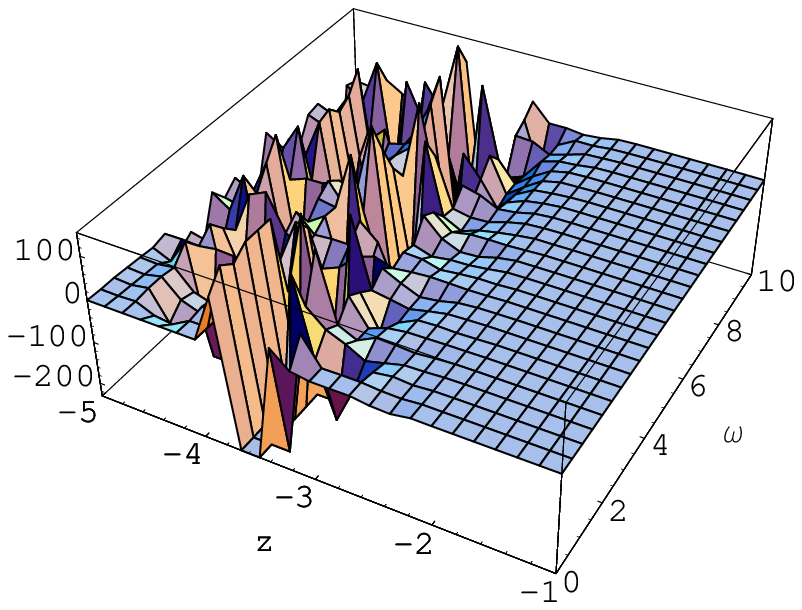,width=0.215\linewidth}
&\epsfig{file=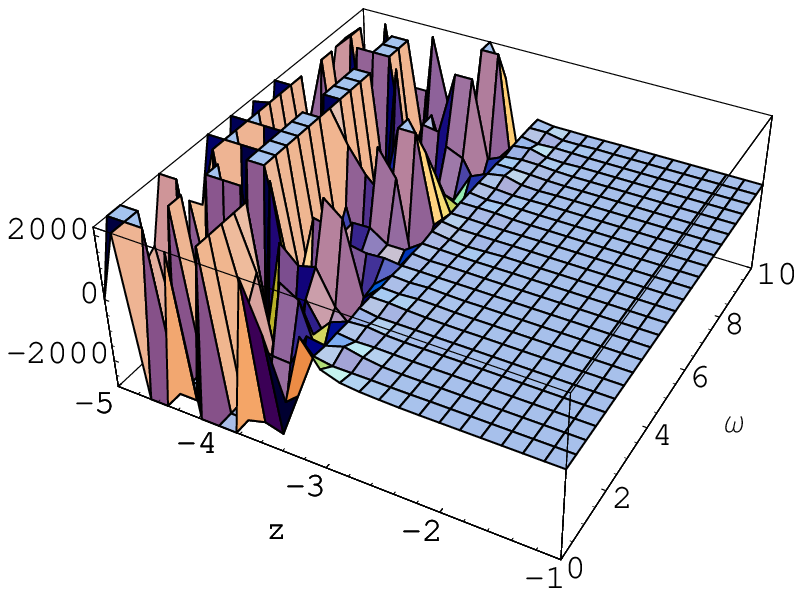,width=0.215\linewidth}
&\epsfig{file=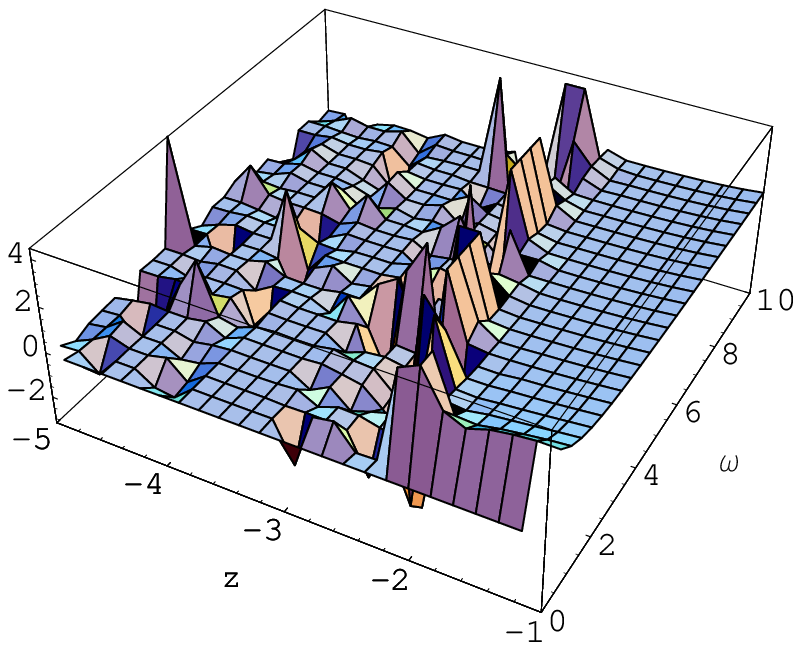,width=0.215\linewidth}
&\epsfig{file=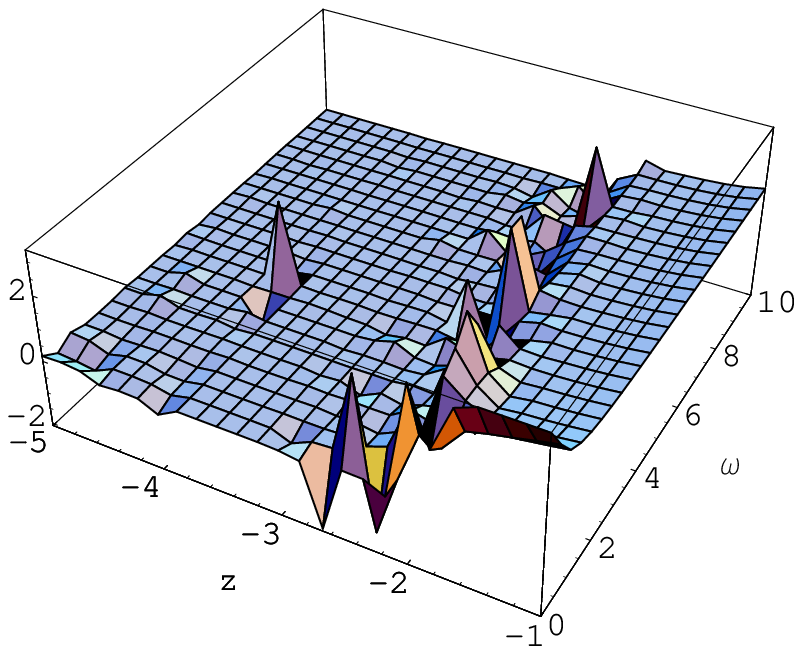,width=0.215\linewidth}\\
3&\epsfig{file=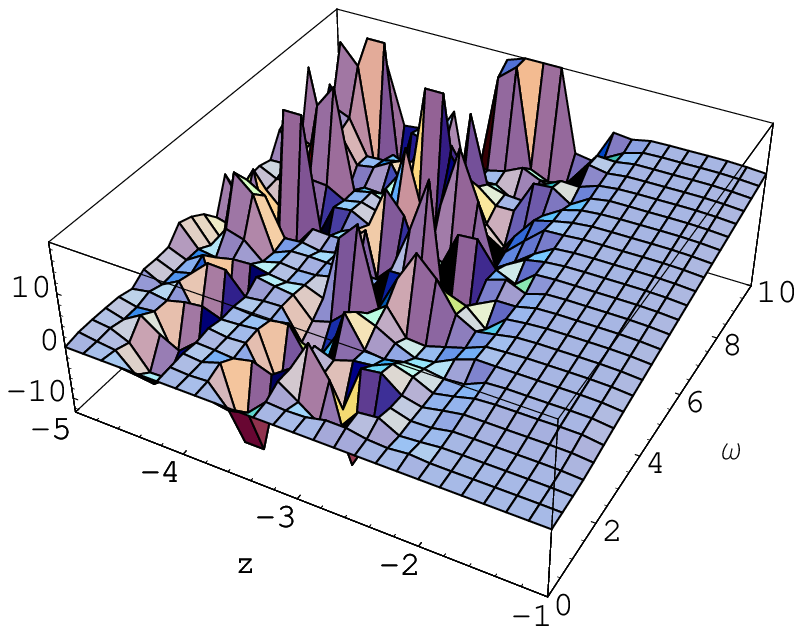,width=0.215\linewidth}
&\epsfig{file=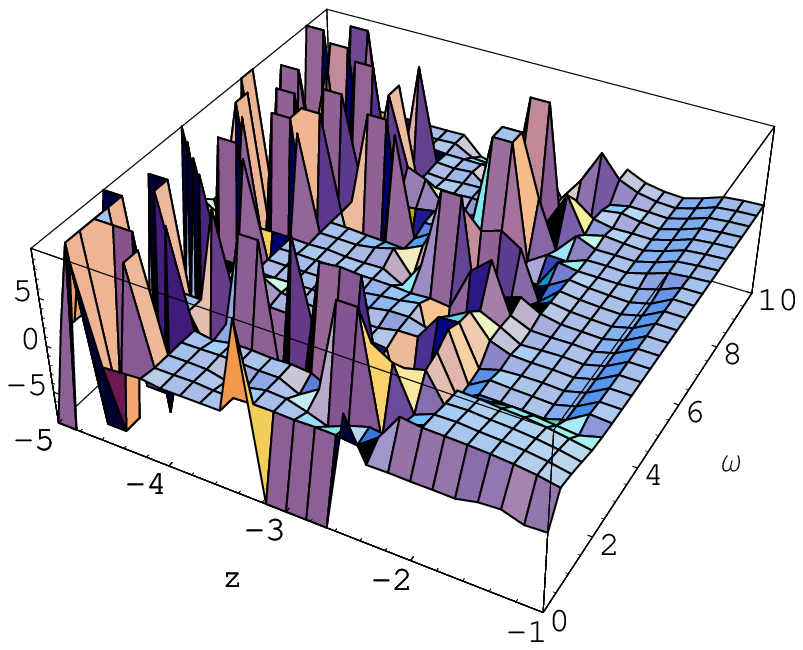,width=0.215\linewidth}
&\epsfig{file=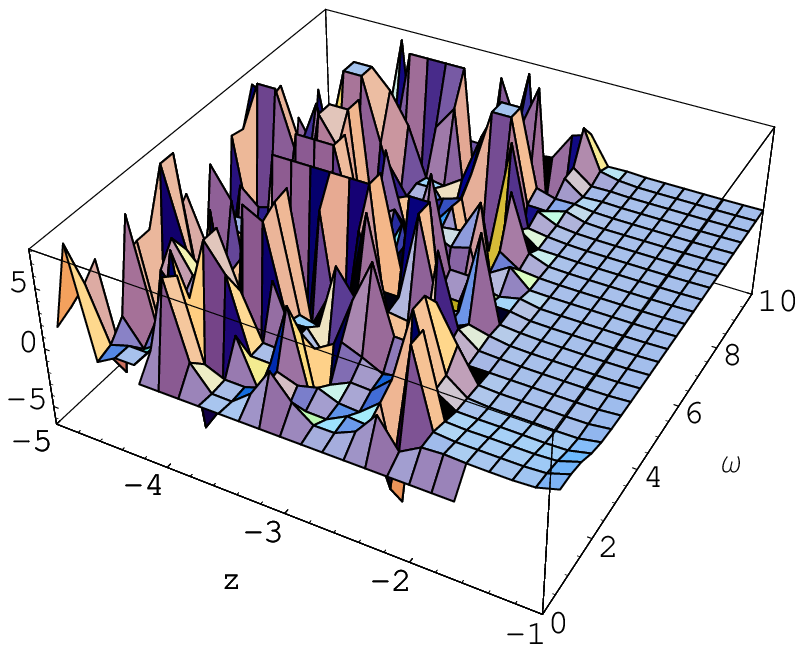,width=0.215\linewidth}
&\epsfig{file=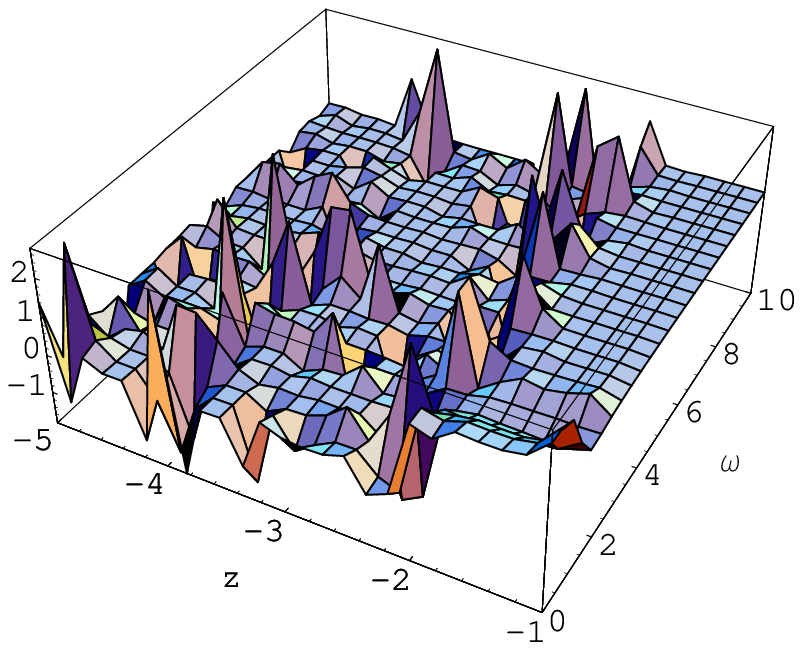,width=0.215\linewidth}\\
4&\epsfig{file=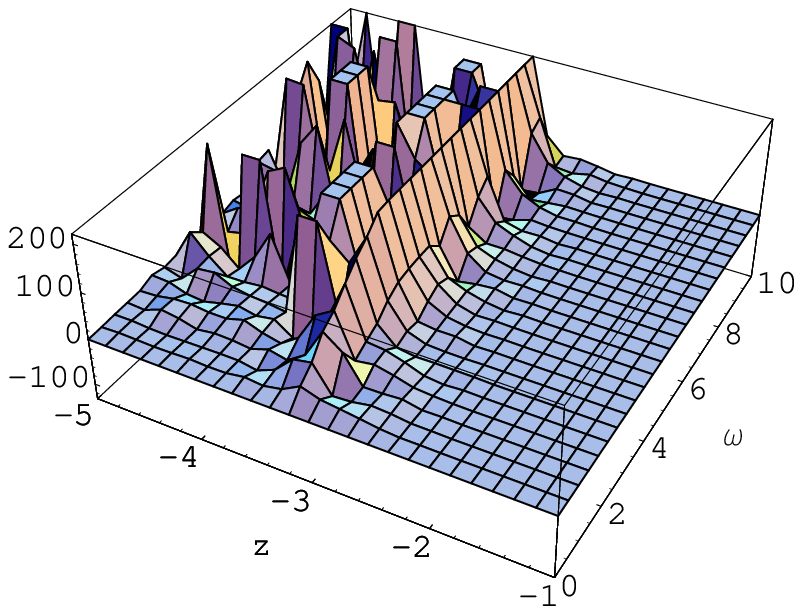,width=0.215\linewidth}
&\epsfig{file=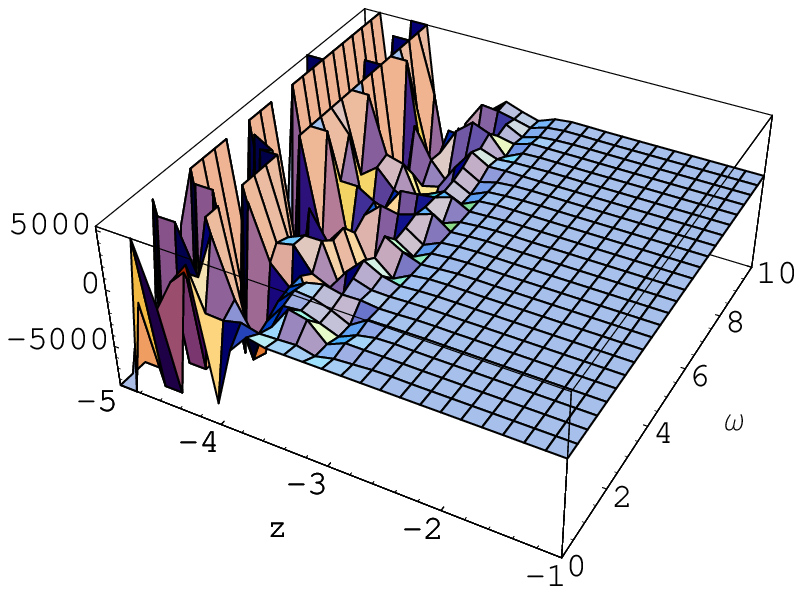,width=0.215\linewidth}
&\epsfig{file=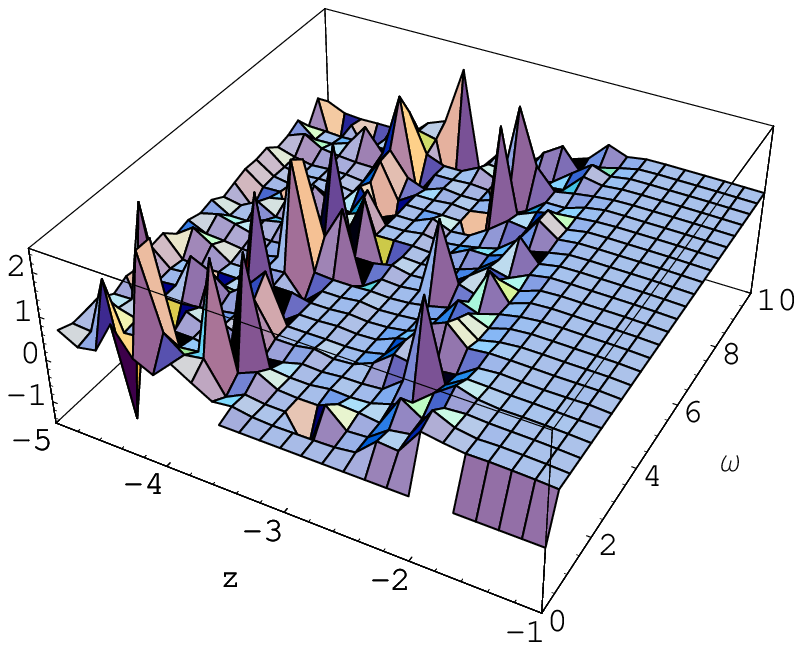,width=0.215\linewidth}
&\epsfig{file=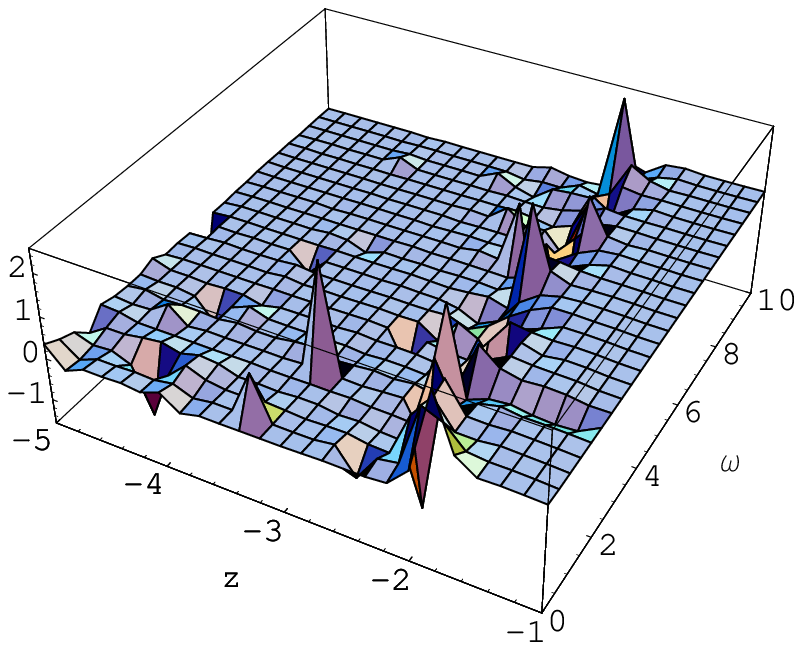,width=0.215\linewidth}\\
5&\epsfig{file=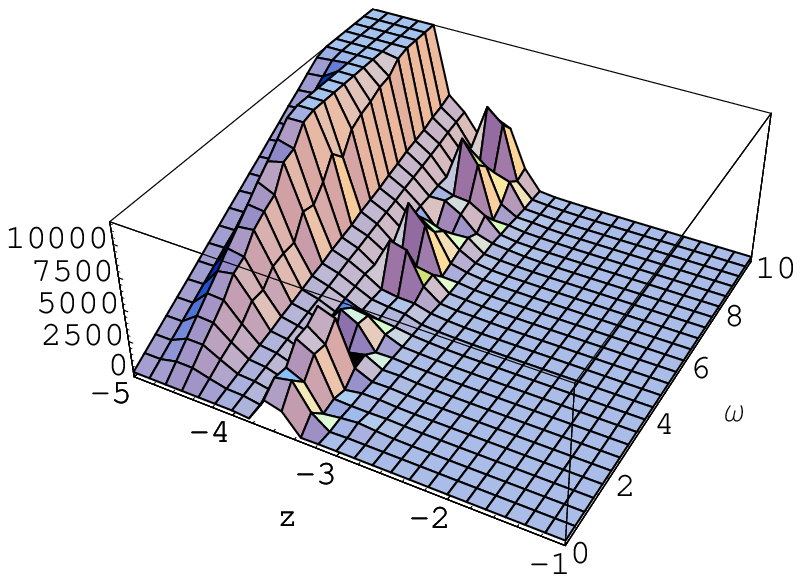,width=0.215\linewidth}
&\epsfig{file=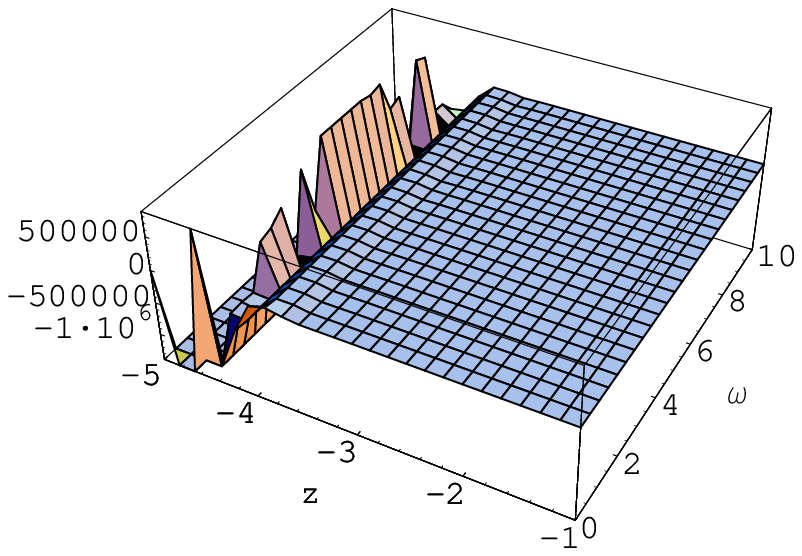,width=0.215\linewidth}
&\epsfig{file=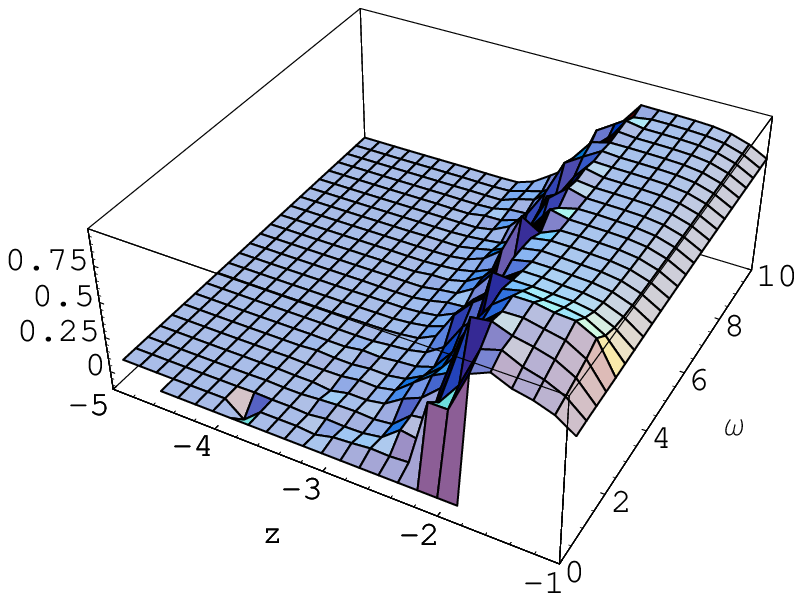,width=0.215\linewidth}
&\epsfig{file=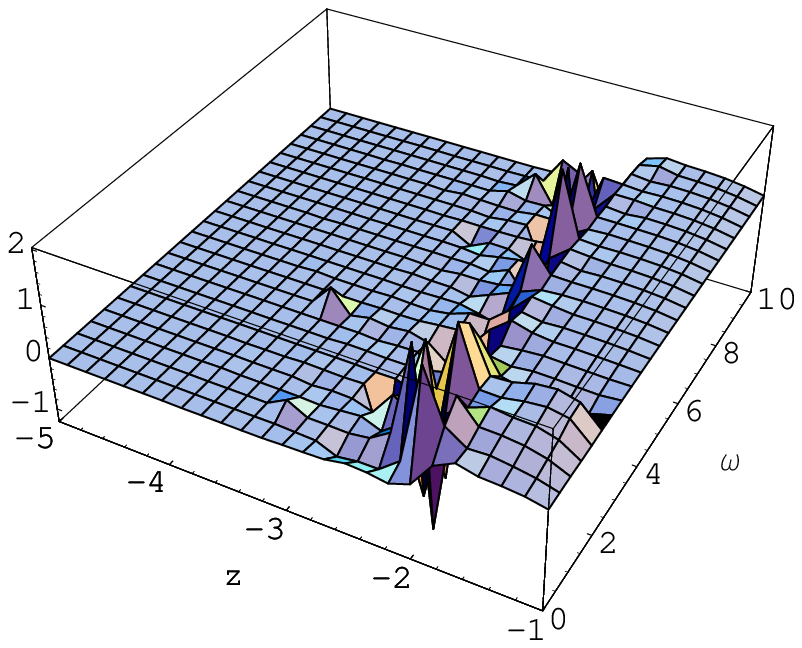,width=0.215\linewidth}\\
6&\epsfig{file=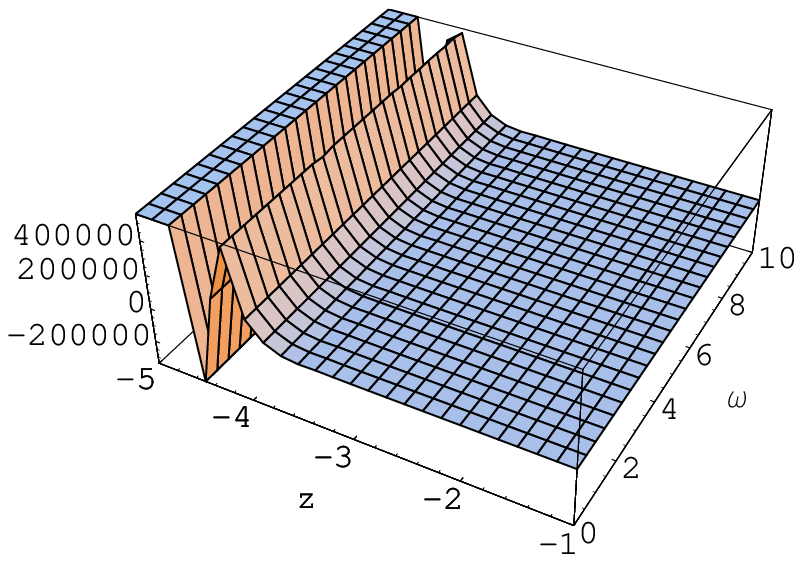,width=0.215\linewidth}
&\epsfig{file=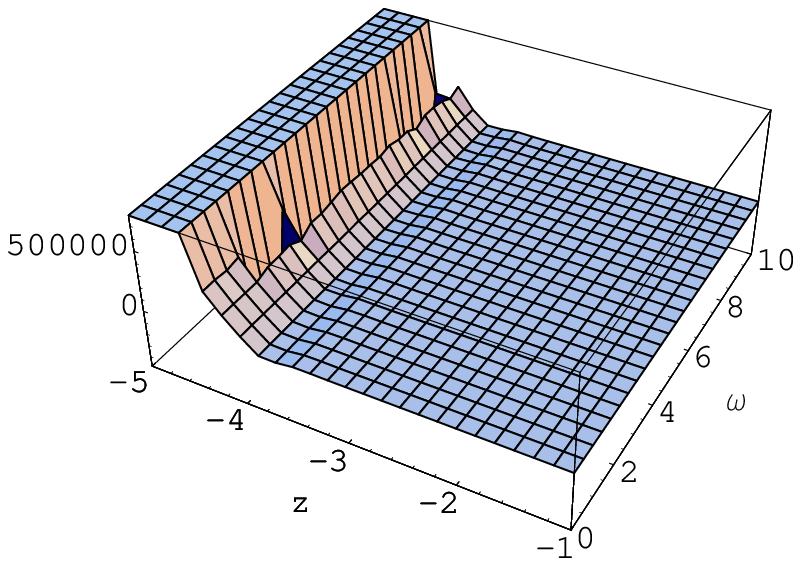,width=0.215\linewidth}
&\epsfig{file=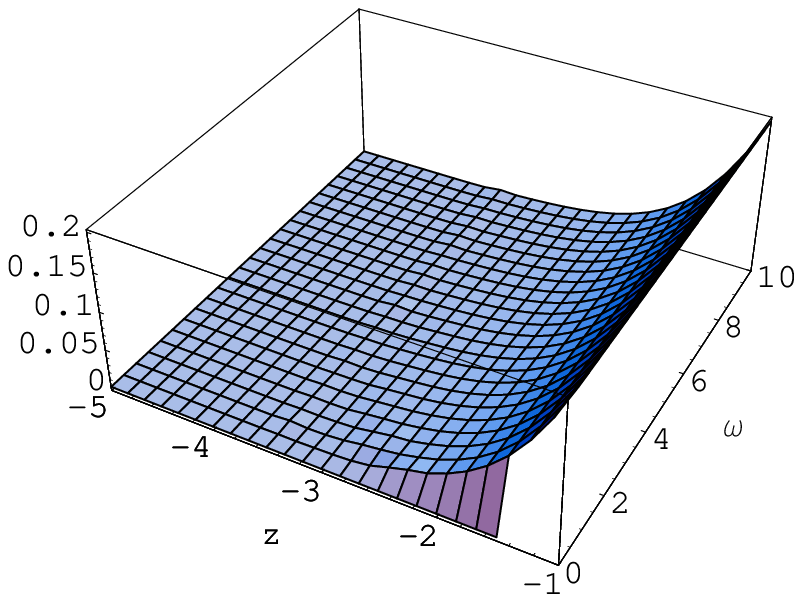,width=0.215\linewidth}
&\epsfig{file=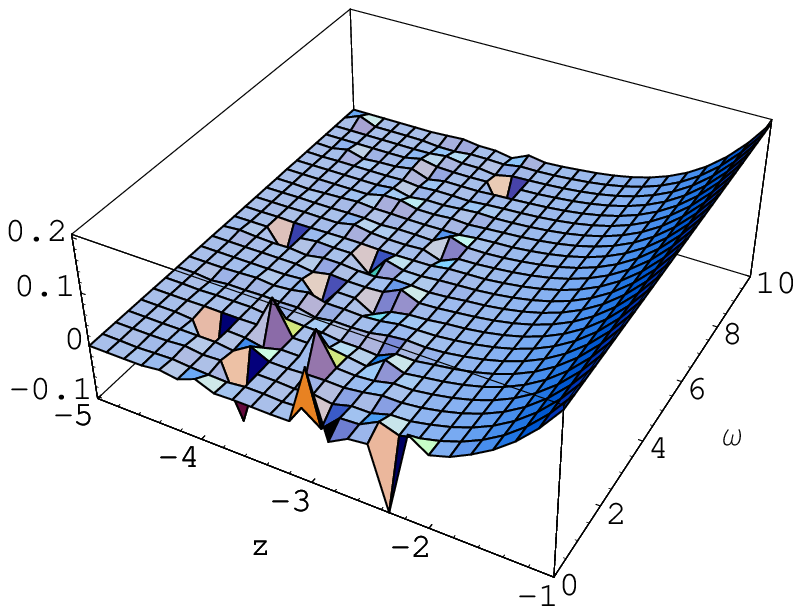,width=0.215\linewidth}
\end{array}$\\
\mbox{Figures \textbf{1-6} show the dispersion relations (related
to the velocity components}\\
\mbox{given by Eq.(\ref{u1})) in the neighborhood of the pair
production region towards}\\
\mbox{the event horizon.}
\\
$\begin{array}{ccccc}
&A&B&C&D\\
7&\epsfig{file=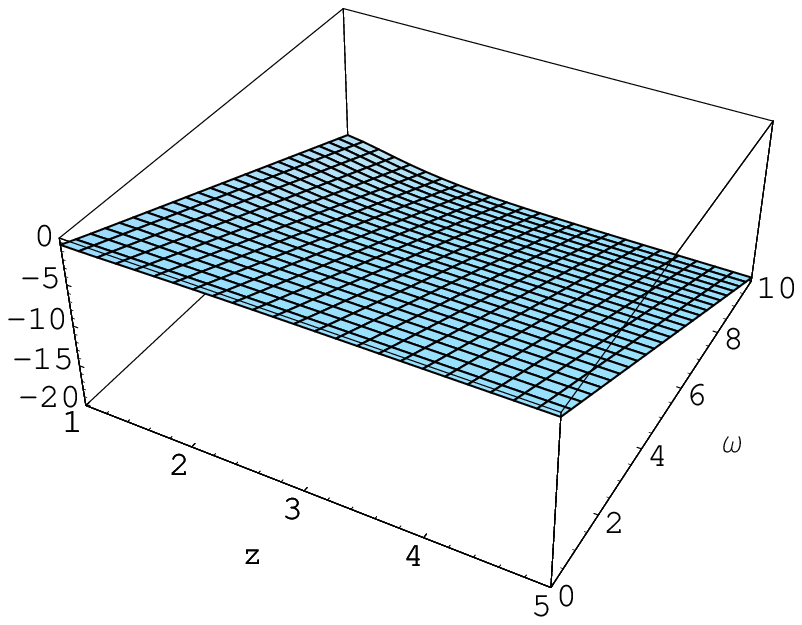,width=0.215\linewidth}
&\epsfig{file=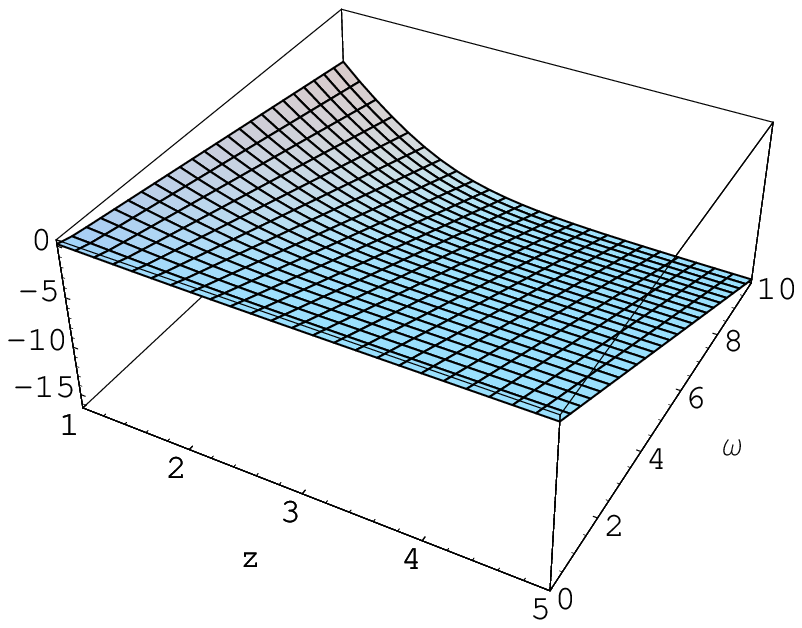,width=0.215\linewidth}
&\epsfig{file=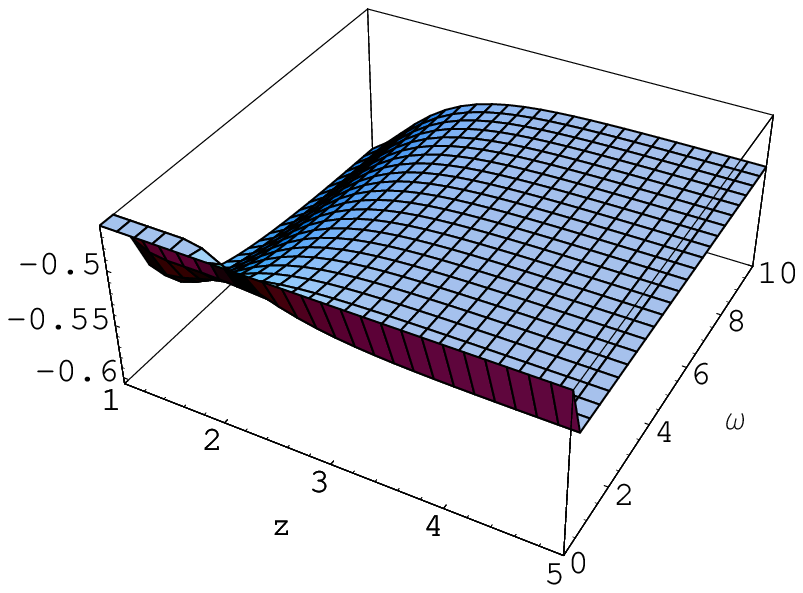,width=0.215\linewidth}
&\epsfig{file=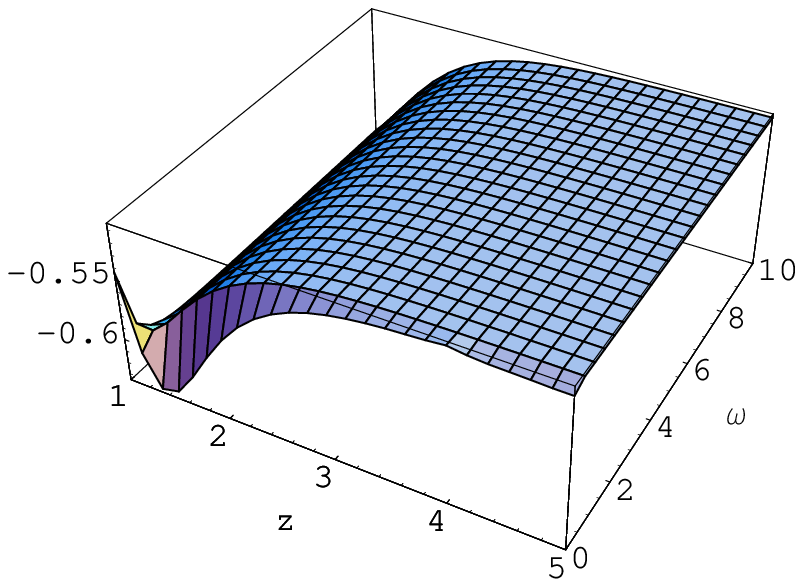,width=0.215\linewidth}\\
8&\epsfig{file=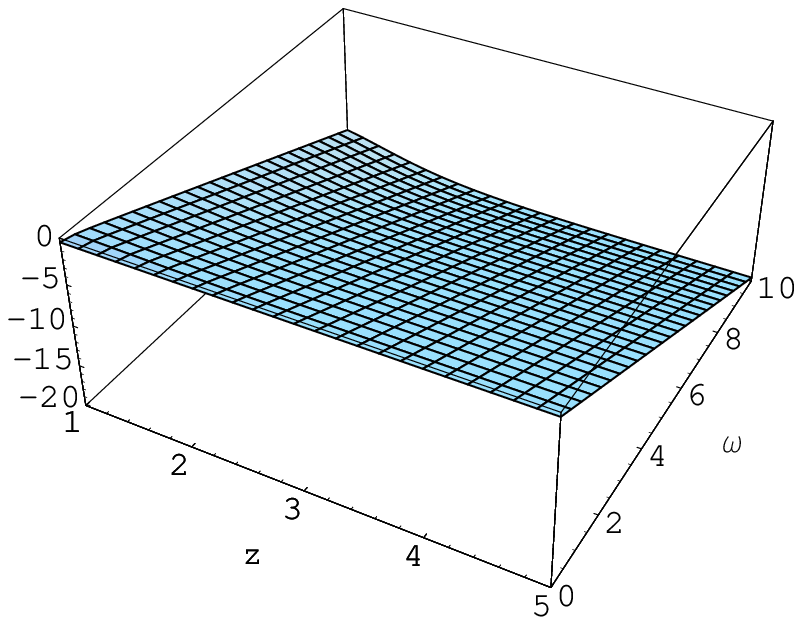,width=0.215\linewidth}
&\epsfig{file=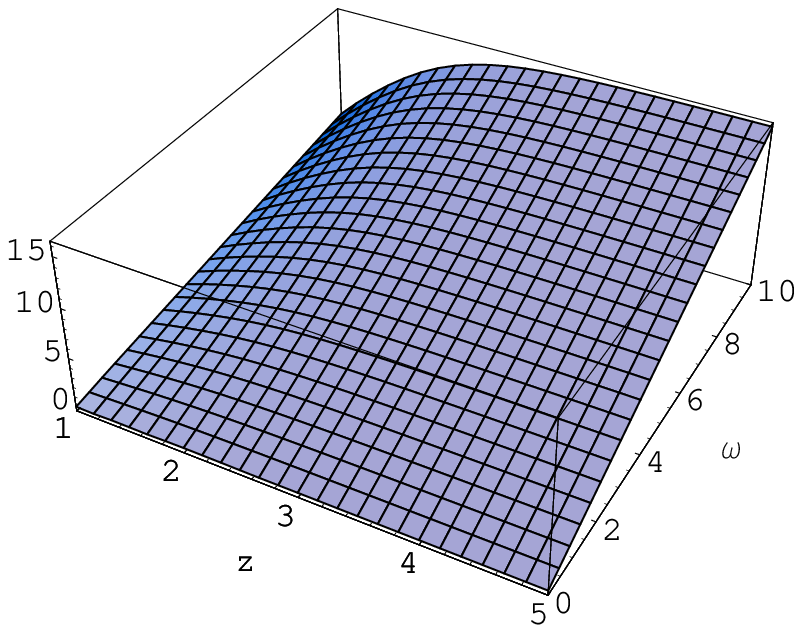,width=0.215\linewidth}
&\epsfig{file=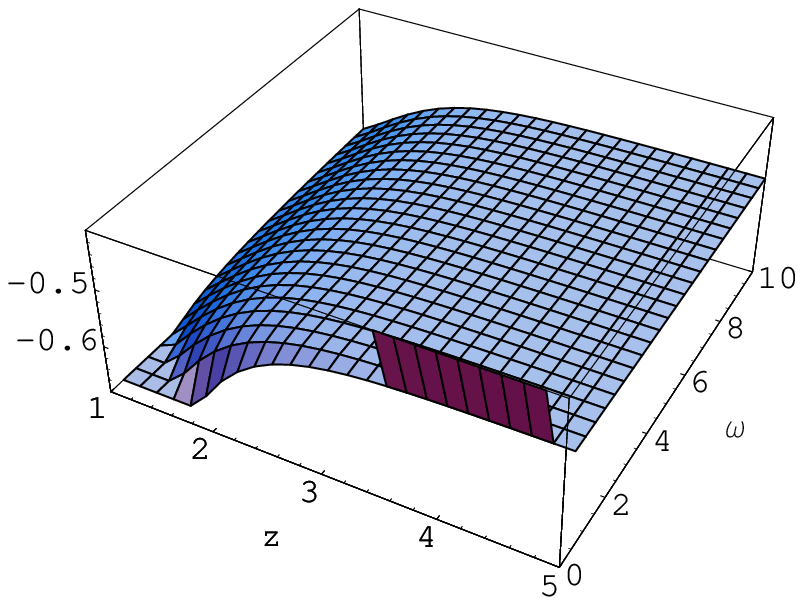,width=0.215\linewidth}
&\epsfig{file=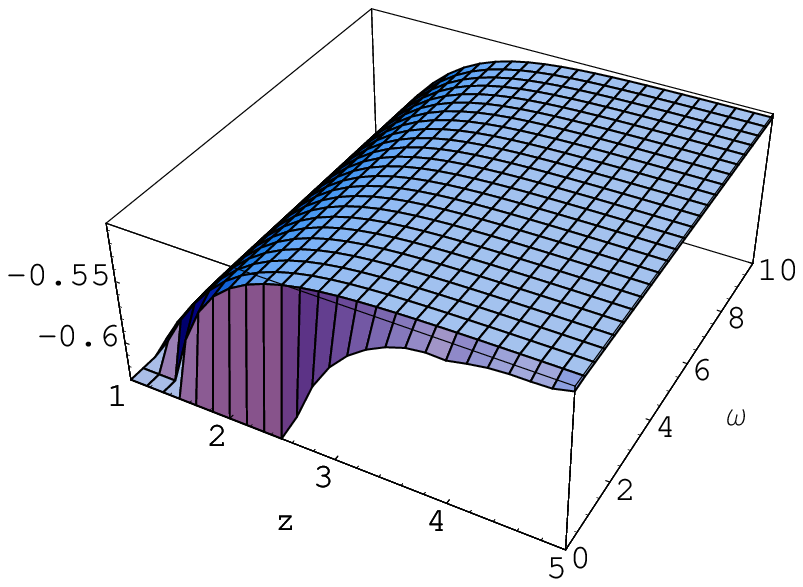,width=0.215\linewidth}\\
9&\epsfig{file=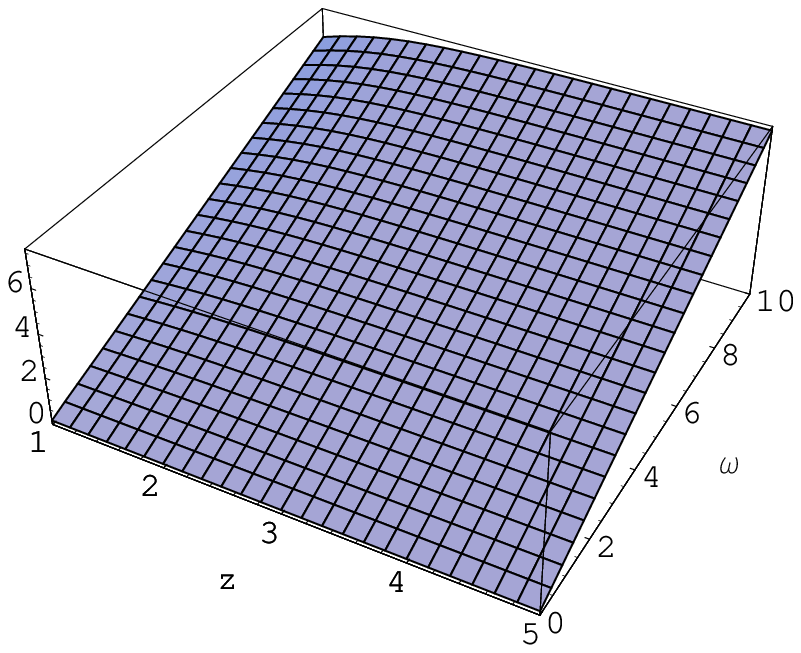,width=0.215\linewidth}
&\epsfig{file=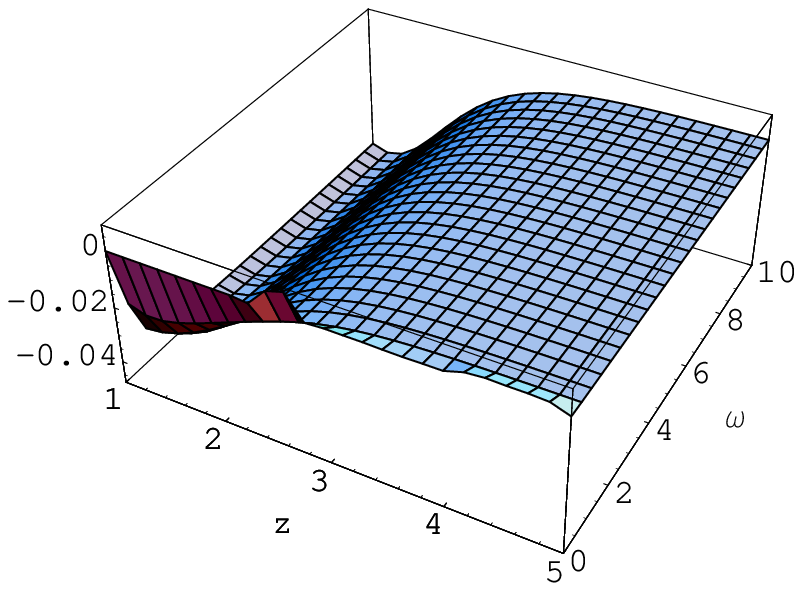,width=0.215\linewidth}
&\epsfig{file=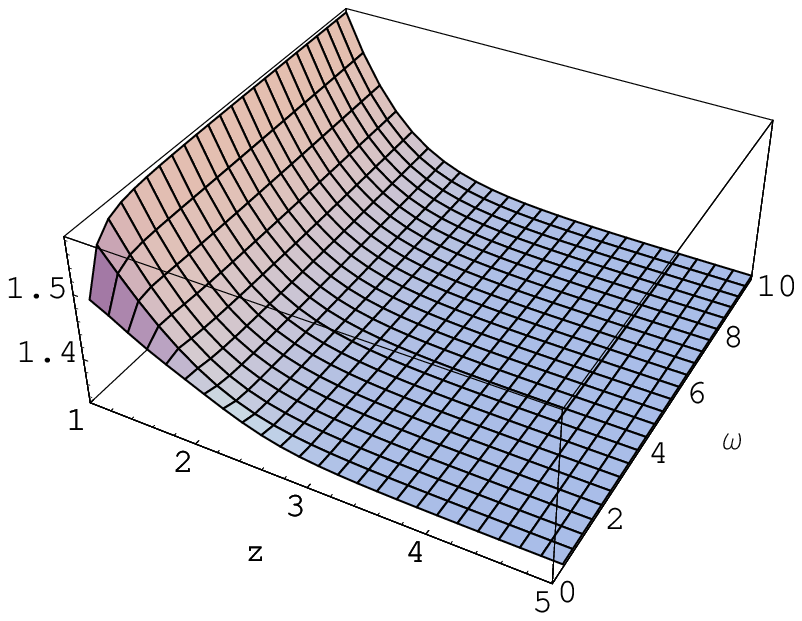,width=0.215\linewidth}
&\epsfig{file=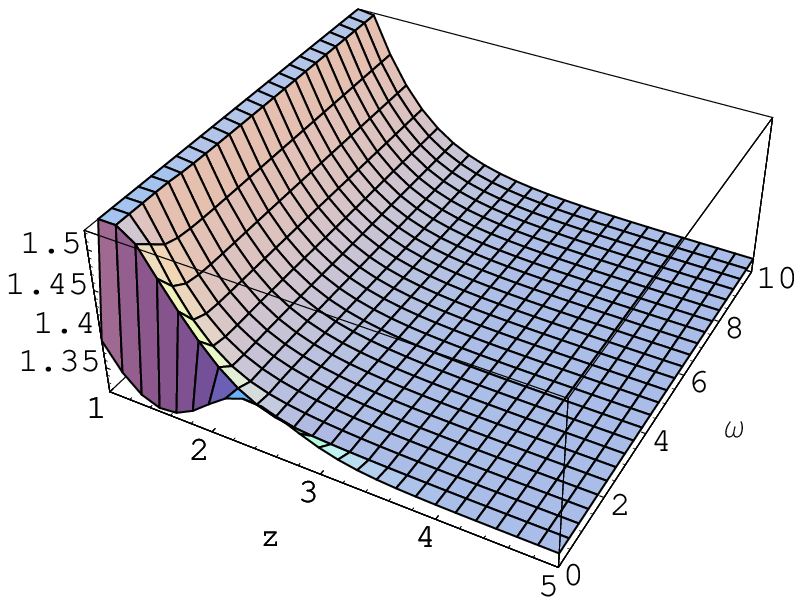,width=0.215\linewidth}\\
10&\epsfig{file=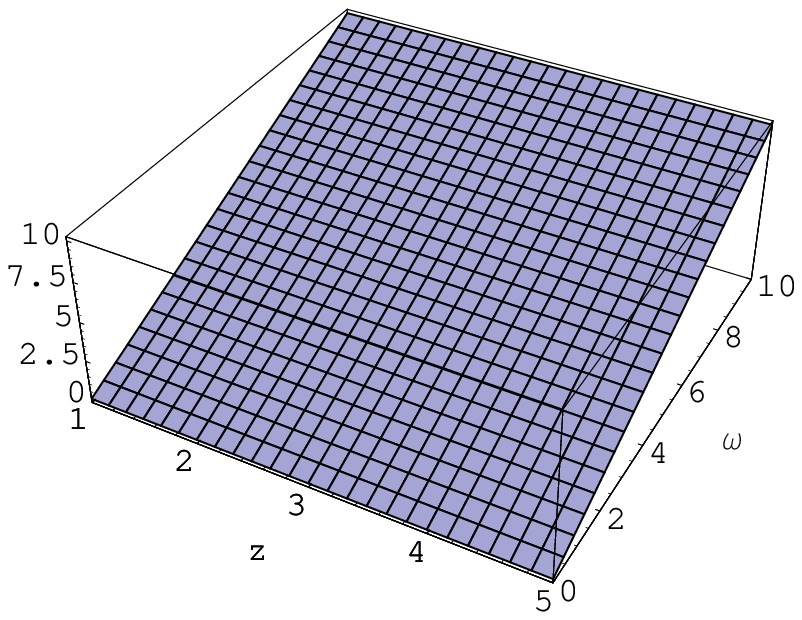,width=0.215\linewidth}
&\epsfig{file=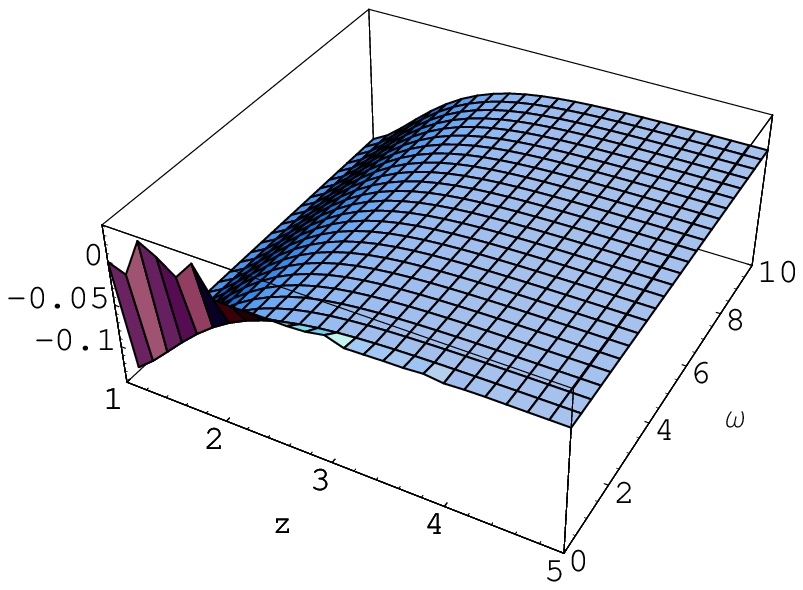,width=0.215\linewidth}
&\epsfig{file=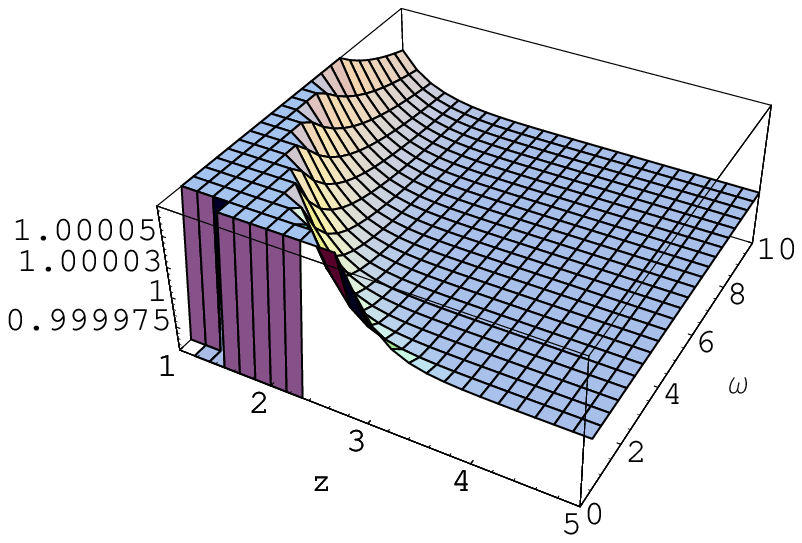,width=0.215\linewidth}
&\epsfig{file=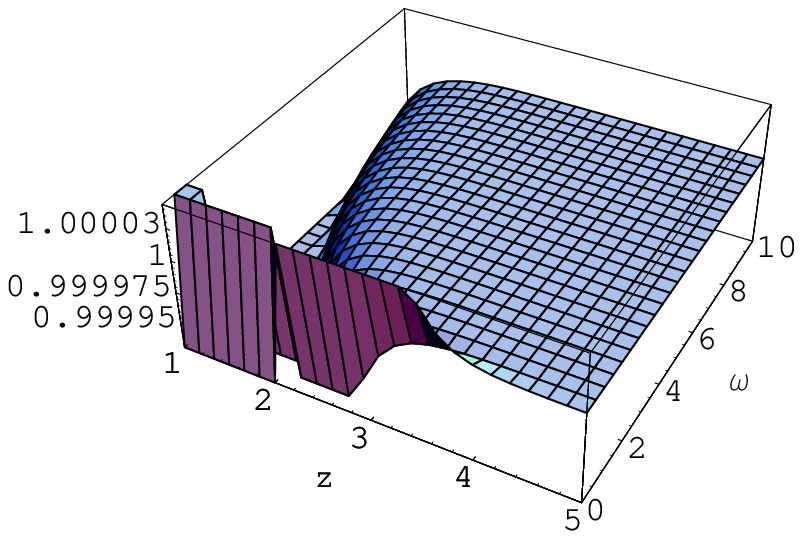,width=0.215\linewidth}\\
11&\epsfig{file=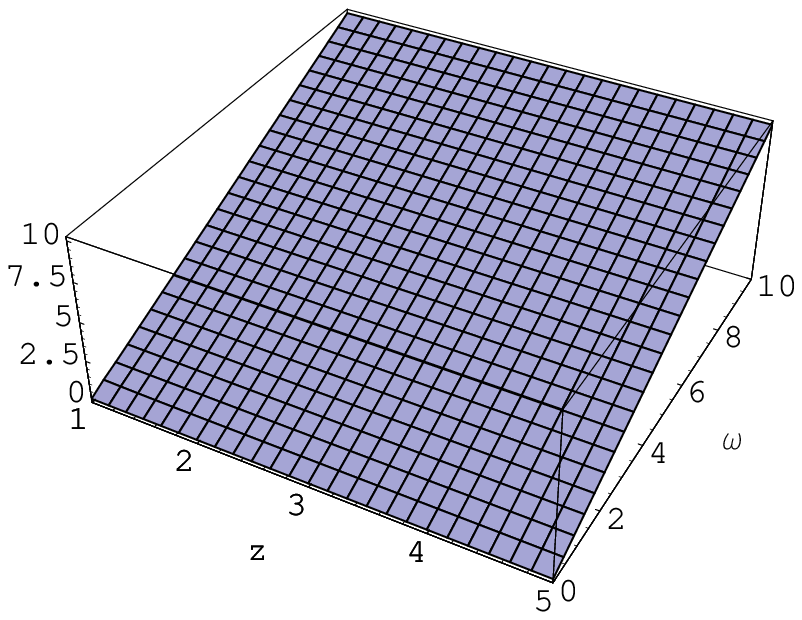,width=0.215\linewidth}
&\epsfig{file=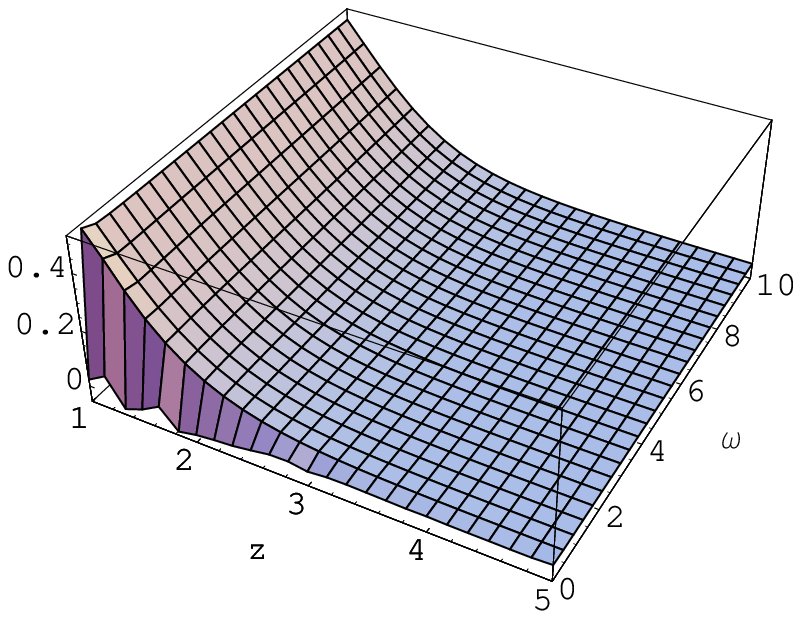,width=0.215\linewidth}
&\epsfig{file=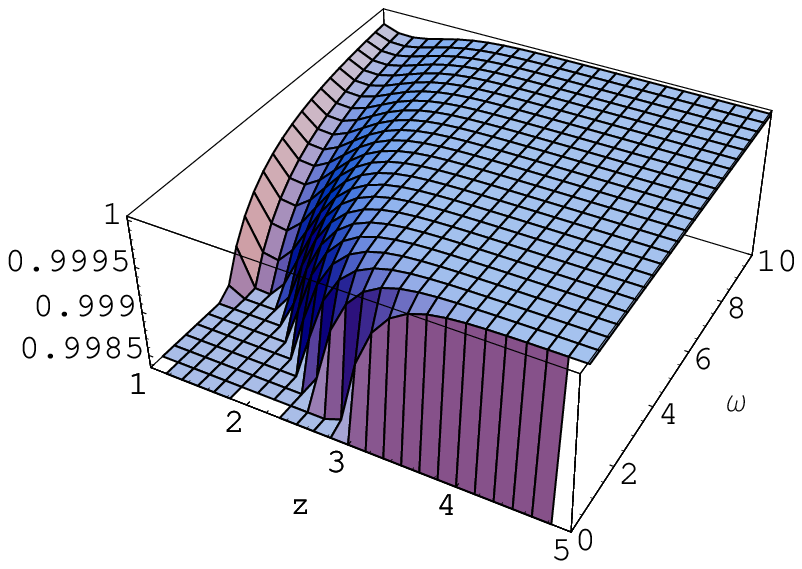,width=0.215\linewidth}
&\epsfig{file=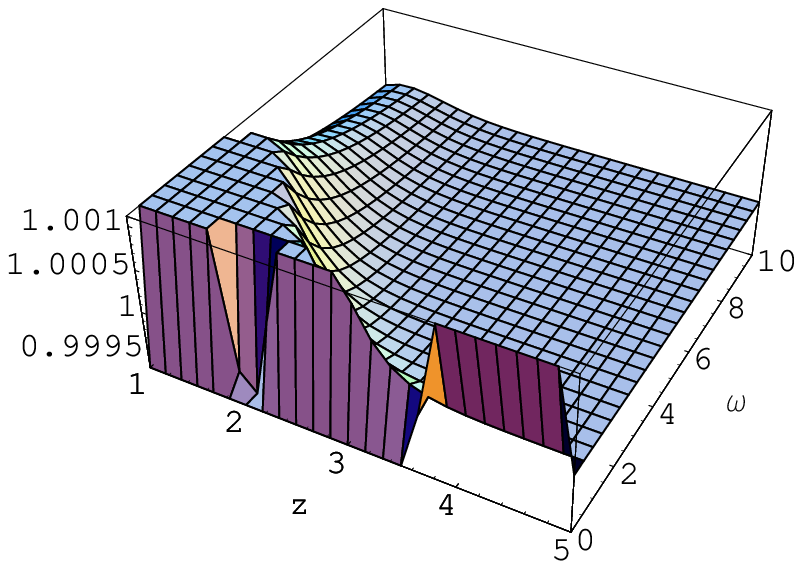,width=0.215\linewidth}\\
12&\epsfig{file=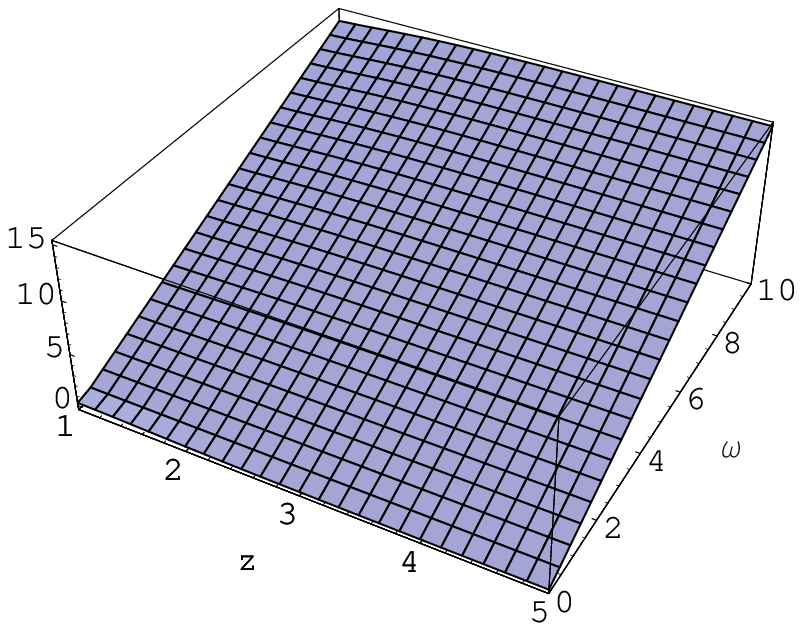,width=0.215\linewidth}
&\epsfig{file=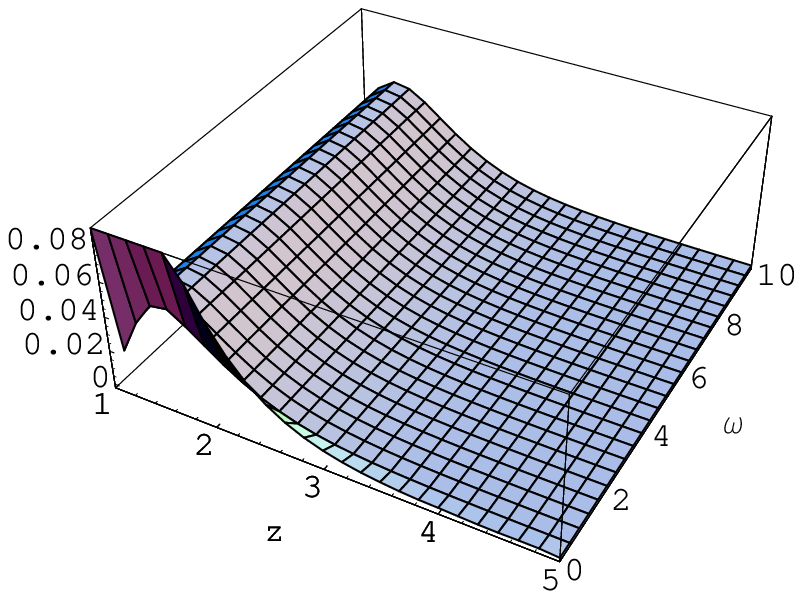,width=0.215\linewidth}
&\epsfig{file=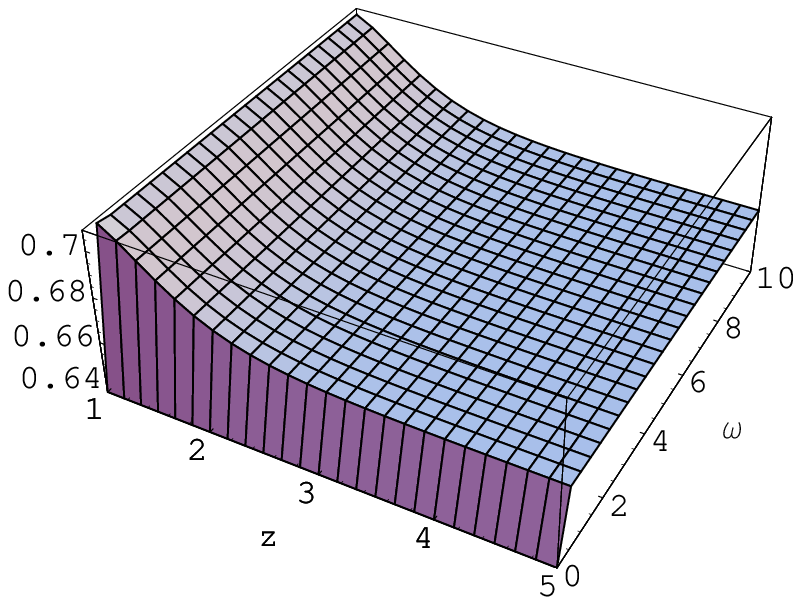,width=0.215\linewidth}
&\epsfig{file=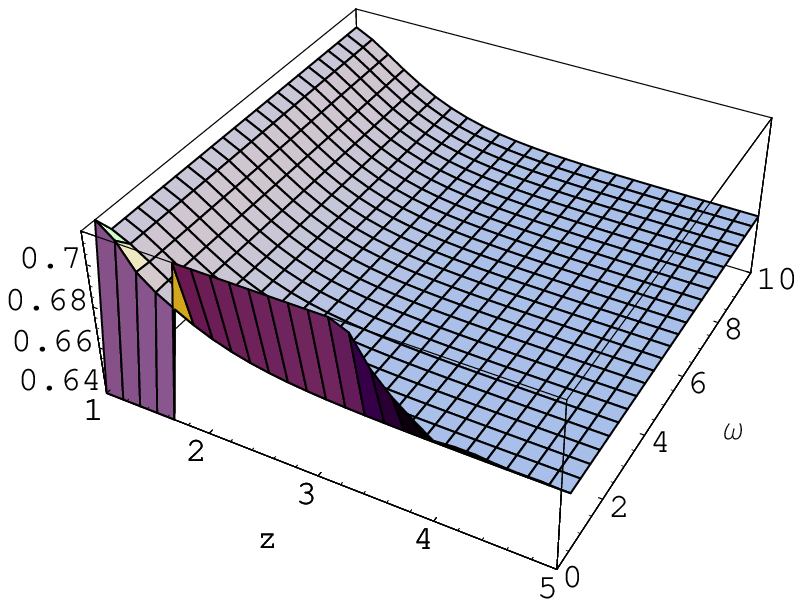,width=0.215\linewidth}
\end{array}$\\
\mbox{Figures \textbf{7-12} indicate the dispersion relations
(related to the velocity comp-}\\
\mbox{onents given by Eq.(\ref{u1})) in the neighborhood of the pair
production region}\\
\mbox{towards the outer end of the magnetosphere.}
\\
$\begin{array}{ccccc}
&A&B&C&D\\
13&\epsfig{file=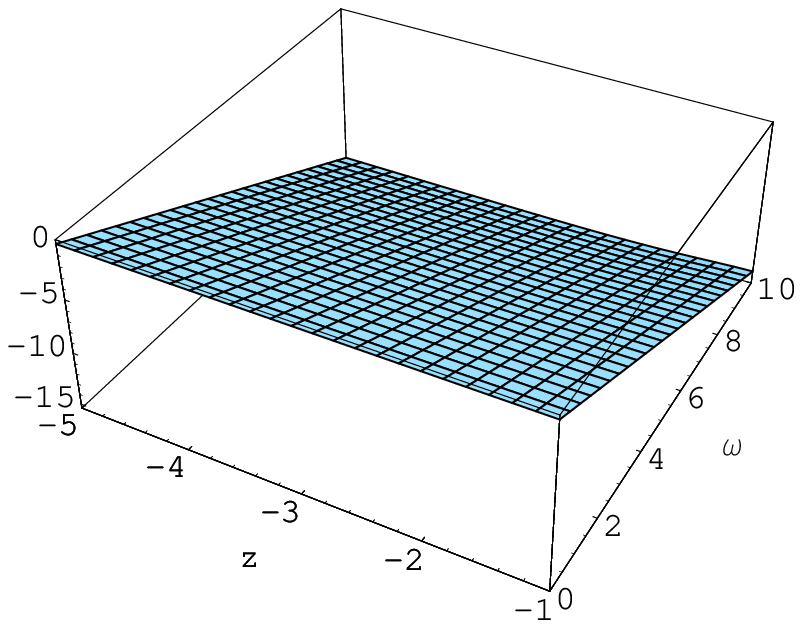,width=0.215\linewidth}
&\epsfig{file=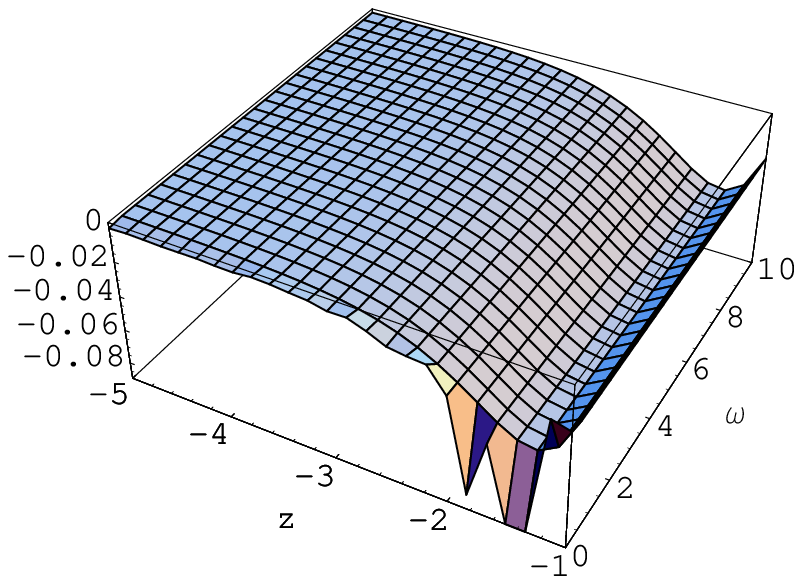,width=0.215\linewidth}
&\epsfig{file=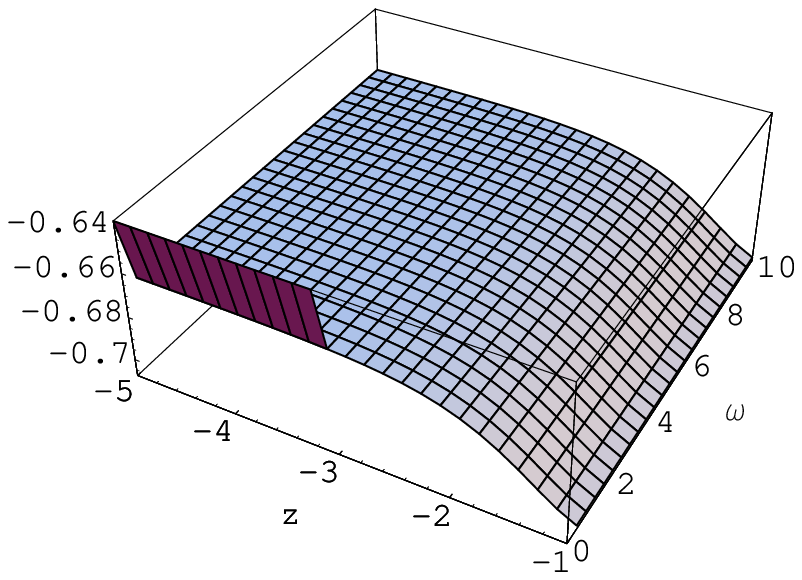,width=0.215\linewidth}
&\epsfig{file=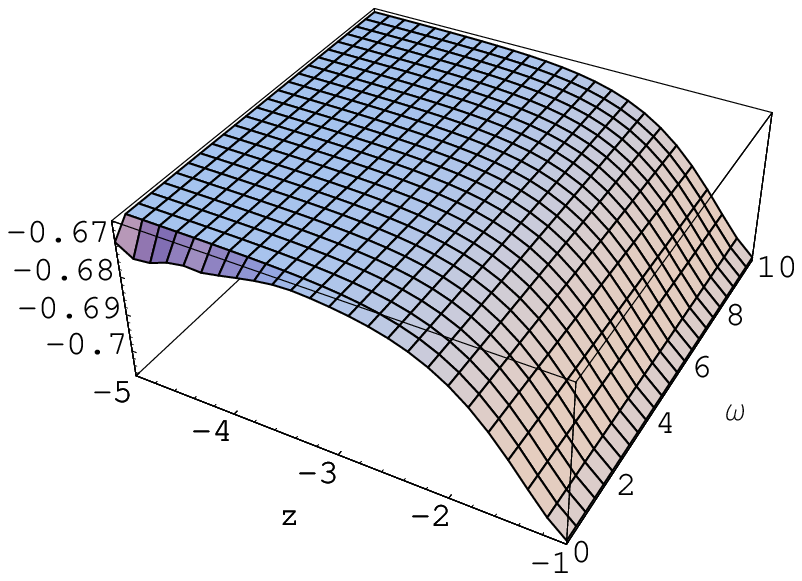,width=0.215\linewidth}\\
14&\epsfig{file=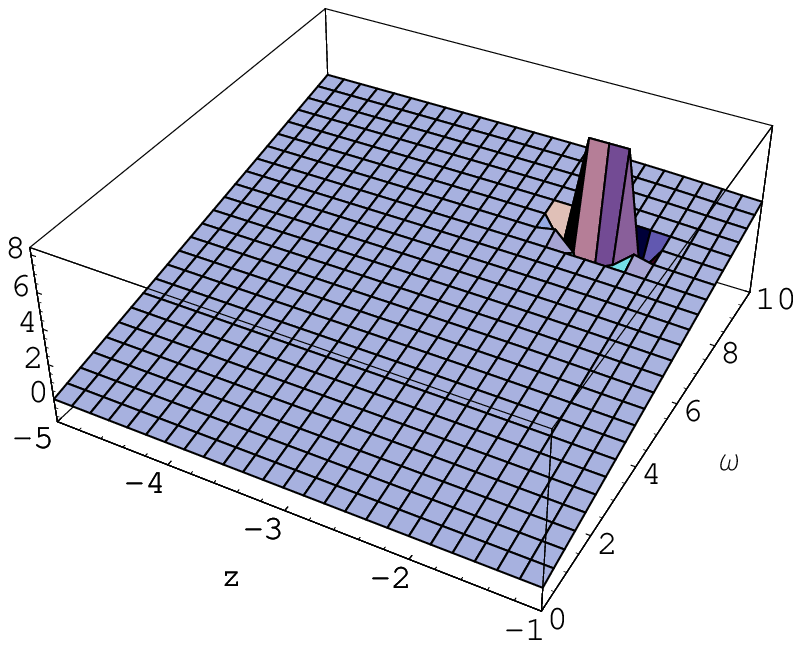,width=0.215\linewidth}
&\epsfig{file=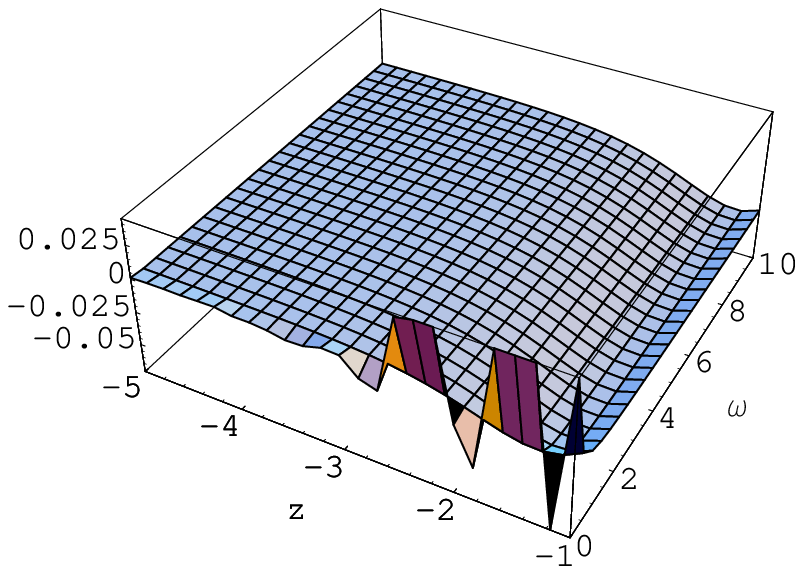,width=0.215\linewidth}
&\epsfig{file=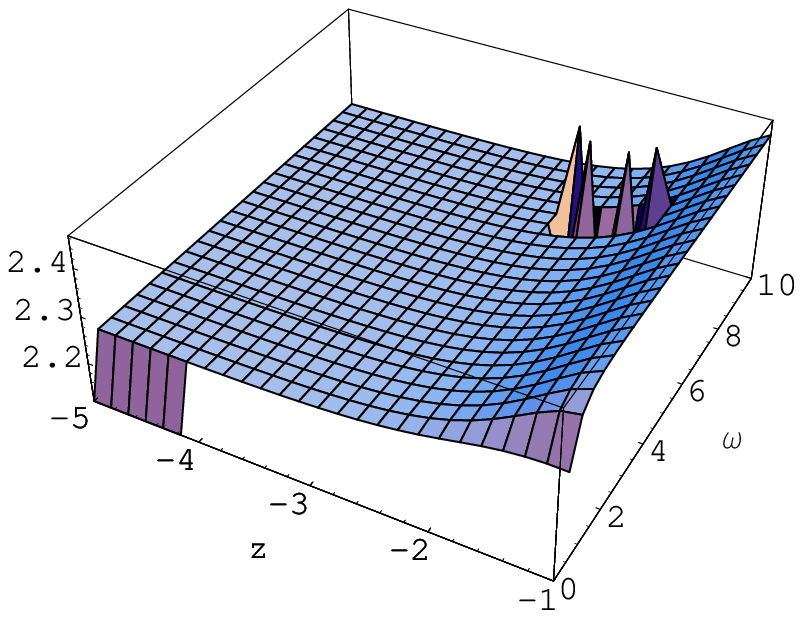,width=0.215\linewidth}
&\epsfig{file=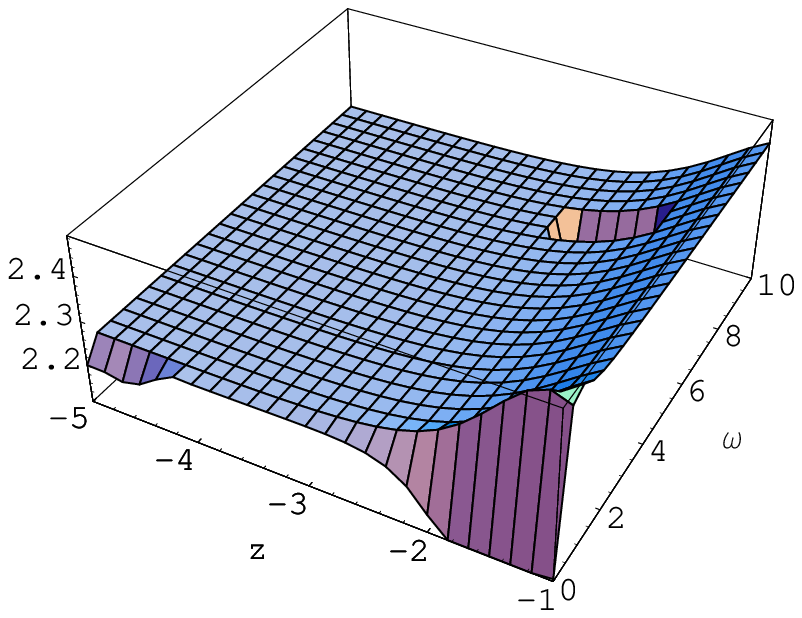,width=0.215\linewidth}\\
15&\epsfig{file=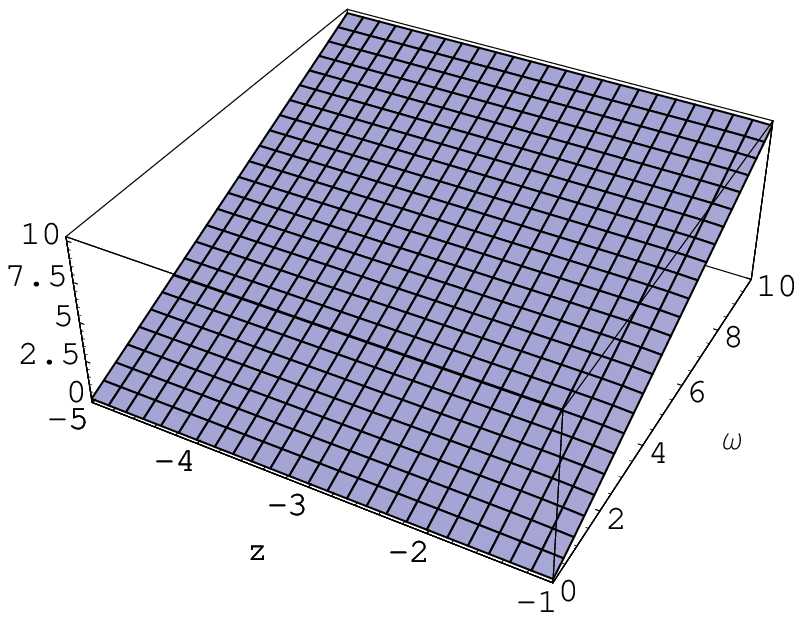,width=0.215\linewidth}
&\epsfig{file=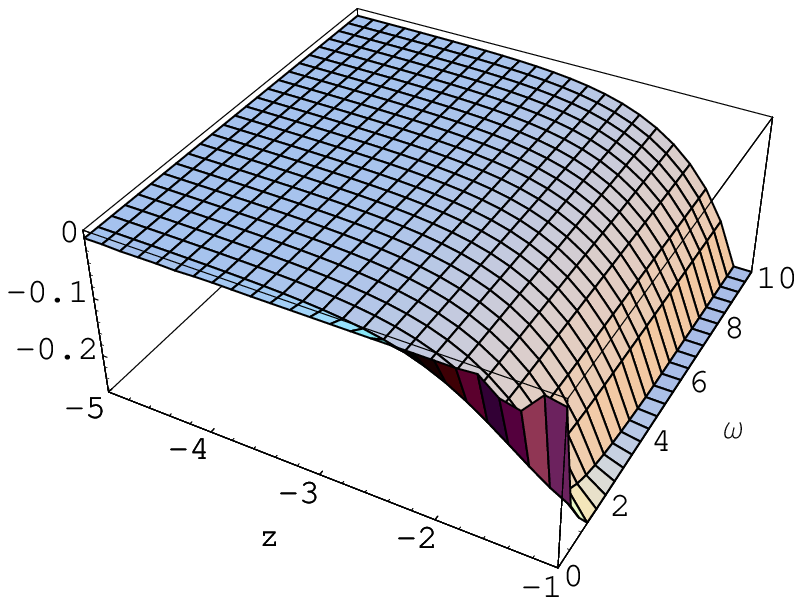,width=0.215\linewidth}
&\epsfig{file=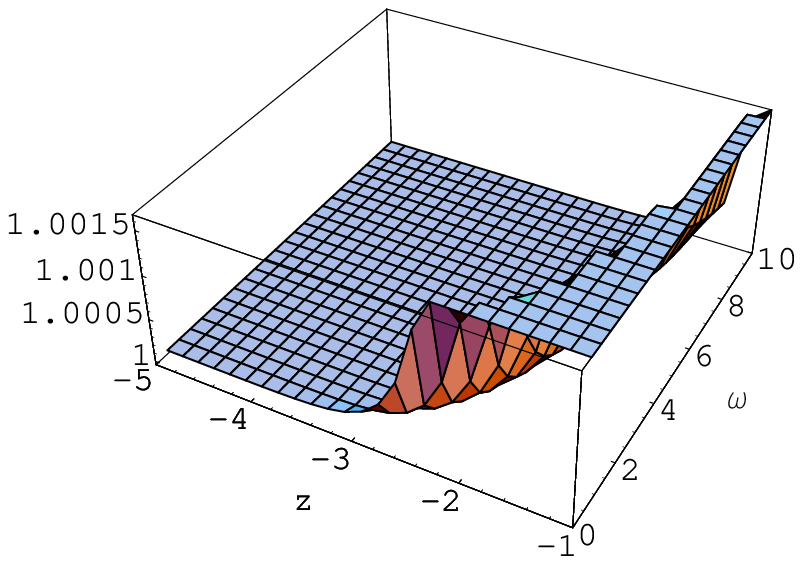,width=0.215\linewidth}
&\epsfig{file=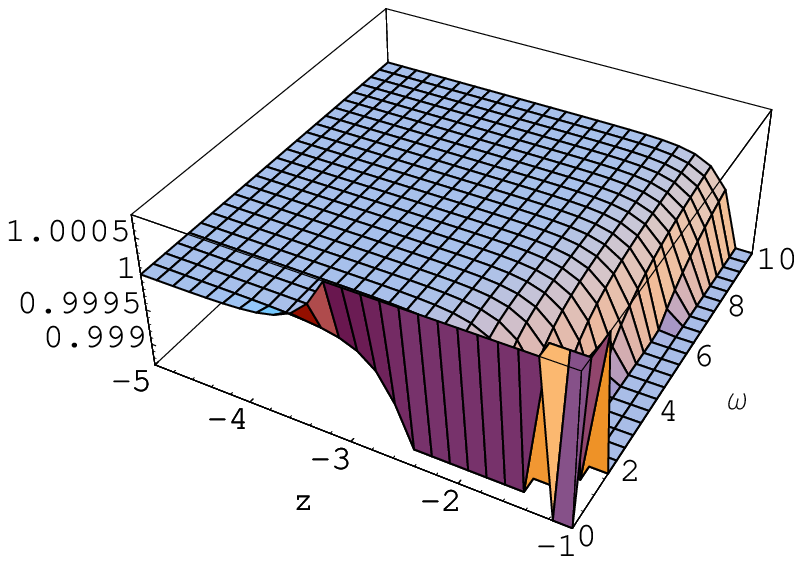,width=0.215\linewidth}\\
16&\epsfig{file=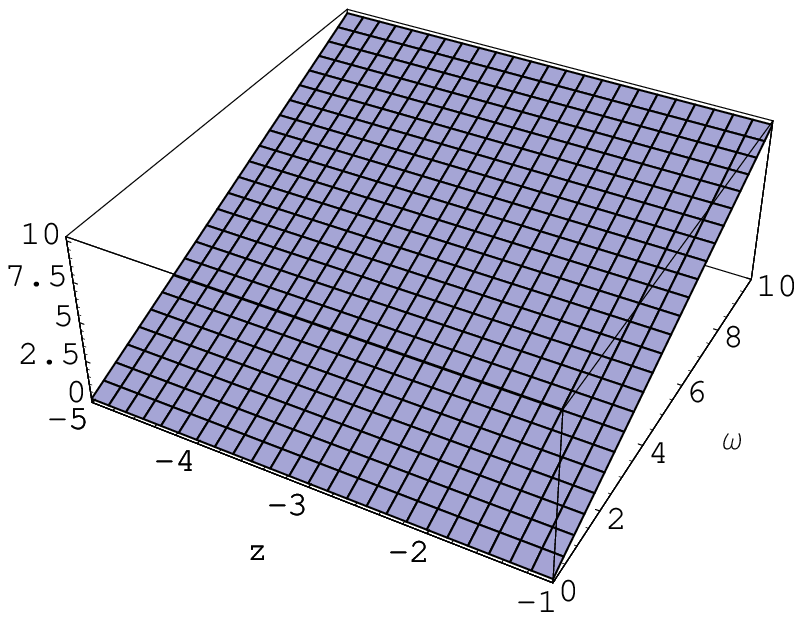,width=0.215\linewidth}
&\epsfig{file=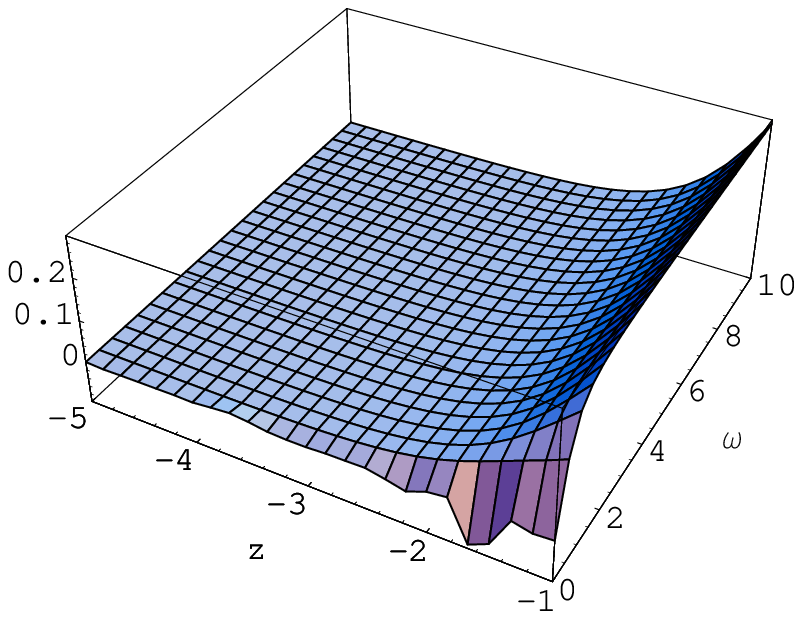,width=0.215\linewidth}
&\epsfig{file=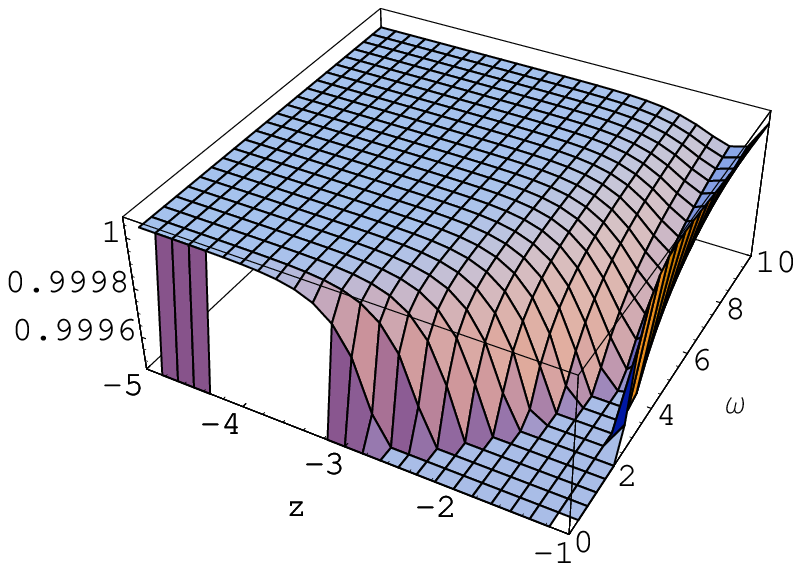,width=0.215\linewidth}
&\epsfig{file=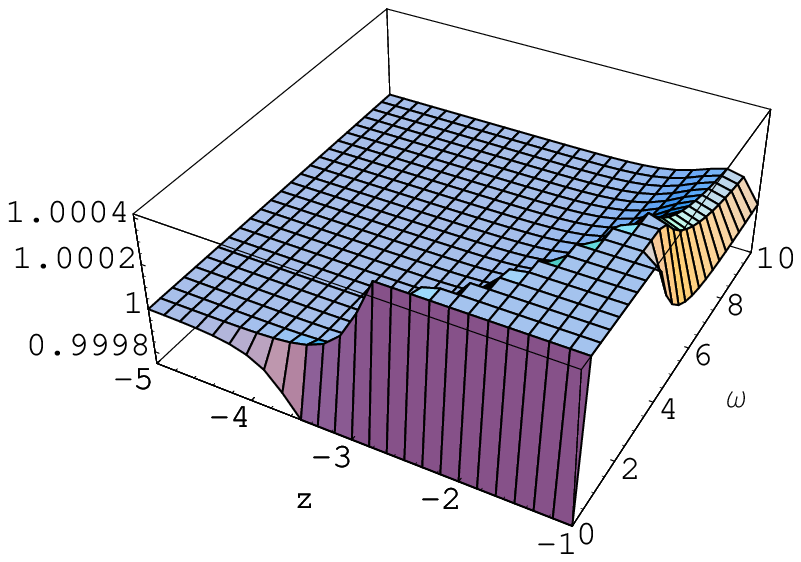,width=0.215\linewidth}\\
17&\epsfig{file=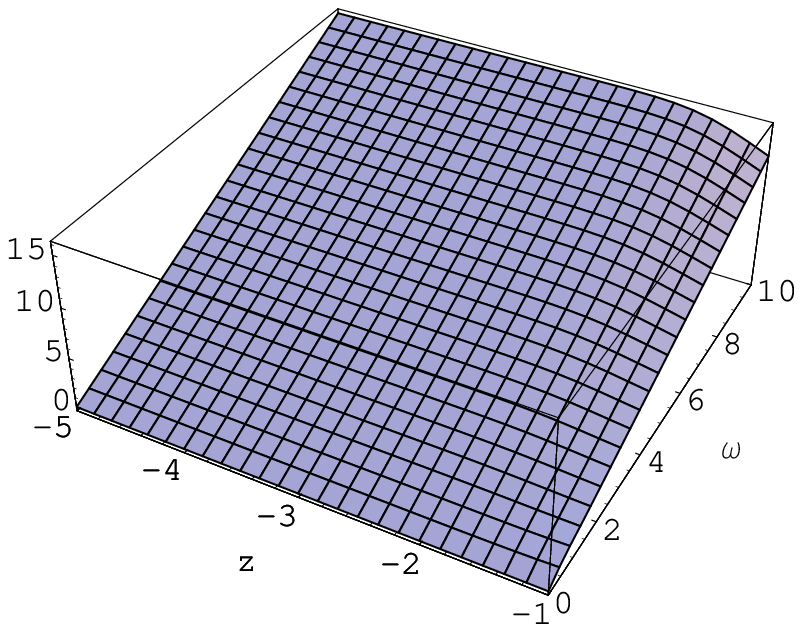,width=0.215\linewidth}
&\epsfig{file=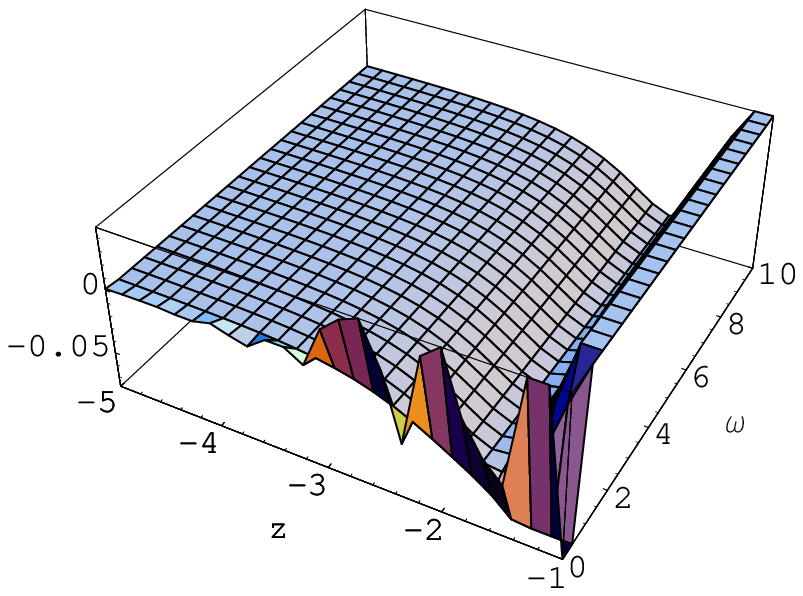,width=0.215\linewidth}
&\epsfig{file=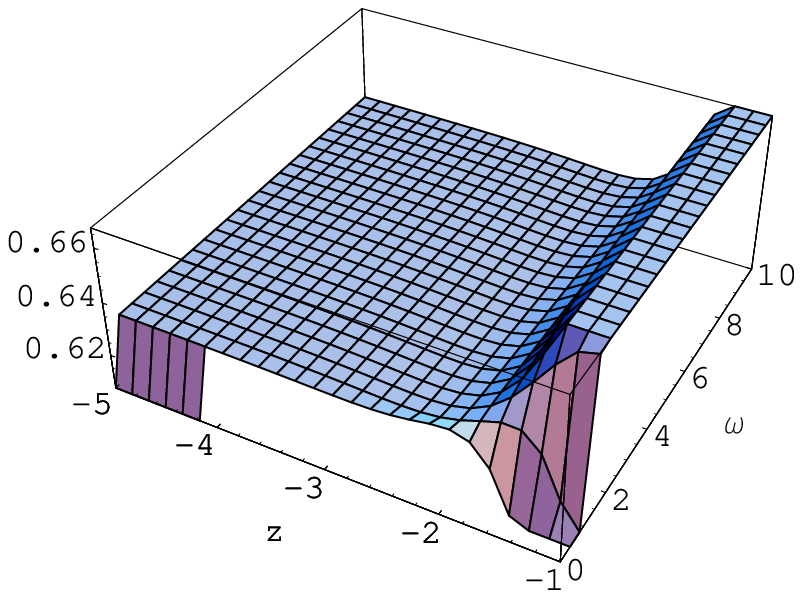,width=0.215\linewidth}
&\epsfig{file=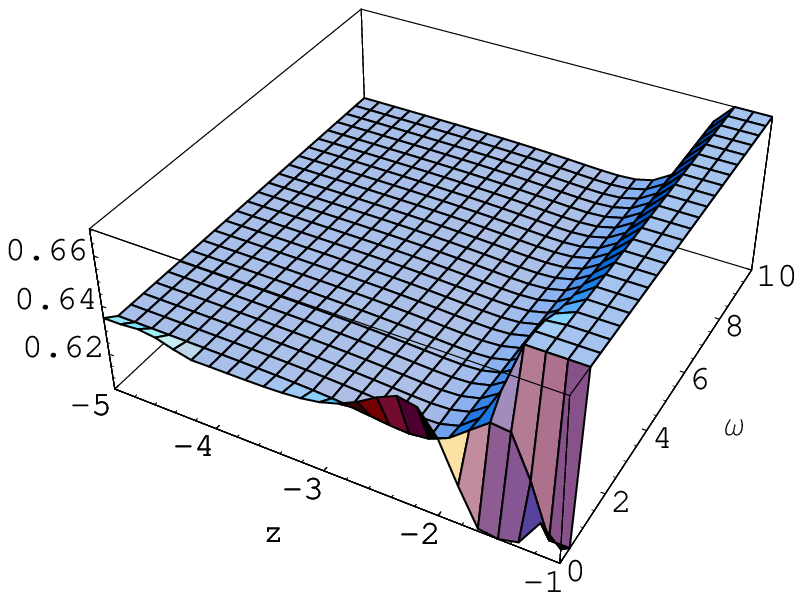,width=0.215\linewidth}\\
18&\epsfig{file=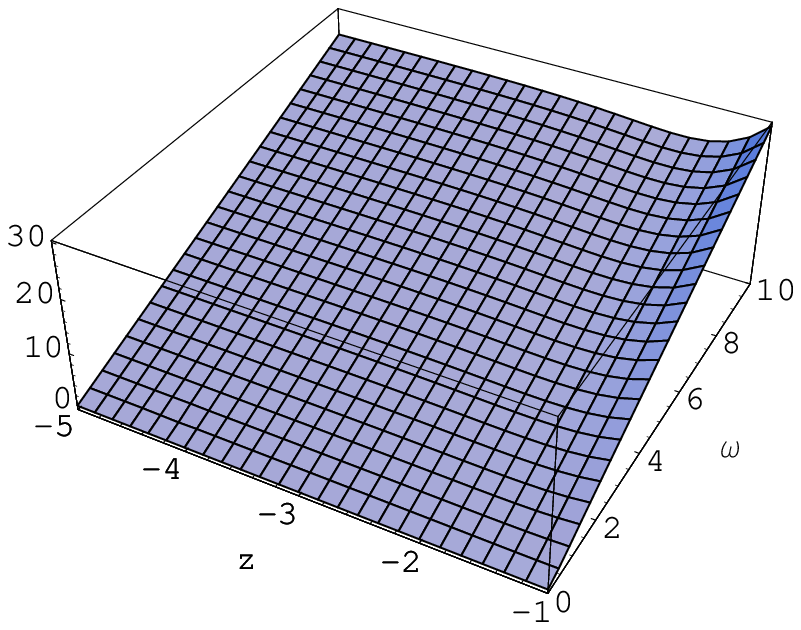,width=0.215\linewidth}
&\epsfig{file=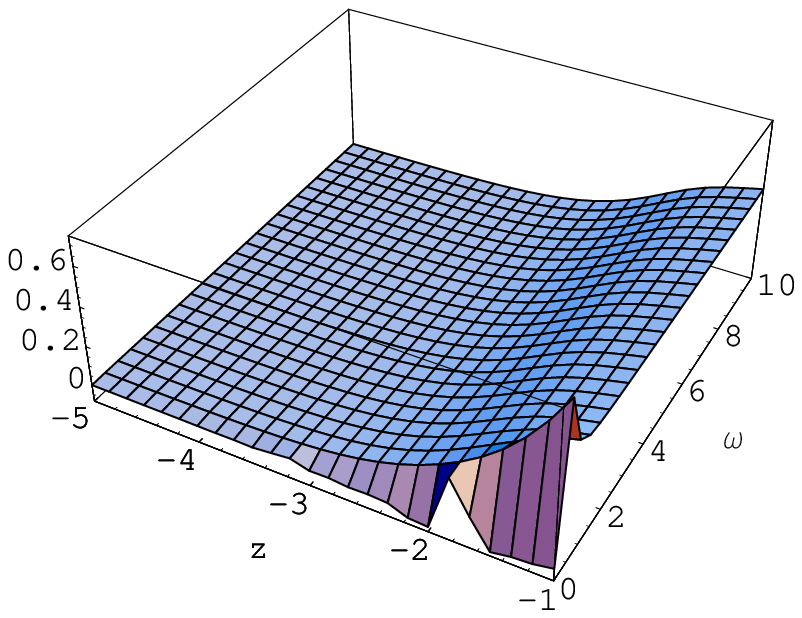,width=0.215\linewidth}
&\epsfig{file=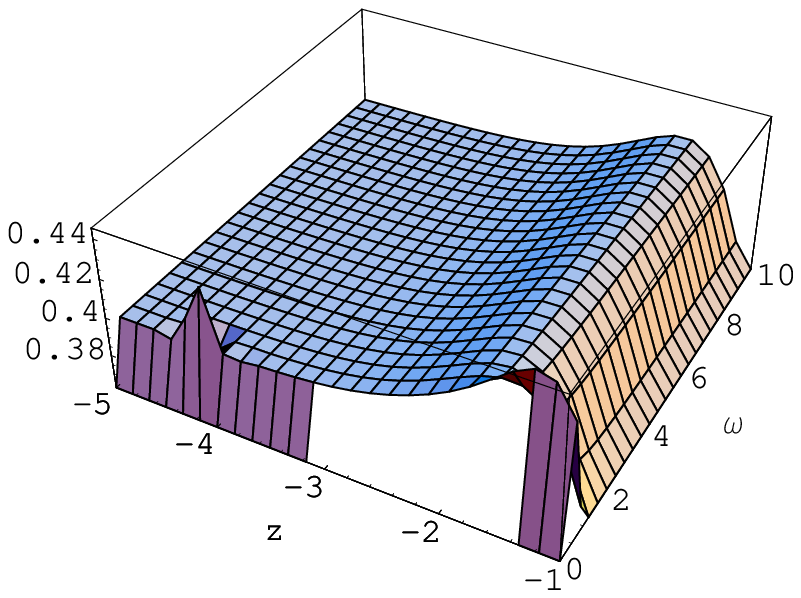,width=0.215\linewidth}
&\epsfig{file=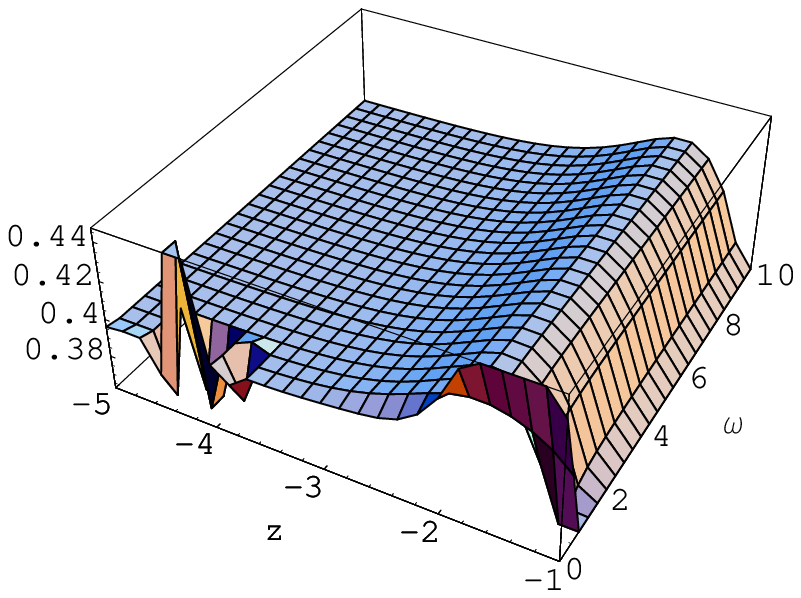,width=0.215\linewidth}
\end{array}$\\
\mbox{Figures \textbf{13-18} represent the dispersion relations
(related to the velocity com-}\\
\mbox{ponents given by Eq.(\ref{u2})) in the neighborhood of the
pair production region}\\
\mbox{towards the event horizon.}
\\
$\begin{array}{ccccc}
&A&B&C&D\\
19&\epsfig{file=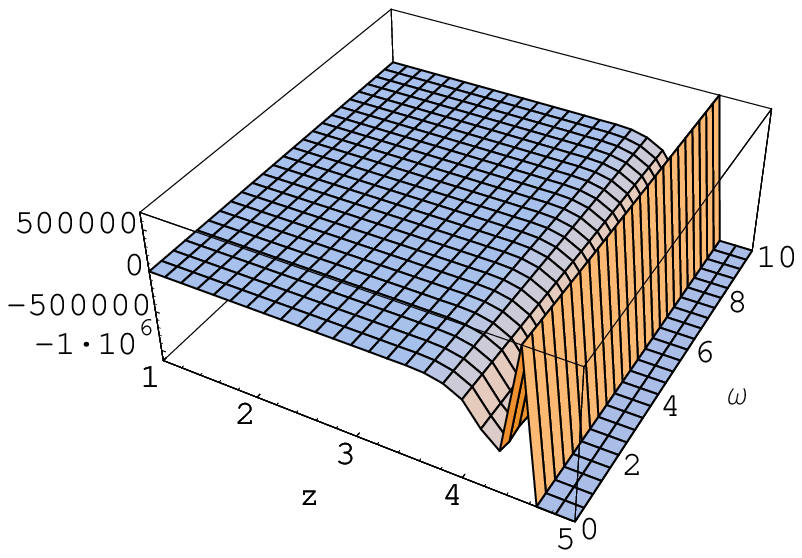,width=0.215\linewidth}
&\epsfig{file=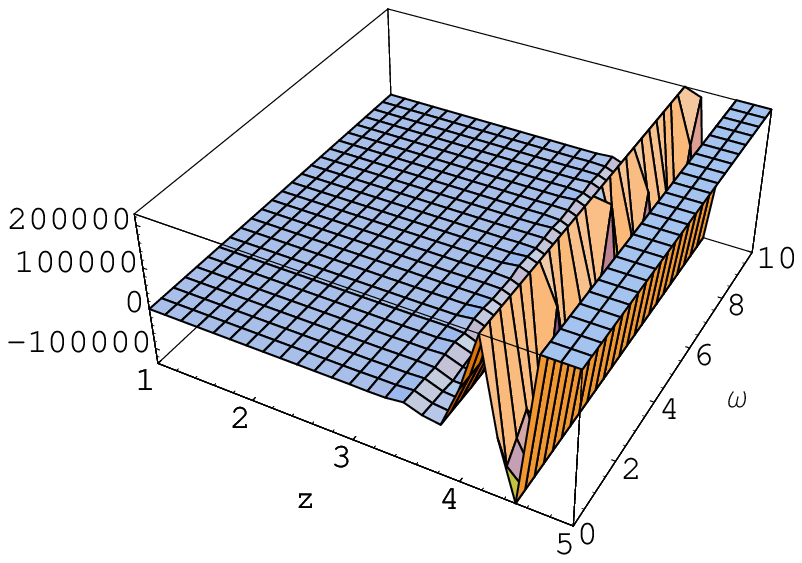,width=0.215\linewidth}
&\epsfig{file=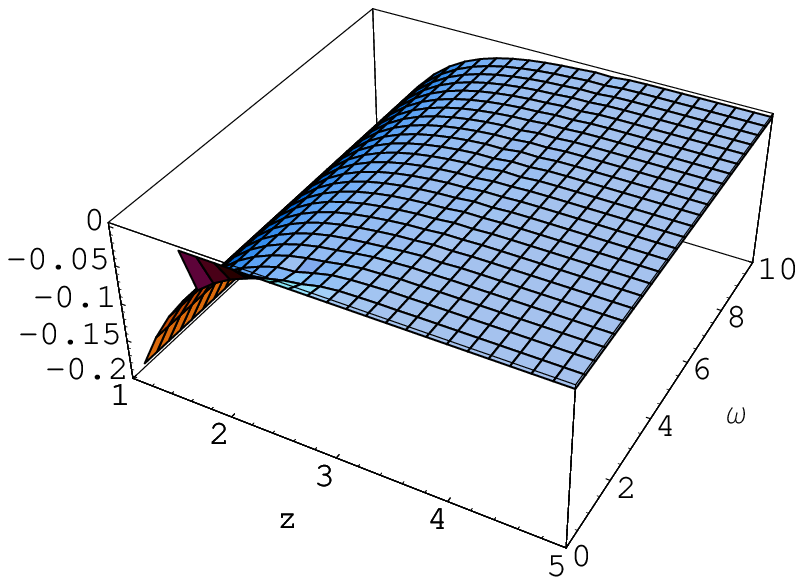,width=0.215\linewidth}
&\epsfig{file=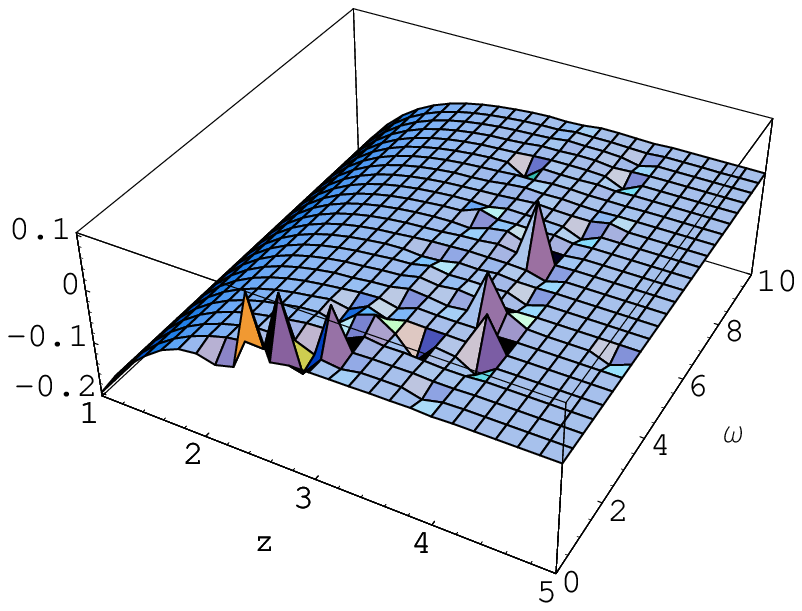,width=0.215\linewidth}\\
20&\epsfig{file=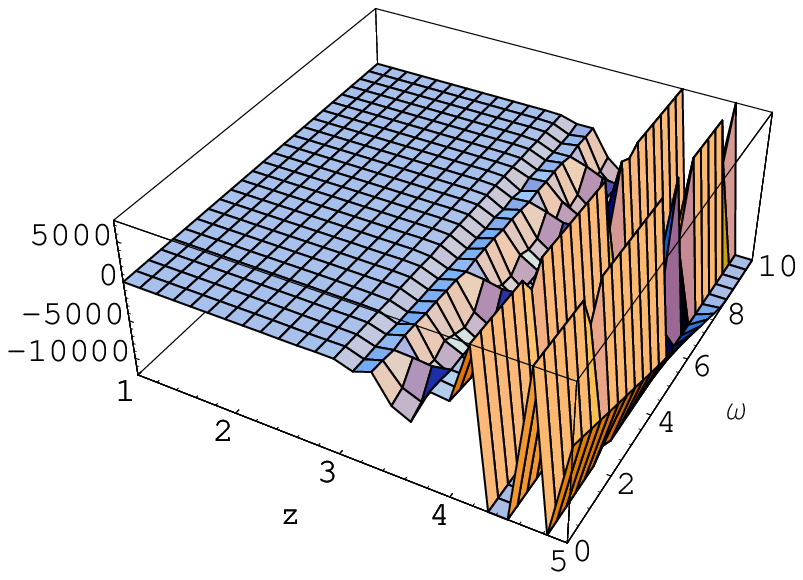,width=0.215\linewidth}
&\epsfig{file=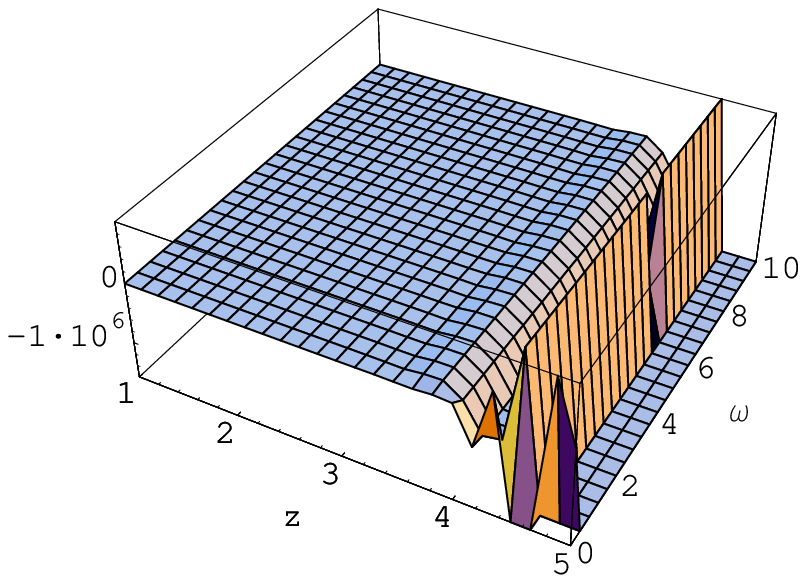,width=0.215\linewidth}
&\epsfig{file=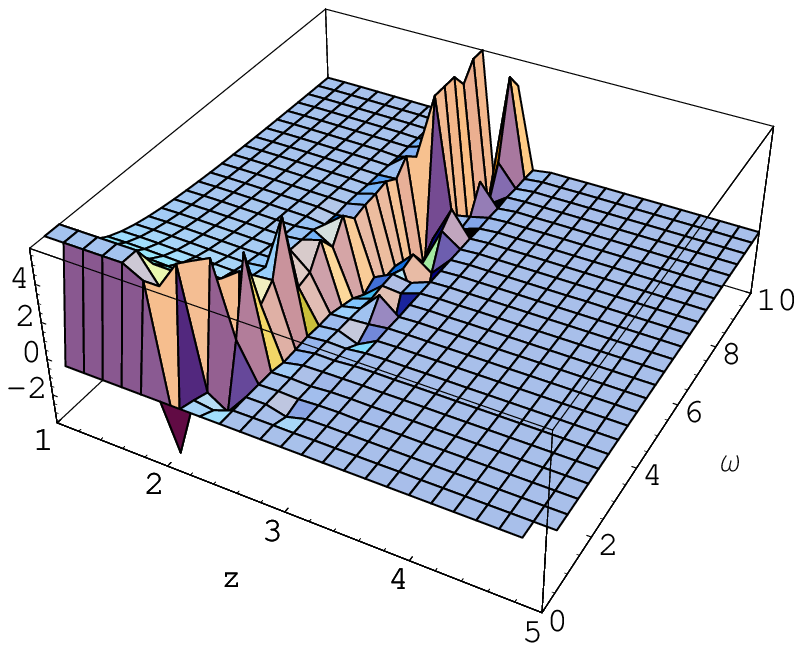,width=0.215\linewidth}
&\epsfig{file=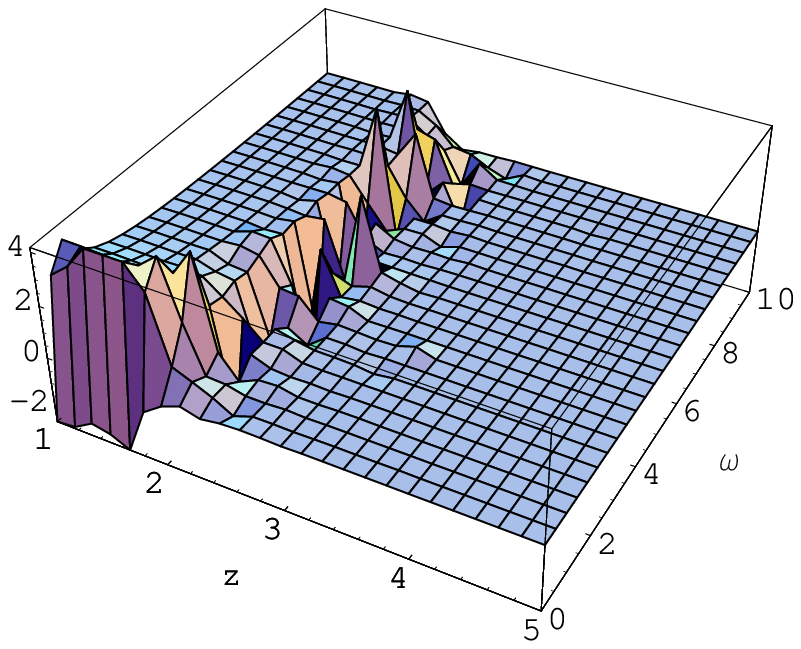,width=0.215\linewidth}\\
21&\epsfig{file=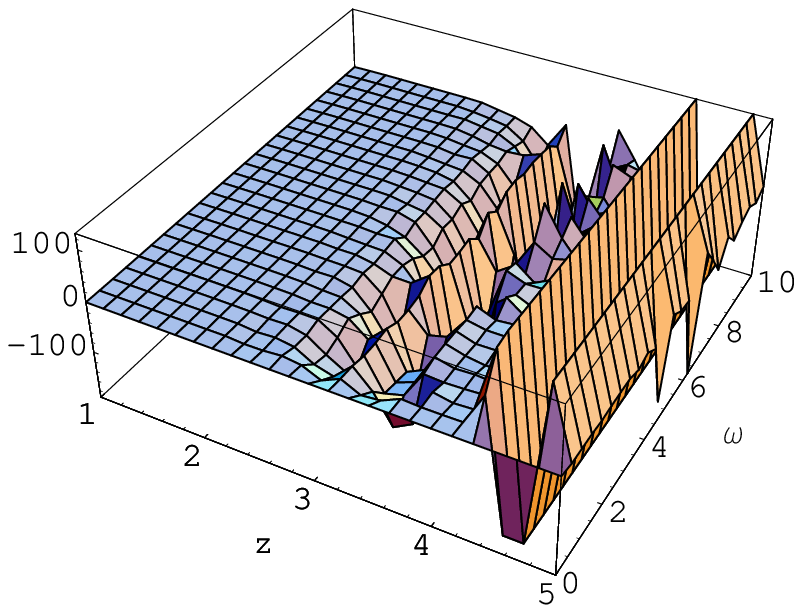,width=0.215\linewidth}
&\epsfig{file=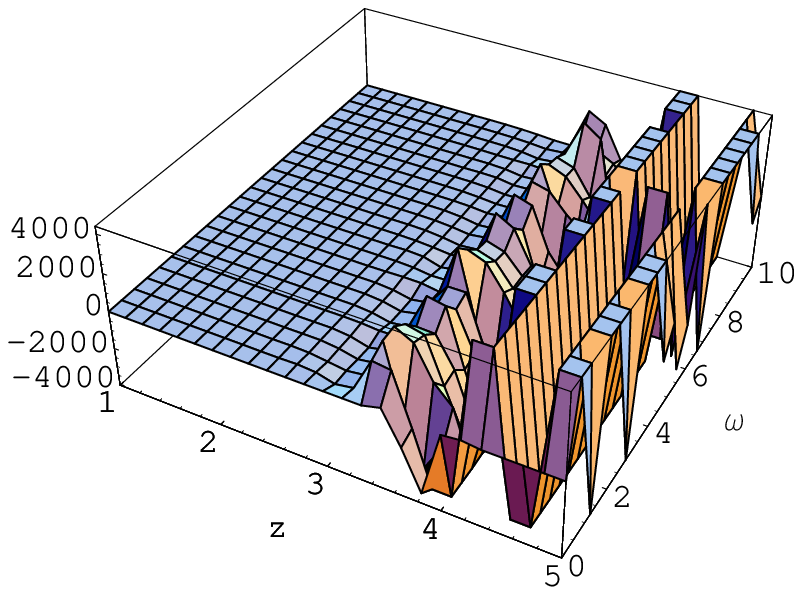,width=0.215\linewidth}
&\epsfig{file=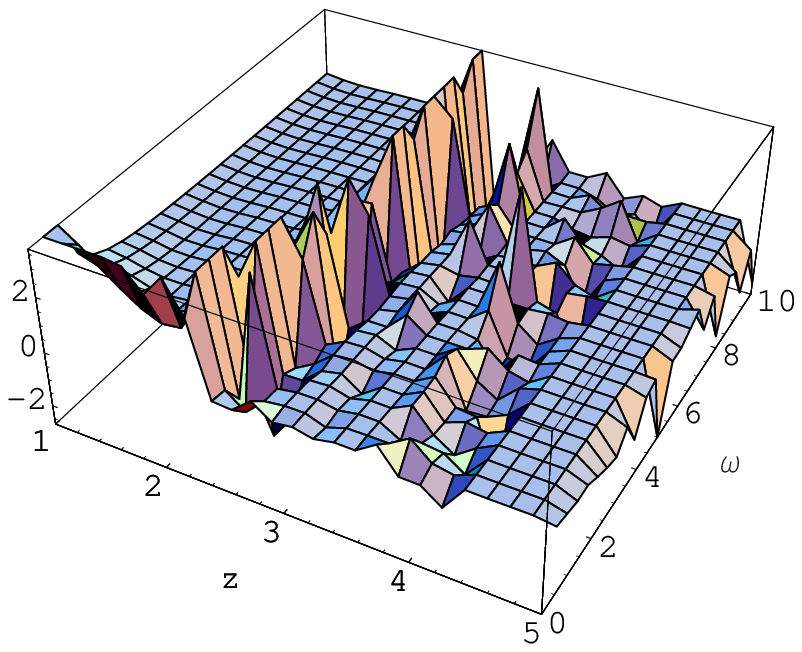,width=0.215\linewidth}
&\epsfig{file=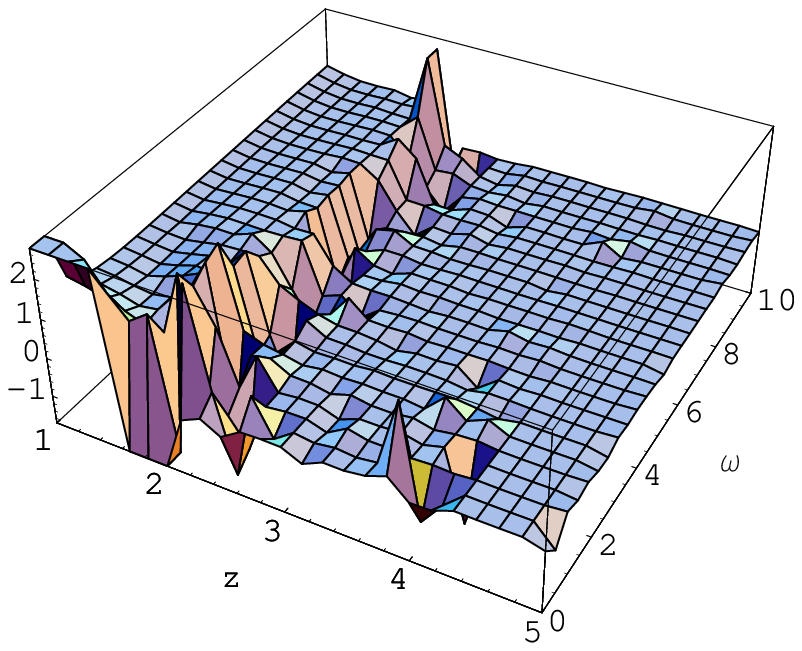,width=0.215\linewidth}\\
22&\epsfig{file=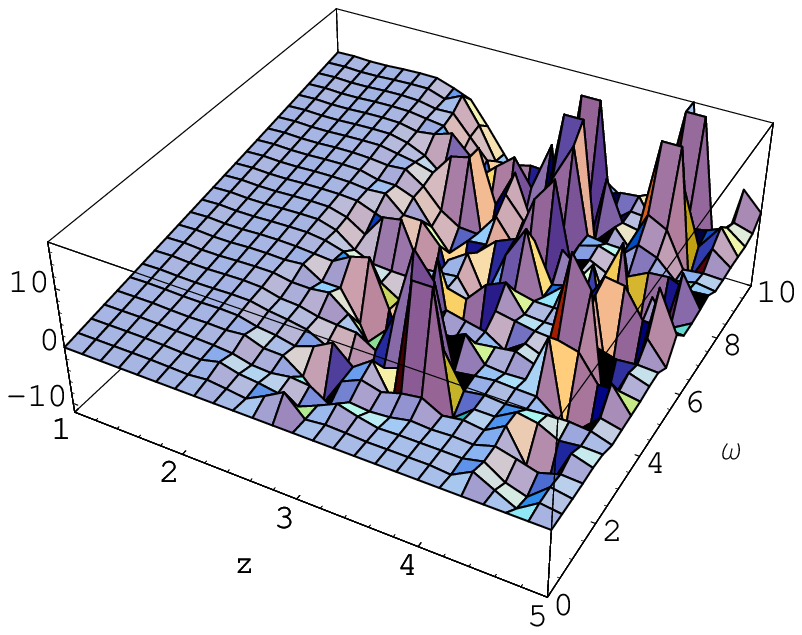,width=0.215\linewidth}
&\epsfig{file=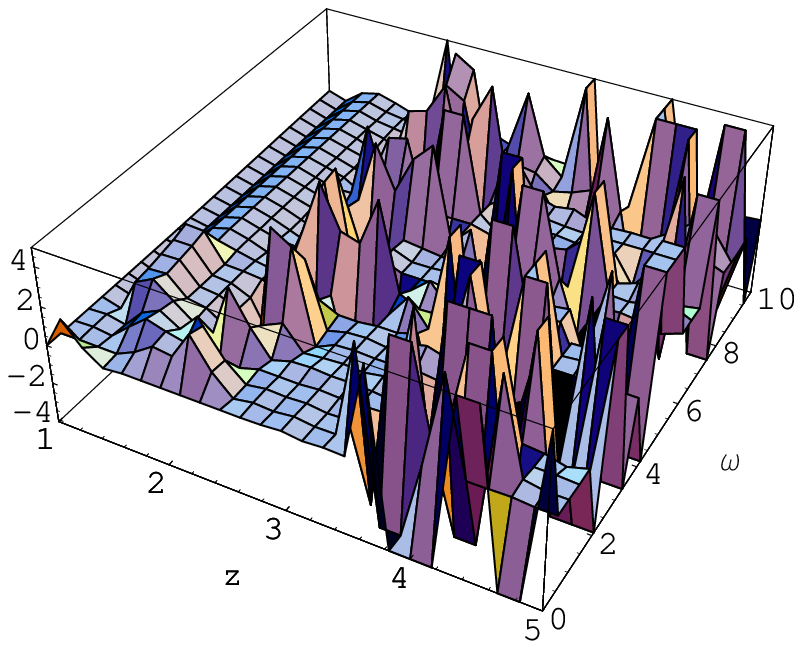,width=0.215\linewidth}
&\epsfig{file=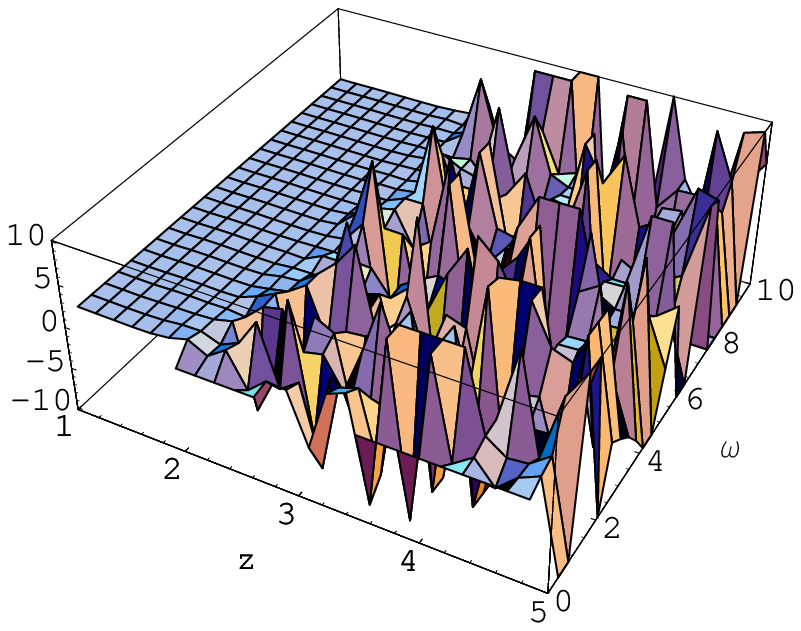,width=0.215\linewidth}
&\epsfig{file=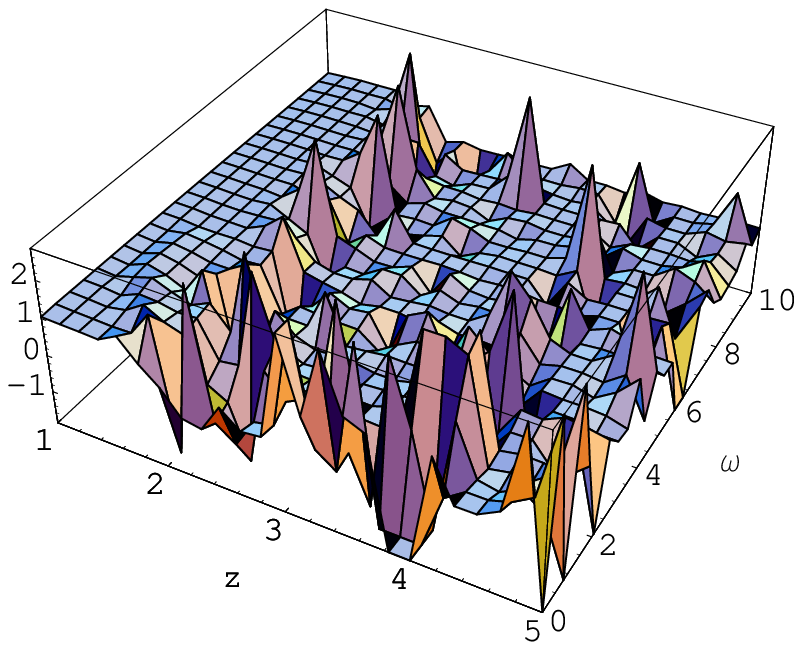,width=0.215\linewidth}\\
23&\epsfig{file=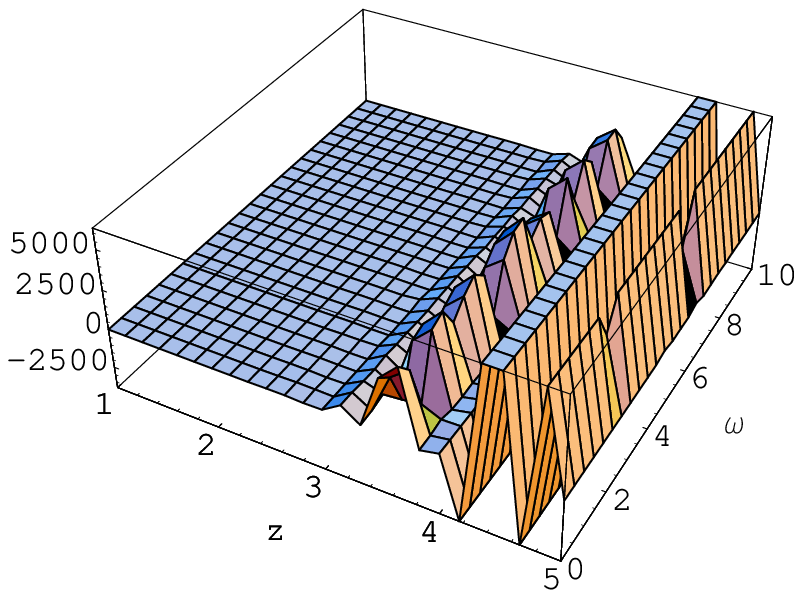,width=0.215\linewidth}
&\epsfig{file=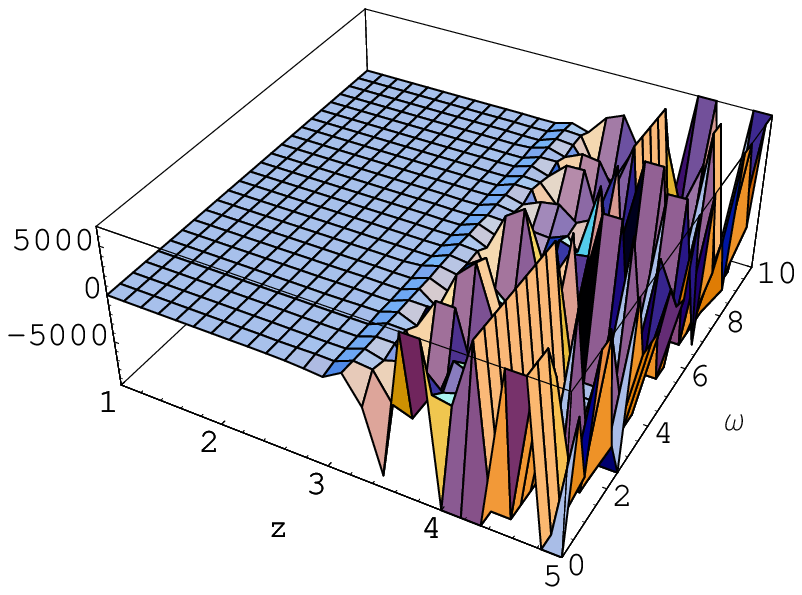,width=0.215\linewidth}
&\epsfig{file=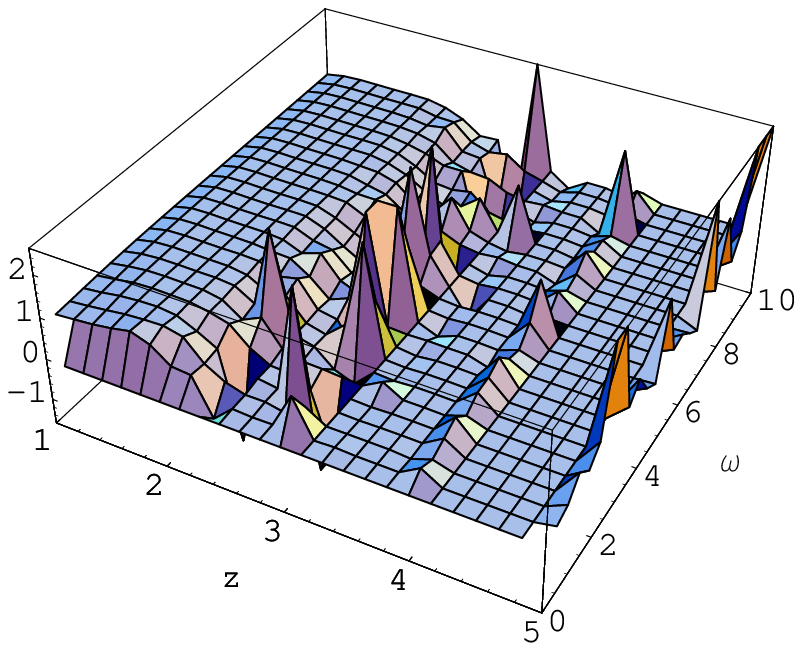,width=0.215\linewidth}
&\epsfig{file=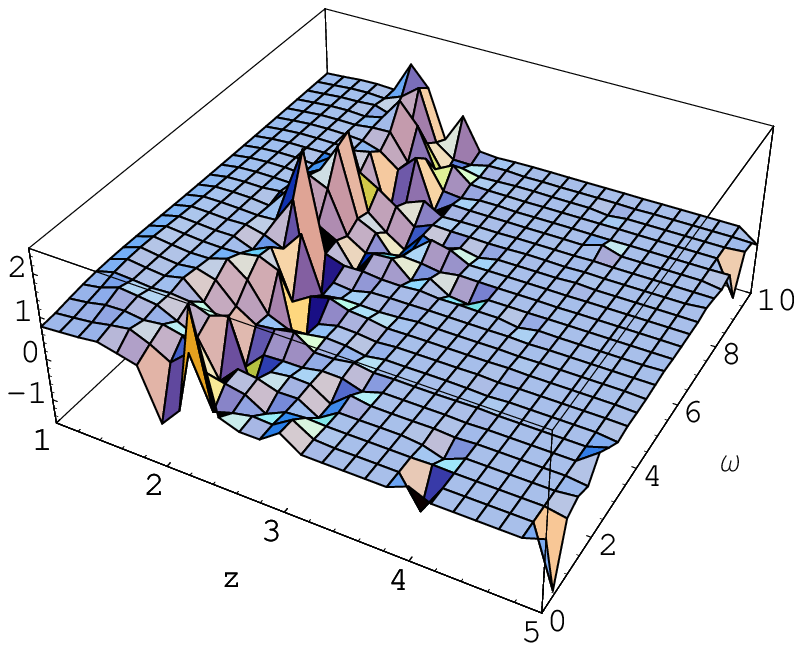,width=0.215\linewidth}\\
24&\epsfig{file=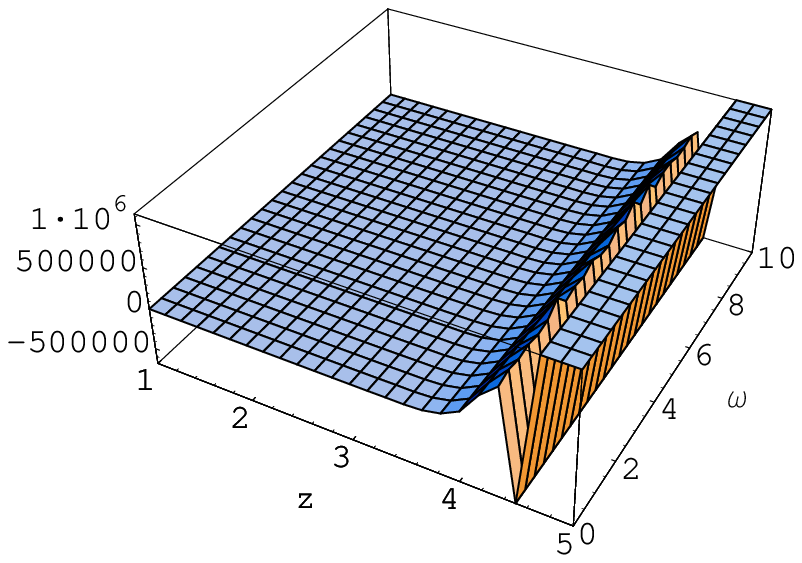,width=0.215\linewidth}
&\epsfig{file=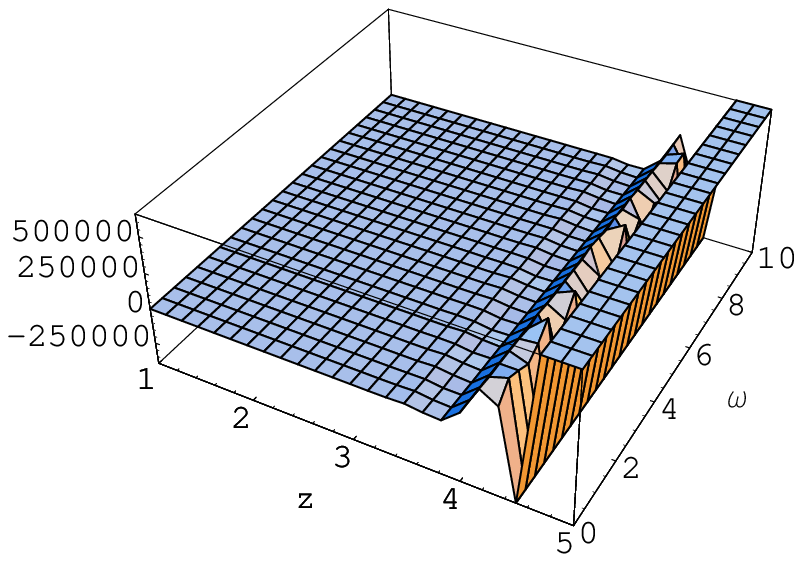,width=0.215\linewidth}
&\epsfig{file=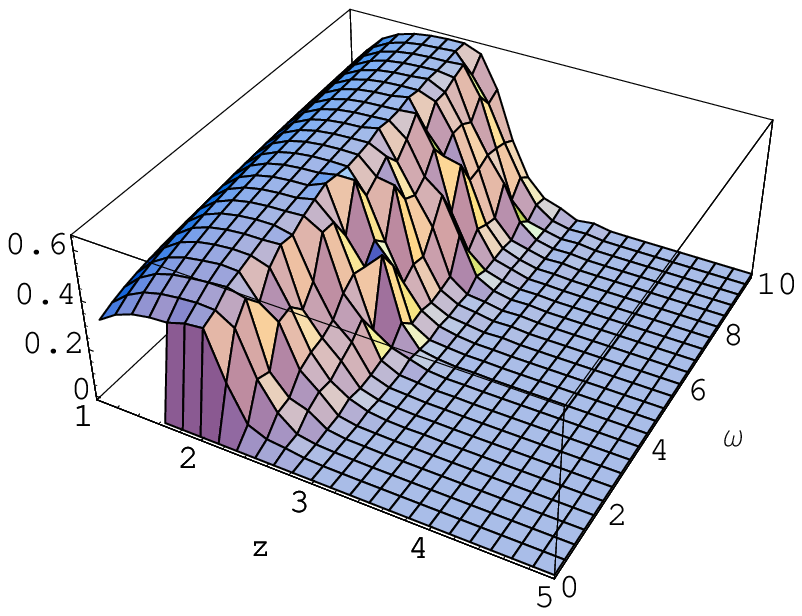,width=0.215\linewidth}
&\epsfig{file=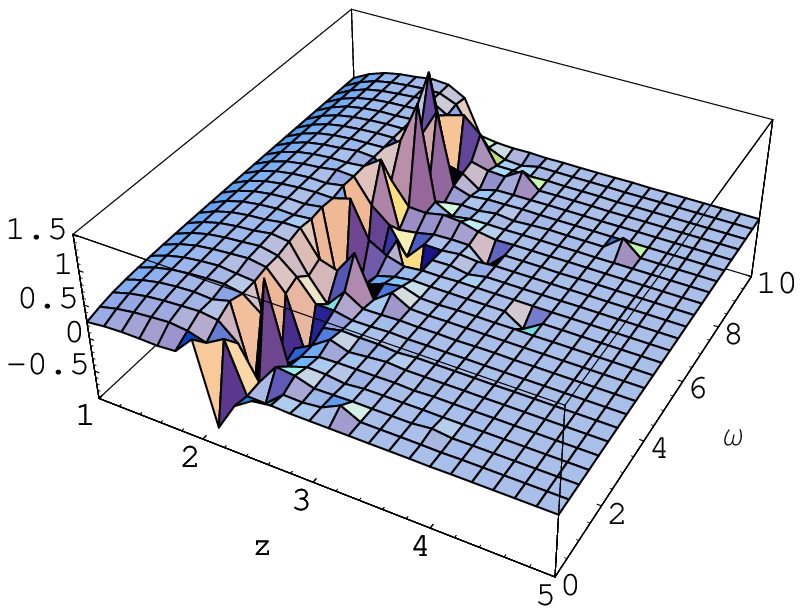,width=0.215\linewidth}
\end{array}$
\mbox{Figures \textbf{19-24} show the dispersion relations
(related to the velocity compo-}\\
\mbox{nents given by Eq.(\ref{u2})) in the neighborhood of the
pair production region}\\
\mbox{towards the outer end of the magnetosphere.}
\newpage
\end{document}